\documentclass[rmp,aps,eqsecnum,letterpaper,twocolumn,showpacs,superscriptaddress,amsfonts,amssymb]{revtex4}

\usepackage{amsmath}
\usepackage{color}

\newcommand{\mc}[1]{{\mathcal #1}}

\newcommand{\mb}[1]{{\mathbb #1}}

\newcommand{\<}{\langle}
\renewcommand{\>}{\rangle}

\renewcommand{\epsilon}{\varepsilon}

\renewcommand{\phi}{\varphi}

\begin{document}

\title{Macroscopic fluctuation theory}

\author{Lorenzo Bertini}
\author{Alberto De Sole}
\affiliation{
Dipartimento di Matematica, Universit\`a di Roma La Sapienza
P.le A.\ Moro 2, 00185 Roma, Italy
}
\author{Davide Gabrielli}
\affiliation{
DISIM, Universit\`a dell'Aquila   67100 Coppito, L'Aquila, Italy
}
\author{Giovanni Jona-Lasinio}
\affiliation{
Dipartimento di Fisica and INFN, Universit\`a di Roma La Sapienza
P.le A.\ Moro 2, 00185 Roma, Italy
}
\author{Claudio Landim}
\affiliation{
IMPA, Estrada Dona Castorina 110, J. Botanico, 22460 Rio
de Janeiro, Brazil}
\affiliation{
CNRS UPRES--A 6085, Universit\'e de Rouen,
76128 Mont--Saint--Aignan Cedex, France
}

\begin{abstract}
  Stationary non-equilibrium states describe steady flows through
  macroscopic systems. Although they represent the simplest
  generalization of equilibrium states, they exhibit a variety of new
  phenomena.  Within a statistical mechanics approach, these states
  have been the subject of several theoretical investigations, both
  analytic and numerical.  The macroscopic fluctuation theory, based
  on a formula for the probability of joint space-time fluctuations of
  thermodynamic variables and currents, provides a unified macroscopic
  treatment of such states for driven diffusive systems.  We give
  a detailed review of this theory including its main predictions and
  most relevant applications.
\end{abstract}

\date{\today}


\maketitle
\tableofcontents

\section{Introduction}
\label{introduction}

Far from equilibrium behavior is ubiquitous.
Indeed, most of the processes that characterize energy flow occur far
from equilibrium; so do typical biological phenomena, and
significant processes in molecules, solids, earth sciences,
astrophysics.
Classical thermodynamics does not cover such processes.
It is a phenomenological theory which deals with states of matter
which either do not change in time (equilibrium) or change very slowly
so that they can be described by a sequence of equilibrium states.

For systems out of equilibrium it does not exist yet a macroscopic
description of a scope comparable with equilibrium thermodynamics. In
non-equilibrium one has to cope with a variety of phenomena much
greater than in equilibrium. From a conceptual point of view the
non-equilibrium situations closest to equilibrium are the stationary
non-equilibrium states which describe a steady flow through some
system. Simple examples are the heat flow in an iron rod whose
endpoints are thermostated at different temperatures or the
stationary flow of electrical current in a given potential difference.
For such states the fluctuations exhibit novel and rich features with
respect to the equilibrium situation. For example, as experimentally
observed \cite{DKS}, the space correlations of the density extend to
macroscopic distances.

Previous formulations of non-equilibrium thermodynamics, notably
Onsager's theory \cite{ons,onsmach}, mostly refer to situations near
equilibrium where some kind of expansion can be made.  Over the last ten
years, a general approach to non-equilibrium diffusive systems known as
\emph{Macroscopic Fluctuation Theory} (MFT) \cite{BDGJL02,BDGJL-sips,D},
making some progress in
far from equilibrium processes and improving on near equilibrium
linear approximations, has been developed.  This theory has been
inspired by stochastic models of interacting particles (stochastic
lattice gases). It is based on the study of rare fluctuations of
macroscopic variables in stationary states and leads to a consistent
definition of non-equilibrium thermodynamic functionals as well as to
significant new results and predictions.

The MFT can be seen as the next stage
beyond Onsager theory which postulates simplified
evolution equations.
In particular, in Onsager theory the space dependence is neglected and
the time derivative of thermodynamic variables are directly identified
with the associated currents.  The currents are assumed to be
proportional to the thermodynamic forces that are identified with the
derivatives of the equilibrium entropy with respect to the
thermodynamic variables.  The entropy is expanded around an
equilibrium (maximum) value up to second order leading to linear
evolution equations. Within Onsager theory the fluctuations are
modeled by Gaussian processes.
The near equilibrium theory has been developed further by Kubo
\cite{MR1295243}.

With respect to Onsager theory, the MFT removes two restrictions.  On
one hand the systems considered may admit nonlinear hydrodynamic
equations.  In the second place the external driving, like the
potential difference, is not assumed small so that stationary states
far from equilibrium are possible. In general, the fluctuations are
not Gaussian.  The main source of new phenomena is the non-linearity
of the underlying evolution equations.

In the context of driven diffusive systems, characterized by an
applied external field and contact with boundary reservoirs, the
MFT allows to define a non-equilibrium
functional which plays a role analogous to the entropy in the Onsager
theory.  The current can be expressed as the sum of two
terms. The first is linear in the thermodynamic force, here identified
with the derivative of this functional, while the second, absent in
equilibrium, plays the role of an effective field and is
orthogonal to the thermodynamic force.

\medskip
In  equilibrium statistical mechanics, the connection
between thermodynamic functionals and fluctuations is provided by
the Einstein theory \cite{E} of equilibrium fluctuations.
Consider a system in contact with an environment; a fluctuation is a
deviation of a thermodynamic variable, e.g.\ the density, from its
equilibrium value.  In the notation of \cite{LL}, the probability of a
fluctuation is given by
\begin{equation}
\label{e}
P \asymp e^{- {\frac {R_\mathrm{min}} {\kappa T_0}}},
\end{equation}
where $\kappa$ is the Boltzmann constant and
\begin{equation}
\label{e2}
R_\mathrm{min} = \Delta U - T_0 \Delta S + P_0 \Delta V
\end{equation}
is the \emph{minimal work} necessary to produce the fluctuation with a
reversible transformation. $\Delta U$, $\Delta S$, $\Delta V$ are the
corresponding variations of energy, entropy, volume, and $T_0$, $P_0$
are the temperature and pressure of the environment.  The exponent
$R_\mathrm{min}$ depends both on the environment and the state of the
system and, with the opposite sign, is equal to the variation of the
\emph{availability}, see e.g.\ \cite{MR0106033}. The Boltzmann-Einstein formula
\eqref{e} is, we believe, the first example in physics of a large
deviation estimate as it is called in modern probabilistic
language. It is derived simply by inverting Boltzmann relationship
between entropy and probability.  In a non-equilibrium situation, like
the case of a system in contact with reservoirs, we may expect a
more complex entanglement between the variables describing the system
and those related to the environment so that it is unlikely that
quantities like $U, S, ... $ can be simply defined. However, as we
shall see, the MFT allows to establish a
formula like \eqref{e}, thus generalizing the notion of availability.

\medskip 
Dynamics plays a major role out of equilibrium. In fact what
distinguishes non-equilibrium is the presence of currents flowing
through the system which have to be considered together with the usual
thermodynamic variables.  
The systems considered by the MFT are connected to several reservoirs
(the environment), possibly distributed continuously on the boundary
surface, characterized by their chemical potentials.  The reservoirs
are assumed to be much larger than the system so that their state will
be essentially constant in time. When the system is put in contact
with the environment, after an initial stage we expect that a
description in terms of diffusive processes may apply for a wide class
of microscopic dynamics. We admit also external fields such that
linear response is valid.  
On the basis of a \emph{local equilibrium} assumption, on macroscopic
scale it is possible to define thermodynamic variables like the
density of mass, electric charge, energy and the corresponding currents, 
which vary smoothly on the same scale. 
Microscopically, this implies that the system reaches a local
equilibrium in a time which is short compared to the times typical of
macroscopic evolution.  So what characterizes situations in which this
description can be applied is a separation of scales both in space
and time.  Furthermore, we assume that the system is
\emph{Markovian}. Namely, the currents at time $t$ depend on the
thermodynamic variables at the same time $t$.  These assumptions are
clearly discussed in \cite{callen,fitts} and are behind the near
equilibrium theories.

\medskip
The proposed theory is based on the following formula for the
probability of joint space-time fluctuations, at constant temperature
$T_0$, of thermodynamic variables and currents
\begin{equation}
  \label{1ff}
  \mb P \asymp \exp\Big\{ - {\frac {1} {\kappa T_0}}  \frac 14 \int\!dt\!\int\!dx \,
  \big(j -J(\rho)\big) \cdot \chi(\rho)^{-1}  \big(j -J(\rho)\big)  \Big\},
\end{equation}
where $\rho$ is the thermodynamic variable, e.g., the local density,
$j$ is the actual value of the current, which is connected to $\rho$
by the continuity equation $\partial_t \rho +\nabla \cdot j=0$, while
$J(\rho)$ is the hydrodynamic current for the given value of $\rho$,
and $\chi$ is the mobility. 
Formula \eqref{1ff} depends only on the relationship between the
thermodynamic variable $\rho$ and the associated hydrodynamic current
$J(\rho)$, that is on the constitutive equations of the system.  With
slight modifications, \eqref{1ff} can be applied also to fluctuations
of the energy as in the heat conduction case.  Its structure and
interpretation is otherwise universal.

According to the reductionist point of view of statistical mechanics,
in the realm of classical physics, \eqref{1ff} should be derived
starting from molecules interacting with realistic forces and evolving
with Newtonian dynamics.  This is beyond the reach of present day
mathematical tools and much simpler models have to be adopted in the
reasonable hope that some essential features are adequately captured.
Formula \eqref{1ff} can be proven, as discussed in
Section~\ref{ld-slg}, for a wide class of stochastic interacting
particle systems.

The exponent on the right hand side of
\eqref{1ff} is proportional to the energy dissipated  by 
the extra current $j-J(\rho)$. 
This can be understood
relying on an active interpretation of the fluctuations.
Namely, given a fluctuation,
we perturb the original system by adding a (deterministic) external field
for which the prescribed fluctuation becomes the trajectory followed by the system.
For the fluctuation $(\rho,j)$ such external field is $F=\chi(\rho)^{-1}  \big(j -J(\rho)\big)$.
Hence, the exponent in the right hand side of \eqref{1ff} is
proportional to $\int\!dt\int\!dx\, F\cdot \chi(\rho)F$.
In the case of an electric
circuit $\chi^{-1}$ is the resistance, $F$ is the electric field,
and the above integral is the energy dissipated by the extra current $j-J(\rho)$
according to Ohm's law.

Within the scheme of fluctuating hydrodynamics \cite{hh,ll2,ippo}, it
is possible to provide an alternative justification of \eqref{1ff}.
Namely, one postulates $j=J(\rho)+\alpha$, where, conditionally on
$\rho$, $\alpha$ is a Gaussian random term with variance
$$
\langle\alpha_i(t,x),\alpha_j(t',x')\rangle
=
2\kappa T_0\,
\chi_{ij}(\rho)\,
\delta(t-t')\delta(x-x')
$$
where the indices $i,j$ label the space directions.
Formula \eqref{1ff} can be inferred from the Gaussian distribution 
of the stochastic field $\alpha$.
However, mathematically the random noise $\alpha$ is singular and
induces, through the non-linear terms in the equation, ultraviolet
divergences that should be properly renormalized.  For Landau--Ginzburg
models with constant mobility, this renormalization has been carried out in
\cite{dedpe}.  
In our setting ultraviolet divergences are not relevant as 
formula \eqref{1ff} is an asymptotic expression for fluctuations of macroscopic
variables defined through coarse--graining.
The role of lattice gases, in particular exactly solvable models, is
important similarly to what happened in the theory of
critical phenomena where special models have provided explicit
illustrations of the more general but often heuristic renormalization
group calculations.

\medskip
As the Boltzmann-Einstein formula \eqref{e}, the \emph{fundamental
formula} \eqref{1ff} provides a quantitative relation between the
probabilities computed in the microscopic ensemble and macroscopic
variables.
With respect to the theory of equilibrium thermodynamic fluctuations, 
formula \eqref{1ff} is an important generalization that holds both in
equilibrium and non-equilibrium.  
Note the difference in notation between formulas \eqref{e} and
\eqref{1ff}: in \eqref{e} $P$ represents the ensemble over the
space of configurations, while in \eqref{1ff} $\mb P$ represents the
ensemble over space-time trajectories.  We will use this same
distinction throughout the paper, both for probabilities and
expectation values.
The Boltzmann-Einstein formula
\eqref{e} and its non-equilibrium analogue will be derived from
\eqref{1ff}.

\medskip
The fundamental formula \eqref{1ff} leads to the prediction of rather
surprising properties of diffusive systems such as the existence of
phase transitions not permitted in equilibrium, the possibility of
states spontaneously breaking the time translation invariance in the
fluctuations of the current, the universality of the cumulants of the
current. Furthermore it predicts generic long range correlations in
stationary non-equilibrium states.

\medskip
The behavior of the fundamental formula \eqref{1ff} under time
reversal plays a crucial role in the MFT. As well known, in the near
equilibrium Onsager theory the symmetry of the transport coefficients
is deduced form a statistical form of time reversal invariance of the
underlying microscopic dynamics \cite{ons}.  Of course in
presence of inhomogeneous boundary conditions and/or external fields,
this time reversal invariance is lost. However in the systems
considered, we assume that given a microscopic path it is
possible, by adding a suitable field, to modify the dynamics in such a
way that the corresponding evolution is given by the time reversed
path.  At the level of stationary ensembles over space-time
trajectories, the time reversed ensemble is then defined by assigning
to a backward path the probability of the forward path under the
original ensemble.  In particular, the stationary macroscopic currents
are inverted under the time reversal operation.  For stochastic
lattice gases the microscopic dynamics is Markovian and the previous
definition can be directly implemented.

The splitting of the hydrodynamic current into two
orthogonal terms discussed before is derived as a simple consequence
of the transformation properties of \eqref{1ff} under time
reversal.  The even part of the current is connected with the work of
thermodynamic forces active within the system in the relaxation or
creation of a non-stationary state while the odd part is connected
with the dissipation necessary to keep the system in a non-equilibrium
state.

\bigskip
For the convenience of the reader we summarize the main achievements
of the MFT.

\medskip
From the fundamental formula \eqref{1ff} for space-time fluctuations
we derive a dynamical variational principle which expresses the
probability of density fluctuations of the stationary ensemble.  This
leads naturally to the definition of a thermodynamic functional called
in the following the \emph{quasi-potential} and denoted $V(\rho)$. 
The argument $\rho$ represents generic thermodynamic variables like the
densities of the different types of matter composing the system.
The quasi-potential is the natural extension to
non-equilibrium of the availability of classical thermodynamics.
In the context of finite dimensional diffusion processes,
the quasi potential was first introduced by Freidlin and Wentzell, \cite{MR2953753}.
A related analysis, motivated by applications to optics, can be found in \cite{graham}.

The quasi-potential satisfies an infinite dimensional
Hamilton-Jacobi equation which is equivalent to the splitting of the
current.
The density correlation functions can be computed by
expanding the Hamilton-Jacobi equation around the stationary density
and exhibit, generically, long range behavior.
In non-equilibrium the quasi-potential $V$ may have singularities not
permitted in equilibrium.  These singularities do occur in a model with
external field and inhomogeneous boundary conditions, \cite{lpt}.

\medskip
The splitting of the current is relevant in the analysis of
non-equilibrium thermodynamic transformations. 
Consider a slow (quasi static) transformation leading from a
stationary state to another one.  
Since to maintain a non-equilibrium stationary state it is necessary
to dissipate a positive amount of energy per unit  time, 
the work associated to the odd part of the current will
diverge when the time diverges. 
The equation expressing the energy balance between the system and the
environment becomes then meaningless. To obtain equations in finite terms
one can subtract the divergent part and define a renormalized work. It
turns out that the renormalized work satisfies a Clausius type
inequality with respect to which quasi-static transformations are
optimal. The idea of renormalizing the work involved in a
non-equilibrium transformation goes back to \cite{op}.

\medskip
The MFT allows a detailed mathematical
description of quasi-static thermodynamic transformations that in
textbooks are discussed only in words.
This is achieved through an expansion of the energy balance and of the
macroscopic evolution (hydrodynamic) equations in terms of the inverse
duration of the transformation. In this expansion the diverging terms
cancel and we obtain new relations among finite quantities which in
principle can be tested experimentally.

\medskip 
One of the most interesting topic in the MFT is provided by current
fluctuations.  From the fundamental formula \eqref{1ff} it is possible
to derive the rate function describing the behavior of the average
current over a long time interval $T$, namely
\begin{equation}
  \label{mphi}
  \mathbb{P} \asymp \exp\big\{  - \beta  T \Phi(J) \big\},
\end{equation}
where $\mathbb{P}$ is the probability of the fluctuation $J$ of the averaged
current and $\beta=1/(\kappa T_0)$. The temperature of the environment
$T_0$ (not to be confused with the time $T$) will be mostly
considered constant.
The functional $\Phi$, first introduced in \cite{noiprlcurrent}, is
defined by a variational principle and it is a genuine (convex)
thermodynamic functional.
As pointed out in the same paper, the singularities of $\Phi$ correspond
to dynamical phase transitions.  For some models it has then been shown
\cite{MR2227084,bdcurrpt} that for suitable values of $J$ these
transitions corresponds to a spontaneous breaking of the time
translation invariance.

\medskip

In the general context of non-equilibrium processes, an important role
is played by the so-called \emph{Fluctuation Theorems} \cite{Jar,
  Crooks, MR1349772, MR1705591, hs, MR1630333, es, MR1705590}.  This
topic will appear only marginally in this paper and we refer the
interested reader to the reviews \cite{Gal, MR2581608, MR2673931}.
There are other relevant topics not covered as each of them would
require an extended review. Among these topics we mention the
so-called \emph{stochastic thermodynamics} \cite{seifert}, the general
approach to non--equilibrium developed in \cite{ottinger}, path
integral approaches \cite{JSP}, algebraic techniques for microscopic
dynamics \cite{gunter}, the analysis of rare fluctuations in finite
dimensional dynamical systems, with the related discussion on current
fluctuations \cite{maescurr}, and in reaction--population systems
\cite{tau}.

\subsection*{Reader's guide}

The aim of this review is to present the MFT as an effective
macroscopic theory, providing a working knowledge of the theory rather
than a chronological exposition of its main results. 
For this reason, some computations are detailed when useful.

The general framework of driven diffusive systems 
and the basic principles are illustrated in Section
\ref{s:2}.

Part of the material in Section \ref{s:3}, which deals with
thermodynamic transformations between non--equilibrium states, is
presented here for the first time. We discuss quantitatively the
interplay between fluctuations and thermodynamics. This section is not
however used in the sequel and can be omitted on first reading.   

Section \ref{s:ft} contains the core of the MFT, deducing the
statistics of density and current from the fundamental formula. 

Section \ref{sec3} discusses  few models where the
quasi-potential can be computed (almost) explicitly. While we use the
terminology of underlying microscopic models, we emphasize that
the computations are macroscopic and require only the knowledge of the
transport coefficients.  

Section \ref{s:6} develops one of the most relevant consequences of
the MFT by deriving the statistics of the time averaged current. 

Section \ref{s:7} contains more specialized material related to
hyperbolic conservation laws. Although these are not diffusive
systems, the fluctuation formula can be obtained from the MFT by a
singular limit procedure. This section can be omitted on first
reading.

Finally, Section \ref{s:8} describes microscopic models of stochastic
lattice gases, showing how the general principles of the MFT can be
analytically derived. 
This Section can be read independently of the others and can be
anticipated if the reader wishes to do so.

\section{Basics of the macroscopic fluctuation theory}
\label{s:2}

We introduce the hydrodynamic description of out of equilibrium
driven diffusive systems which are characterized by conservation
laws. We then introduce the fundamental formula of the MFT and discuss
its behavior under time-reversal together with its main implications. 
We restrict to the case of a single conservation law, e.g., the
conservation of the mass.

\subsection{Hydrodynamic description}
\label{s:2.1}

 We denote by $\Lambda \subset \mb R^d$ the
bounded region occupied by the system, by $\partial \Lambda$ the
boundary of $\Lambda$, by $x$ the macroscopic space coordinates and by
$t$ the macroscopic time. The system is in contact with boundary
reservoirs, characterized by their chemical potential $\lambda (t,x)$,
and under the action of an external field $E(t,x)$.

At the macroscopic level the system is completely described
by the local density $\rho(t,x)$ and the local current $j(t,x)$. Their
evolution is given by the continuity equation together with the constitutive
equation which expresses the current as a function of the
density. Namely,
\begin{equation}
\label{2.1}
\begin{cases}
\partial_t \rho (t) + \nabla\cdot j (t) = 0,\\
j (t)= J(t,\rho(t)),
\end{cases}
\end{equation}
where we omit the explicit dependence on the space variable $x\in\Lambda$.
For driven diffusive systems the constitutive equation takes the form
\begin{equation}
\label{2.2}
J(t,\rho)  = - D(\rho) \nabla\rho + \chi(\rho) \, E(t),
\end{equation}
where the \emph{diffusion coefficient} $D(\rho)$ and the \emph{mobility}
$\chi(\rho)$ are $d\times d$ symmetric and positive definite matrices.
Equation \eqref{2.2} relies on the diffusive approximation and on the
linear response to the external field.
The evolution of the density is thus given  by the driven diffusive
equation
\begin{equation}
\label{r01}
\partial_t \rho (t) + \nabla\cdot \big( \chi(\rho)  E(t) \big)
= \nabla\cdot \big( D(\rho) \nabla\rho \big).
\end{equation}
We emphasize that the diffusion coefficient and the mobility do depend
on the value of the local density. Accordingly, equation \eqref{r01}
is nonlinear and this is the source of interesting phenomena.  In
contrast, the near equilibrium approximation can be obtained by
expanding $\rho$ around some (constant) equilibrium value
so that \eqref{r01} becomes linear.

The transport coefficients $D$ and $\chi$ are not arbitrary
matrices. The characterization of equilibrium states implies (see
Section~\ref{s:eq}) that they satisfy the local Einstein relation
\begin{equation}
\label{r29}
D(\rho) = \chi(\rho) \, f''(\rho),
\end{equation}
where $f$ is the equilibrium free energy per unit  volume.

Equations \eqref{2.1}--\eqref{2.2} have to be supplemented by the
appropriate boundary condition on $\partial\Lambda$ due to the
interaction with the external reservoirs. If $\lambda(t,x)$,
$x\in\partial \Lambda$, is the chemical potential of the external
reservoirs, the boundary condition reads
\begin{equation}
\label{2.3}
f'\big(\rho(t,x) \big) = \lambda(t,x), \qquad x\in\partial \Lambda.
\end{equation}
While  in the near equilibrium approximation the variation of $\lambda$ on
$\partial \Lambda$ is required to be small, we shall not restrict to
this case.

\medskip
One of the achievements of mathematical physics is the derivation of
the hydrodynamic equations \eqref{2.1}--\eqref{2.3} 
as laws of large numbers from an underlying microscopic stochastic
dynamics in the diffusive scaling limit \cite{ELS,ippo,MR1707314}. 
This means taking the limit of infinitely many degrees of freedom and
rescaling space and time keeping $x^2/t$ fixed.

\medskip
We now restrict the discussion to time-independent chemical potential
$\lambda(x)$ and external field $E(x)$. We denote by
$\bar\rho=\bar\rho_{\lambda, E}$ the stationary solution of
\eqref{r01},\eqref{2.3},
\begin{equation}
\label{05}
\begin{cases}
\nabla \cdot J(\bar\rho)= \nabla \cdot \big( -D(\bar\rho)
\nabla\bar\rho + \chi(\bar\rho) \, E  \big) = 0,  
\\
 f' (\bar\rho(x)) = \lambda (x), 
\qquad x\in\partial \Lambda.
\end{cases}
\end{equation}
We will assume that this stationary solution is unique.
The stationary density profile $\bar\rho$ is characterized by the
vanishing of the divergence of the associated current, $\nabla\cdot
J(\bar\rho)=0$.  A special situation is when the current itself
vanishes, $J(\bar\rho)=0$; if this is the case we say that the system
is in an equilibrium state.
Uniqueness to \eqref{05} can be proven in one dimension when the
external field $E$ is constant in space. 
It can also be proven in general, by a
perturbation argument, near equilibrium.
In the case of several conserved quantities, i.e.\ when $\rho$ is not
a scalar, uniqueness may fail.

Homogeneous equilibrium states correspond to the case in which the
external field vanishes and the chemical potential is constant. The
stationary solution is then constant and satisfies
$f'(\bar\rho_{\lambda,0}) =\lambda$.  Inhomogeneous equilibrium states
correspond to the case in which the external field is gradient,
$E=-\nabla U$, and it is possible to choose the arbitrary constant in
the definition of $U$ such that $U(x)=-\lambda(x)$,
$x\in\partial\Lambda$. 
By the Einstein relation \eqref{r29}, the stationary solution satisfies
$-f'\big(\bar\rho_{\lambda,E}(x)\big)=U(x)$ and the stationary current
vanishes, $J(\bar\rho_{\lambda,E})=0$.  Examples of inhomogeneous
equilibrium states in presence of an external field are provided by a
still atmosphere in the gravitational field or by sedimentation in a
centrifuge.

\subsection{Fundamental formula}

In the context of equilibrium systems, the Einstein theory of
thermodynamic fluctuations establishes a connection between the
thermodynamic functionals and the probability of observing a
fluctuation. The extension of this theory to non-equilibrium
stationary states is provided by the  fundamental formula \eqref{1ff},
describing the joint fluctuations of thermodynamic variables and
currents at the level of space-time paths.

Let $\mb P_{\rho_0}$ be the statistical ensemble on microscopic
trajectories such that at time $t=T_0$ the density profile is
$\rho_0$.  
Consider a path $(\rho,j)$ satisfying the continuity equation in
\eqref{2.1}, the boundary condition \eqref{2.3}, and $\rho(T_0) =
\rho_0$. The fundamental formula \eqref{1ff} can be written in detail as 
\begin{equation}
\label{r02}
\begin{split}
& \mb P_{\rho_0} \big( (\rho_\epsilon (t), j_\epsilon (t))
\approx (\rho (t), j(t))\,,\, t\in [T_0, T_1] \, \big) \\
& \quad \asymp \exp\big\{ - \epsilon^{-d} \,
\mathcal I_{[{T_0},{T_1}]}(\rho,  j) \big\}
\end{split}
\end{equation}
where the \emph{rate functional} $\mathcal I$ is 
\begin{equation}
\label{r05}
\begin{split}
& \mathcal I_{[{T_0},{T_1}]}(\rho,  j)
\\
&\quad=
\frac 14 \int_{{T_0}}^{{T_1}} \! \!dt \int_\Lambda
\! dx\, [j-J(t,\rho)] \cdot  \chi(\rho)^{-1} [j-J(t,\rho)].
\end{split}
\end{equation}
In \eqref{r02}  $\epsilon\ll 1$ is a dimensionless scaling factor, e.g., the
ratio between the microscopic length scale (say the typical
inter-molecular distance) and the macroscopic one, and the
symbol $\asymp$ denotes logarithmic equivalence as $\epsilon \to 0$.
We denote by $\rho_\epsilon$ the \emph{empirical density}, that is, $\rho_\epsilon(x)$ is the
density of particles in a macroscopically small volume around $x$.  
Analogously, $j_\epsilon$ denotes the \emph{empirical current}, that is $j_\epsilon (t,x)
\cdot \hat n d\sigma dt$ is the flow of mass across a macroscopically
small surface $d\sigma$ centered at $x$ and orthogonal to the unit
vector $\hat n$ in the macroscopic time interval $[t,t+dt]$. 
The factor $\epsilon^{-d}$ is proportional to the number of particles in the
macroscopic volume. It plays the role of Avogadro's number (implicit
in the Boltzmann constant) in \eqref{1ff}.

The interpretation of \eqref{r02}--\eqref{r05} is quite intuitive and
already discussed in the Introduction. In Section~\ref{s:8} these 
formulas will be derived
from an underlying microscopic dynamics in the case of stochastic
lattice gases.

\medskip
Assume that the external drivings do not depend on time and let $P$ be
the \emph{stationary ensemble}, that is the invariant measure of the
underlying microscopic dynamics. Note that $P$ is an ensemble on 
the configuration space.
The probability of observing a fluctuation $\rho$
of the density profile can be written in the form
\begin{equation}
\label{ld}
P (\rho_\epsilon \approx \rho)
\asymp \exp\big\{ - \epsilon^{-d} \, V(\rho)\big\}.
\end{equation}
While for equilibrium states $V$ is given by the Boltzmann-Einstein
formula \eqref{e}-\eqref{e2}, in non--equilibrium we do not have a
general formula for $V$. In the following $V$ will be called the 
\emph{quasi--potential}, according to the terminology of
\cite{MR2953753}. 
As we show in the following sections, the MFT provides characterizations
of the quasi-potential that can be used either for exact computations
or for perturbation expansions.

\medskip
Let $\mb P$ be the \emph{stationary process}, that is
the ensemble on paths for which the initial conditions are sampled
according to the stationary ensemble $P$.
If we make a  Markovian assumption on the
microscopic dynamics, combining \eqref{ld} and \eqref{r02}, the fluctuation formula 
for the stationary process $\mb P$ is
\begin{equation}
\label{r21}
\begin{split}
& \mb P \big( (\rho_\epsilon (t), j_\epsilon (t))
\approx (\rho (t), j(t))\,,\, t\in [T_0, T_1] \, \big) \\
& \quad \asymp \exp\big\{ - \epsilon^{-d} \,
\mc R_{[{T_0},{T_1}]}(\rho,  j) \big\},
\end{split}
\end{equation}
where
\begin{equation}
\label{r09}
\mc R_{[{T_0},{T_1}]}(\rho,  j) = V(\rho(T_0)) +
\mathcal I_{[{T_0},{T_1}]}(\rho,  j)\,.
\end{equation}
Formula \eqref{r09} states that at time $T_0$ the density profile
$\rho(T_0)$ is sampled according to the stationary ensemble 
and the corresponding asymptotic probability is given by
\eqref{ld}. Then, the probability of following the path $(\rho,j)$
in the time interval $[T_0,T_1]$ with initial condition $\rho(T_0)$
is given by  \eqref{r02}.
We point out that \eqref{r09} may hold also when the
microscopic dynamics is not Markovian provided the Markov property is
recovered at the macroscopic level.

\subsection{Time reversal and its consequences}
\label{s:tr}

We analyze the behavior of the fundamental formula
\eqref{r21}--\eqref{r09} under time reversal and deduce, in
particular, the orthogonal decomposition of the hydrodynamic current.

The time reversal operator $\theta$ on density and
current paths is defined as $[\theta \rho](t)=\rho(-t)$, $[\theta
j](t)=-j(-t)$. Denote by $\mb P^{*}$ the time reversal of $\mb P$,
that is $\mb P^{*} = \mb P \circ \theta^{-1}$. 
In words, $\mb P^*$ is defined by assigning
to a backward path the probability of the forward path under $\mb P$.

Then $\mb P^*$ is the stationary processes associated to some dynamics
that we call adjoint. When $\mb P$ is Markov then $\mb P^*$ is also
Markov and has the same stationary ensemble. By definition,
\begin{equation}
\label{r03}
\begin{split}
& \mb P \Big( \rho_\epsilon \approx \rho \,,\, j_\epsilon \approx j
\,,\, t\in [T_0,T_1] \Big)
\\
&\qquad =  \mb P^{*} \Big( \rho_\epsilon \approx \theta\rho \,, \,
j_\epsilon \approx \theta j  \,, \, t \in [-T_1,-T_0]\Big).
\end{split}
\end{equation}
At the level of large deviations the identity \eqref{r03} implies 
\begin{equation}
\label{r04}
\mc R_{[T_0,T_1]}(\rho,j)= \mc{R}^*_{[-T_1,-T_0]}(\theta\rho,\theta j),
\end{equation}
where $\mc{R}^*$ is the large deviation functional for the
stationary adjoint process.

Since the stationary ensembles of process and its time reversal
coincide, the functional $\mc{R}^*$ can be written as
\begin{equation}
\label{r06}
\mc{R}^*_{[T_0,T_1]}(\rho, j) \;=\; V(\rho(T_0)) + \mathcal I^*_{[T_0,T_1]}(\rho,
j).
\end{equation}
From equations \eqref{r09}, \eqref{r04} and \eqref{r06}, we have
\begin{equation}
\label{r28}
V(\rho (T_0)) + \mathcal I_{[T_0,T_1]}(\rho, j)=V(\rho(T_1)) +
\mathcal I_{[-T_1,-T_0]}^*(\theta \rho, \theta j).
\end{equation}

We now assume that the adjoint dynamics admits a hydrodynamic
description of the form \eqref{2.1}--\eqref{2.3} with a suitable
external field. This assumption is very natural from the
physical point of view. It expresses the fact that empirically by
acting on a system with suitable external fields we can invert the
evolution of a process. For example, we can arrange the action on the
system in such a way that heat flows from a
lower temperature to a higher temperature reservoir.  
In view of this assumption, the adjoint process satisfies a
dynamical large deviations
principle of the same form as \eqref{r02} with $\mathbb P$
replaced by $\mathbb P^*$ and $\mathcal{I}$ replaced
by $\mathcal{I}^*$ where
\begin{equation}
\label{r10}
\mathcal I^*_{[T_0,T_1]}(\rho, j)= \frac 14 \int_{T_0}^{T_1} \!dt
\int_{\Lambda} \!dx\,  [j - J^*(\rho) ] \cdot \chi(\rho)^{-1}
[ j - J^*(\rho) ]
\end{equation}
in which $J^*(\rho)$ expresses the constitutive relationship of the
adjoint hydrodynamics. 

Relation \eqref{r28} has far reaching consequences.  By choosing
$[T_0,T_1]=[-T,T]$, dividing both sides by $2T$, and taking the limit
$T \rightarrow 0$, we find
\begin{equation}
\label{bb123}
\begin{split}
& \int_{\Lambda} dx\,  \frac{\delta V}{\delta\rho} \nabla \cdot j
= \frac 12 \int_{\Lambda} dx\, [J(\rho)+ J^*(\rho)] \cdot \chi(\rho)^{-1} j \\
&\quad - \frac 14 \int_{\Lambda} dx\,
[J (\rho)+  J^*(\rho)] \cdot \chi(\rho)^{-1} [J(\rho)-J^*(\rho) ],
\end{split}
\end{equation}
which has to be satisfied for any $\rho$ and $j$.  
Since the path $\rho(t)$ satisfies the boundary condition \eqref{2.3},
in \eqref{bb123} we can restrict to profiles $\rho$ satisfying
\eqref{2.3}. For such profiles $\delta V/\delta \rho$ vanishes at the
boundary (see Section \ref{s:df}), integrating by parts the
left hand side of \eqref{bb123}  we obtain that
\begin{equation}
\label{r12}
\left\{
\begin{split}
& J(\rho)+J^*(\rho)=-2\chi(\rho)\nabla\frac{\delta V}{\delta \rho} \\
&  \int_{\Lambda}dx\,  J (\rho) \cdot \chi(\rho)^{-1} J(\rho)
= \int_{\Lambda}dx\, J^* (\rho) \cdot \chi(\rho)^{-1} J^*(\rho).
\end{split}
\right.
\end{equation}
These two equations are  symmetric in $J$ and $J^*$. The first
equation may be considered as a fluctuation--dissipation relation
for the currents.

We now define the \emph{symmetric current} $J_S$ by 
\begin{equation}
  \label{Ons-form}
  J_S(\rho)=-\chi(\rho)\nabla\frac{\delta V}{\delta\rho}.
\end{equation}
Since the stationary density $\bar\rho$ is a minimum for $V$, then
$(\delta V/\delta\rho)(\bar\rho)=0$.  The symmetric current thus
vanishes at the stationary profile,
\begin{equation}
\label{01}
J_S(\bar\rho) \;=\;0.
\end{equation}
We rewrite the hydrodynamic current as
\begin{equation}
\label{split}
J(\rho)=J_S(\rho)+J_A(\rho), 
\end{equation}
which defines the \emph{antisymmetric current} $J_A$.

In view of these definitions, equations \eqref{r12} become
\begin{equation}
\label{r18}
\left\{
\begin{split}
&  J^*(\rho) \;=\; J_S(\rho)-J_A(\rho)\;,\\
& \int_{\Lambda}dx\,  J_S (\rho) \cdot \chi(\rho)^{-1} J_A(\rho)
= 0.
\end{split}
\right.
\end{equation}
In this way we see that the splitting of the currents and the orthogonality property
are a consequence of the existence of a time reversed dynamics admitting
an hydrodynamic behavior.
Moreover, inserting the first of the two equations \eqref{r12} into
the second we obtain the equation for $V$
\begin{equation}
\label{r15-2}
\int_{\Lambda}dx\,  \nabla \frac{\delta V}{\delta \rho} \cdot \chi(\rho)
\nabla \frac{\delta V}{\delta \rho}
-  \int_{\Lambda}dx\, \frac{\delta V}{\delta \rho} \nabla \cdot J(\rho)
= 0\,.
\end{equation}
This will be interpreted as a Hamilton-Jacobi equation. As shown in
\cite{BDGJL02}, $V$ is the maximal positive 
solution to \eqref{r15-2} which vanishes when $\rho=\bar \rho$.
Since $J(\rho)$ and $J^*(\rho)$ play a symmetric role in \eqref{r12},
the Hamilton-Jacobi equation \eqref{r15-2} holds replacing $J(\rho)$
with $J^*(\rho)$.

In view of the fluctuation-dissipation relation \eqref{r12}, we may
write the hydrodynamic equation and the adjoint hydrodynamic equation
as
\begin{equation*}
\begin{split}
\partial_t \rho &=
\nabla \cdot
\Big( \chi(\rho) \nabla \frac {\delta V}{\delta \rho} \Big)
- \nabla\cdot J_A (\rho),
\\
\partial_t \rho &=
\nabla \cdot
\Big( \chi(\rho) \nabla \frac {\delta V}{\delta \rho} \Big)
+ \nabla\cdot J_A(\rho)
\end{split}
\end{equation*}
respectively.
Another way of writing the adjoint hydrodynamic equation is
\begin{equation}
\label{r20}
  \partial_t \rho = - \nabla \cdot D(\rho)\nabla \rho +
  \nabla \cdot \chi (\rho)
\, \Big(E+2\nabla\frac{\delta V}{\delta \rho} \Big).
\end{equation}
In spite of its appearance, the forward evolution of this equation is
well posed. Indeed, the added external field $2\nabla (\delta V/\delta
\rho)$ produces a second order term which makes the equation of
parabolic type. In the case of equilibrium states 
the adjoint hydrodynamics coincides with the original one.

\medskip
We illustrate the decomposition of the current by giving three simple
examples.

\smallskip
\noindent\emph{Equilibrium states.} \
Equilibrium states, homogeneous and
inhomogeneous, are characterized by $J(\bar\rho)=0$. In this case 
the quasi-potential $V$ is given by 
(see Section \ref{s:eq})
\begin{equation}
  \label{a1}
  V(\rho) = \int_\Lambda \!dx \, \big[ f(\rho) - f(\bar\rho) -
  f'(\bar\rho) (\rho -\bar\rho) \big].
\end{equation}
Observe that, due to the convexity of $f$, $V$ is convex, positive and is
minimal on the stationary density profile $\bar\rho$.  
The Einstein relation \eqref{r29} and $J(\bar\rho)=0$ imply that  
\begin{equation}
\label{a2}
J(\rho) = - \chi(\rho) \nabla \frac {\delta V}{\delta \rho}.
\end{equation}
Hence, the antisymmetric current vanishes, $J_A=0$, and the current,
as in Onsager theory, is proportional to the thermodynamic force. 
In geometrical terms, the hydrodynamic evolution can thus be
viewed as the flow along the steepest descent of $V$ with an intensity
given by the mobility.

\smallskip
\noindent\emph{Circulation of a fluid in a ring.} \
In absence of an external field, we have an equilibrium state that
fits in the scheme just discussed: the density $\bar\rho$ is constant,
and the current $J(\bar\rho)$ is zero.  Moreover, if we start with an
arbitrary density profile $\rho$, the system evolves to the
equilibrium according to the hydrodynamic equation
$$
\partial_t\rho=-\nabla\cdot J(\rho)
=\nabla\cdot\Big(\chi(\rho)\nabla\frac{\delta V}{\delta\rho}\Big),
$$
where $V(\rho)$ is given by \eqref{a1}. In this case, by conservation of
mass, the expression \eqref{a1} simplifies since the last term does
not contribute.

When we switch on a constant weak driving field $E$ tangent to the
ring, in the stationary regime the particle density $\bar\rho$ is
still constant, but there is a non-zero current
$J(\bar\rho)=\chi(\bar\rho) E$.  The corresponding hydrodynamic
equation is
\begin{equation}\label{modello}
\partial_t\rho=-\nabla\cdot J(\rho)=
\nabla\cdot\Big(\chi(\rho)\nabla\frac{\delta V}{\delta\rho}
-\chi(\rho)E\Big).
\end{equation}

The stationary non-equilibrium situation, with density $\bar\rho$ and
current $J(\bar\rho)$, is not invariant under time reversal.  In fact,
time reversal corresponds to inverting the current, namely to changing
$E$ with $-E$.  Therefore, the hydrodynamic equation for the time
reversed system will be
\begin{equation}\label{modellotr}
\partial_t\rho=-\nabla\cdot J^*(\rho)=
\nabla\cdot\Big(\chi(\rho)\nabla\frac{\delta V}{\delta\rho}
+\chi(\rho)E\Big),
\end{equation}
which corresponds to 
\begin{equation}\label{modello2}
J_S(\rho)=-\chi(\rho)\nabla\frac{\delta V}{\delta\rho}
\,\,,\,\,\,\,
J_A(\rho)=\chi(\rho)E\,.
\end{equation}
A simple computation shows that these two components satisfy the
orthogonality condition in \eqref{r18}.

\smallskip
\noindent\emph{Rarefied gas with boundary reservoirs.} \
For simplicity we consider again to the one-dimensional case.  When we
can neglect the interaction among the particles the transport
coefficients are $D(\rho)= D_0$, $\chi(\rho)= \chi_0 \rho$ where
$D_0,\chi_0$ are constants, and the equilibrium free energy per unit
 volume is given by $f(\rho) = (D_0/\chi_0) \rho \log \rho$. 
The quasi-potential $V(\rho)$ is again given by \eqref{a1}.

Letting $\Lambda = (0,L)$, $\lambda(0)=\lambda_0$,
$\lambda(L)=\lambda_1$, the stationary density profile is $\bar\rho(x)
= \rho_0 (1-x/L) + \rho_1 x/L$ where $\rho_0$ and $\rho_1$ are the
densities associated to $\lambda_0$ and $\lambda_1$ by \eqref{2.3}. In
particular, $(\nabla \bar\rho)(x) = (\rho_1 - \rho_0)/L$.  In this
case the hydrodynamic equation reduces to the heat equation and the
constitutive equation to
\begin{equation*}
J(\rho) \,=\, - D_0 \nabla \rho
\;=\; -\, \chi(\rho)\nabla\frac{\delta V}{\delta\rho}
\,-\, D_0 \frac{\nabla \bar\rho} {\bar\rho} \, \rho\;.
\end{equation*}
To reverse the current in the stationary state we have to add a
suitable external field. The unique expression for the adjoint current
such that $J^*(\bar\rho)=-J(\bar\rho)$ is 
\begin{equation*}
\begin{split}
J^*(\rho) \, & =\, - D_0 \nabla \rho\;+\; 2 D_0
\frac{\nabla\bar\rho}{\bar\rho}\, \rho, 
\end{split}
\end{equation*}
which corresponds to 
\begin{equation}\label{modello2bis}
J_S(\rho)=  D_0\Big( \frac {\nabla\bar\rho}{\bar\rho}  \, \rho - \nabla\rho\Big),
\,\,\,\,\,\,
J_A(\rho)=  -D_0\, \frac {\nabla\bar\rho}{\bar\rho}  \, \rho,
\end{equation}
which satisfy the orthogonality in \eqref{r18}.

\medskip
The decomposition \eqref{split} of the current is entirely general for driven
diffusive systems.
It depends on the chemical potential $\lambda$ and
the external field $E$ and can not be inferred by inspection as in the
simple examples discussed above.  
We mention the general approach to non-equilibrium introduced in
\cite{ottinger} that is based on a separation of the evolution
equations into dissipative and conservative terms which may remind this
decomposition.

\section{Thermodynamic transformations}
\label{s:3}

As clearly stated in classical textbooks, e.g.\ \cite{callen, LL}, in
a transformation between equilibrium states a system necessarily goes
through deviations from equilibrium which are small if the
transformation is quasi-static. Classical thermodynamics is unable to
describe this intrinsically dynamic aspect.  The aim of this section
is to develop a coherent dynamical approach to thermodynamic
transformations covering both equilibrium and non-equilibrium states
\cite{MR2999559,prlqs}.

\subsection{Non-Equilibrium Clausius inequality}
\label{s:rci}

The second law of thermodynamics can be expressed as follows. Consider
a system in an equilibrium state in thermal contact with an
environment at a given tempera\-ture. The system then undergoes an
isothermal transformation to a final state. By denoting with $W$ the
mechanical work done on the system,
\begin{equation}
  \label{ci}
  W \ge \Delta F
\end{equation}
where $\Delta F$ is the difference of the free energy between
the final and the initial state. If equality holds the transformation is said
to be reversible. It can be implemented by performing very slow
variations so that the system goes through a sequence of equilibrium
states.  With a slight abuse of terminology, we shall refer to
\eqref{ci} as the Clausius inequality.

We present a dynamical derivation of the Clausius inequality based on
the hydrodynamic description and the local Einstein relation
\eqref{r29}.  Consider a system in a time dependent environment, that
is, $E$ and $\lambda$ depend on time, as described in Section
\ref{s:2.1}. The work done by the environment on the system in the
time interval $[0,T]$ is
\begin{equation}
  \label{W=}
  \begin{split}
    W_{[0,T]} = & \int_{0}^{T}\! dt \, \Big\{ \int_\Lambda \!dx\, j(t)
    \cdot E(t)
    - \int_{\partial\Lambda} \!d\sigma \, \lambda
    (t) \: j(t) \cdot \hat{n} \Big\},
  \end{split}
\end{equation}
where $\hat n$ is the outer normal to $\partial \Lambda$ and $d\sigma$
is the surface measure on $\partial \Lambda$.  The first term on the
right hand side is the energy provided by the external field
while the second is the energy provided by the reservoirs.

Fix time dependent paths $\lambda(t)$ of the chemical potential and
$E(t)$ of the driving field. Given a density profile $\rho_0$, let
$\rho(t)$, $j(t)$, $t \ge 0$, be the solution of
\eqref{2.1}--\eqref{2.3} with initial condition $\rho_0$.  
By using the Einstein relation \eqref{r29} and the boundary condition 
$f'(\rho(t)) = \lambda(t)$, an application of the
divergence theorem yields
\begin{equation}
\label{04}
\begin{split}
W_{[0,T]} \,&=\,  F(\rho(T)) - F(\rho(0)) \\
& +\,  \int_{0}^{T} \!dt  \int_\Lambda \!dx \;
j(t)\cdot \chi(\rho(t) )^{-1} j(t),
\end{split}
\end{equation}
where $F$ is the equilibrium free energy functional,
\begin{equation}
\label{10}
F(\rho) = \int_\Lambda \!dx \: f (\rho(x)).
\end{equation}
Equation \eqref{04} is not simply a rewriting of
\eqref{W=}, as it depends on a physical principle, the local
Einstein relationship.

Since the second term on the right hand side of \eqref{04} is
positive, we deduce the Clausius inequality \eqref{ci} with $\Delta F
= F(\rho_1) - F(\rho_0)$ for arbitrary density profiles
$\rho_0=\rho(0)$, $\rho_1=\rho(T)$.  Note that this derivation holds
both for equilibrium and non-equilibrium systems.

\medskip
For equilibrium states, the former dynamical derivation of Clausius
inequality allows to discuss precisely in which sense quasi-static
transformations approximate reversible transformations.
We consider the simpler case of spatially homogeneous equilibrium
states.  As we mentioned before, such states are characterized by a
vanishing external field $E$ and by a constant chemical potential
$\lambda$.  In this case the stationary solution $\bar \rho$ of the
hydrodynamic equations \eqref{2.1}--\eqref{2.3} is the constant $\bar\rho$
satisfying $f'(\bar\rho) = \lambda$.

Fix two constant chemical potentials $\lambda_0$, $\lambda_1$.
Consider a system initially in the state $\bar\rho_0$ which is driven
to a new state $\bar\rho_1$ by changing the chemical potential in time
in a way that $\lambda(t)=\lambda_0$ for $t\le 0$ and
$\lambda(t)=\lambda_1$ for $t\ge t_0$, where $t_0$ is some fixed
positive time.

Let $\rho(t)$, $j(t)$, $t \ge 0$, be the solution of
\eqref{2.1}--\eqref{2.3} with initial condition $\bar\rho_0$.  Since
the chemical potential is equal to $\lambda_1$ for $t\ge t_0$,
$\rho(t)\to\bar\rho_1$ as $t\to \infty$. Moreover, as $\bar\rho_1$ is
an equilibrium state, the current $j(t)$ relaxes to
$J(\bar\rho_1)=0$. We deduce that the integral in \eqref{04} is
finite as $T\to\infty$ and that
\begin{equation}
\label{11}
\begin{split}
W \,&=\,  F(\bar\rho_1) - F(\bar\rho_0) \\
& +\,  \int_{0}^{\infty} \!dt  \int_\Lambda \!dx \;
j(t)\cdot \chi(\rho(t) )^{-1} j(t)\;,
\end{split}
\end{equation}
where $W= \lim_{T\to +\infty} W_{[0,T]}$.

It remains to show that in the quasi-static limit equality in
\eqref{ci} is achieved.  For any fixed transformation the inequality
\eqref{ci} is strict because the second term on the right hand side of
\eqref{11} cannot be identically zero. Therefore, reversible
transformations cannot be achieved exactly.  We can however exhibit a
sequence of transformations for which the second term on the right
hand side in \eqref{11} can be made arbitrarily small.  This sequence
of transformations is what we call a quasi-static transformation.  Fix
a smooth function $\lambda(t)$ such that $\lambda(0)=\lambda_0$ and
$\lambda(t)=\lambda_1$ for $t\ge t_0$. Given $\tau >0$ we set
$\lambda^\tau (t) = \lambda (t/\tau)$.  Since $E=0$, the last
term on the right hand side of \eqref{11} is given by
\begin{equation*}
\int_{0}^{\infty}\!dt \int_\Lambda\!dx \,
\nabla f' (\rho^\tau(t))  \cdot
\chi(\rho^\tau(t)) \nabla f' (\rho^\tau(t)),
\end{equation*}
where $\rho^\tau$ is the solution to \eqref{2.1}--\eqref{2.3} with
initial condition $\bar\rho_0$ and boundary conditions
$\lambda^\tau(t)$. For each $t\ge 0$, let
$\bar\rho_{\lambda^\tau(t)}$ be the equilibrium state associated to
the constant chemical potential $\lambda^\tau (t)$.  Since $\nabla
f' (\bar\rho_{\lambda^\tau(t)}) =0$, we can rewrite the previous
integral as
\begin{equation*}
\begin{split}
\int_{0}^{\infty}\!dt \int_\Lambda \!dx\,
\nabla \big[ & f' (\rho^\tau(t)) - f' (\bar\rho_{\lambda^\tau(t)}) \big] \\
& \cdot \chi(\rho^\tau(t))
\nabla \big[ f' (\rho^\tau(t)) - f' (\bar\rho_{\lambda^\tau(t)} ) \big].
\end{split}
\end{equation*}
The difference between the solution of the hydrodynamic equation
$\rho^\tau(t)$ and the stationary profile
$\bar\rho_{\lambda^\tau(t)}$ is of order $1/\tau$ uniformly in time,
and so is the difference $f' (\rho^\tau(t)) - f' (\bar
\rho_{\lambda^\tau(t)})$.  As the integration over time essentially
extends over an interval of length $\tau$, the previous
expression vanishes for $\tau\to \infty$.  This implies that equality in
\eqref{ci} is achieved in this limit.

\medskip
For non-equilibrium states, the inequality \eqref{ci} does not carry
any significant information when we consider transformations over
long time intervals. In fact, as non-equilibrium stationary states
support a non vanishing current, to maintain such a current one needs
to dissipate a positive amount of energy per unit time.  If we
consider a transformation between non-equilibrium stationary states,
the energy dissipated along such transformation will necessarily
include the contribution needed to maintain such states and therefore
the amount of energy exchanged in an unbounded time window is
unbounded.  In this case, the left hand side of \eqref{ci} is infinite
while the right hand side is finite.

To transform \eqref{ci} into a meaningful inequality, by using the
decomposition \eqref{split} of the current, we give a natural
definition of renormalized work performed along any given
transformation. This definition has been inspired by the point of view
in \cite{op} further developed in \cite{MR2269563,MR2749712}.
We then show that the renormalized work satisfies a Clausius
inequality and prove that equality is achieved in the quasi-static
limit.

The idea to define a renormalized work is to subtract 
the energy needed to maintain the system out of equilibrium.
For time independent drivings, by 
the orthogonal decomposition \eqref{split} and \eqref{01},
$J(\bar\rho)=J_{\mathrm{A}} (\bar\rho)$ is the macroscopic current in
the stationary state.  In view of the general formula for the total
work \eqref{04}, the amount of energy per unit time needed to maintain
the system in the stationary profile $\bar\rho$ is
\begin{equation}
\label{toman}
\int_\Lambda \!dx \:
J_\mathrm{A}(\bar\rho) \cdot \chi(\bar\rho)^{-1}
J_\mathrm{A}(\bar\rho).
\end{equation}

Fix now $T>0$, a density profile $\rho_0$, and space-time dependent
chemical potentials $\lambda(t)$ and external field $E(t)$, $t\in
[0,T]$.  Let $(\rho(t),j(t))$ be the corresponding solution of
\eqref{2.1}--\eqref{2.3} with initial condition $\rho_0$.  
We define the renormalized work
$W^\textrm{ren}_{[0,T]}$ done by the reservoirs and the external field
in the time interval $[0,T]$ as
\begin{equation}
\label{Weff}
\begin{split}
&W^\textrm{ren}_{[0,T]} = W_{[0,T]}
\\
&\quad
- \int_{0}^{T}\! dt \int_\Lambda \!dx\, J_\mathrm{A}(t,\rho(t)) \,
\cdot \chi(\rho(t))^{-1} J_\mathrm{A}(t,\rho(t))
\end{split}
\end{equation}
where $J_\mathrm{A}(t,\rho)$ is the antisymmetric current for the
system with the time independent external driving obtained by freezing
the time dependent chemical potential $\lambda$ and external field $E$
at time $t$.  Observe that the definition of the renormalized
work involves the antisymmetric current $J_\mathrm{A}(t)$ computed not
at density profile $\bar\rho_{\lambda(t), E(t)}$ but at the solution
$\rho(t)$ of the time dependent hydrodynamic equation.
The definition \eqref{Weff} is natural within MFT  and leads to a
Clausius inequality.

In view of \eqref{04} and the orthogonality in \eqref{r18} between the
symmetric and the antisymmetric part of the current, 
\begin{equation}
\label{Weff2}
  \begin{split}
    & W^\textrm{ren}_{[0,T]}  = F(\rho(T)) - F(\rho_0)
    \\ &\quad
    + \int_{0}^{T}\! dt \int_\Lambda \!dx\, J_\mathrm{S}(t,\rho(t)) \cdot
    \chi(\rho(t))^{-1} J_\mathrm{S}(t,\rho(t)).
  \end{split}
\end{equation}
Consider a space-time dependent chemical potential and external field
$(\lambda(t),E(t))$, $t\ge 0$, 
with  $(\lambda(0),E(0))= (\lambda_0,E_0)$ and
$(\lambda(\infty),E(\infty))= (\lambda_1,E_1)$. 
Let $\bar\rho_0 = \bar \rho_{\lambda_0,E_0}$, 
$\bar\rho_1 = \bar \rho_{\lambda_1,E_1}$ 
be the corresponding stationary profiles and let
$(\rho(t),j(t))$, $t \ge 0$  be the solution of
\eqref{2.1}--\eqref{2.3} with initial condition $\bar\rho_0$.  Since
$\rho(T)$ converges to $\bar\rho_1$, the symmetric part of the
current, $J_\mathrm{S}(\rho(T))$, relaxes as $T\to +\infty$ to
$J_\mathrm{S}(\bar\rho_1) = 0$.  
By letting $W^\textrm{ren}=
\lim_{T\to\infty} W^\textrm{ren}_{[0,T]}$, we thus get
\begin{equation}
\label{16}
\begin{split}
  & W^\textrm{ren} =  F(\bar\rho_1) - F(\bar\rho_0)
  \\
  &\quad + \int_{0}^{\infty} \!dt \int_\Lambda \!dx \:
  J_\mathrm{S}(t,\rho(t)) \cdot \chi(\rho(t))^{-1} J_\mathrm{S}(t,\rho(t))
\end{split}
\end{equation}
where $F$ is the equilibrium free energy functional \eqref{10}.
In particular,
\begin{equation}
\label{15}
W^\textrm{ren} \ge  F (\bar\rho_1) - F(\bar\rho_0),
\end{equation}
which is a meaningful version of the Clausius inequality for
non-equilibrium states.

Arguing as in the equilibrium case, we can exhibit a sequence of
transformations $(\lambda^\tau(t), E^\tau(t))$ which vary appreciably
on a time scale $\tau$, such that in the quasi-static limit $\tau \to
\infty$ equality in \eqref{15} is achieved,
\begin{equation}
  \label{cu}
  W^\textrm{ren} =  F (\bar\rho_1) - F(\bar\rho_0).
\end{equation}

\subsection{Excess work}

Consider a homogeneous equilibrium state with vanishing external field
and constant chemical potential $\lambda_0$ and let $\bar\rho_0$ be
the corresponding homogeneous density, $\lambda_0 = f'(\bar\rho_0)$.
The system is put in contact with a new environment with chemical
potential $\lambda_1$. In this case, recalling that $f$ is the free
energy per unit volume and that the temperature of the system is
the same of the environment, the \emph{availability} per unit 
volume is defined by $a= f(\bar\rho_0) -\lambda_1 \bar\rho_0$,
\cite[Ch.~7]{MR0106033}.  The function $a$, which depends on the state
of the system and on the environment, can be used to compute the
maximal useful work that can be extracted from the system in the given
environment. More precisely, by letting $\bar\rho_1$ be such that
$f'(\bar\rho_1)=\lambda_1$, then
\begin{equation}
\label{da}
-\Delta a = f(\bar\rho_0) - f(\bar\rho_1) - \lambda_1 (\bar\rho_0
-\bar\rho_1)\ge 0
\end{equation}
is the the maximal useful work per unit volume that can be
extracted from the system in the given environment.
Comparing $-\Delta a$ with the large
deviations functional $V$ in \eqref{a1} which expresses the
probability of density fluctuations in the equilibrium ensemble 
corresponding to the chemical potential $\lambda_1$, we realize that
\begin{equation}
\label{V=a}
V(\bar\rho_0)= - |\Lambda| \Delta a.
\end{equation}
An analogous relationship can be easily obtained for spatially
inhomogeneous equilibrium states.

In order to discuss the thermodynamic role of the quasi-potential for
non-equilibrium states, we introduce the \emph{excess work} with
respect to a quasi-static transformation.  Consider a stationary
non-equilibrium state with density profile $\rho$ for $t\le 0$, while
at time $t=0$ the external driving is abruptly changed to new values
$(\lambda,E)$ so that for $t>0$ the evolution is given by the
hydrodynamic equation with initial condition $\rho$ and time
independent driving $(\lambda,E)$. We define the excess of work as the
difference between the renormalized work and the renormalized work
involved in a quasi-static transformation from $\rho$ to $\bar
\rho_{\lambda,E}$, namely $W_\mathrm{ex} = W^\textrm{ren} - \min
W^\textrm{ren}$.  According to the discussion in Section~\ref{s:rci},
the excess work is given by
\begin{equation}
\label{17bis}
\begin{split}
W_\mathrm{ex} & =  W^\textrm{ren}(\rho) - \big[ F(\bar\rho) -F(\rho)\big]
\\
&= \int_0^\infty \!dt \int_{\Lambda} \!dx \, J_\mathrm{S}(\rho(t))
\, \cdot \chi(\rho(t))^{-1} J_\mathrm{S}(\rho(t)),
\end{split}
\end{equation}
where $\bar\rho$ is the stationary density corresponding to
$(\lambda,E)$.  By using the orthogonality in \eqref{r18} and
the formula \eqref{Ons-form} for the symmetric part of the current in
terms of the quasi-potential, straightforward computations yield
\begin{equation}
  \label{V=Wex}
  W_\mathrm{ex}  =   V_{\lambda,E}(\rho),
\end{equation}
where we have made explicit the dependence of the quasi-potential on
the driving.  Therefore, while a definition of thermodynamic
potentials, that is functionals of the state of the system
(in this case of the density $\rho$), does not
appear possible in non-equilibrium thermodynamics, the quasi-potential
is the natural extension of the availability.

\subsection{Finite time thermodynamics}
\label{FTT}

We next discuss the energy balance along slow transformations from a
quantitative point of view taking into account that quasi-static
transformations are an idealization and real transformations take
place on a finite time window whose duration is denoted by $\tau$.

For $s\in [0,1]$ a \emph{protocol} is defined by a choice of the external drivings
$E(s,x)$, $x\in\Lambda$, and $\lambda(s,x)$, $x\in \partial\Lambda$.
The slow transformation is then realized, for $\tau$ large, by
\begin{equation*}
\left\{
\begin{array}{l}
E^\tau(t)=E\left(t /\tau \right), \\
\lambda^\tau(t)=\lambda\left( t/\tau \right),
\end{array}
\right. t\in [0,\tau].
\end{equation*}
Let $\rho^\tau(t)$ and $j^\tau(t)$, $0\leq t\leq \tau$, be the
solution to the hydrodynamic equations
with the slow external field $E^\tau$ and chemical potential $\lambda^\tau$,
\begin{equation}
\left\{
\begin{array}{l}
\partial_t \rho^\tau + \nabla \cdot J(t/\tau,\rho^\tau(t))=0,\\
j^\tau(t)=J(t/\tau,\rho^\tau(t))\\
f'(\rho^\tau(t))\big|_{\partial \Lambda}=\lambda^\tau(t)\\
\rho^\tau(0)=\bar \rho(0)
\end{array}
\right.
\label{eq-delta}
\end{equation}
where we recall that
$J(t,\rho)=-D(\rho)\nabla \rho+\chi(\rho)E\left(t\right)$.

For $s\in[0,1]$, let $\bar \rho(s)$
be the unique stationary solution of the hydrodynamics
with external field $E(s)$ and chemical potential $\lambda(s)$.
When $\tau$ is large the solution $(\rho^\tau,j^\tau)$ has an
expansion of the type (recall $s\in[0,1]$)
\begin{equation}\label{davide}
\begin{array}{l}
\displaystyle{
\rho^\tau (\tau s)=\bar \rho(s)+\tfrac 1\tau \, r(s)
+ o\big(\tfrac1{\tau}\big)\,,
} \\
\displaystyle{
j^\tau(\tau s)
=J(s,\bar\rho(s))+\tfrac 1\tau \, g(s) + o\big(\tfrac 1{\tau} \big)
\,.}
\end{array}
\end{equation}
By \eqref{eq-delta} we get the corresponding linear evolution
equations for the first order corrections $(r,g)$,
\begin{equation}\label{davide2bis}
\left\{
\begin{array}{l}
\partial_s  \bar \rho(s) + \nabla \cdot g(s) =0 \\
g(s)= -
\Big[
D(\bar\rho(s))  \nabla r(s)
+ r(s) D'(\bar\rho(s))  \nabla \bar\rho(s) \Big]
\\
\phantom{g(s)=}
+ r(s) \chi'(\bar \rho(s)) E(s)
\\
r(s, x)=0, \; x\in\partial \Lambda
\end{array}
\right.
\end{equation}
which has the form of a Poisson equation for $r(s)$.

Evaluating equations \eqref{04} along the
transformation $(\rho^\tau, j^\tau)$, we obtain
\begin{equation}
\label{lunga}
  \begin{split}
&F\big(\rho^\tau (\tau)\big)-F\big(\bar\rho(0)\big)
=
\tau \int_0^1\!ds \int_\Lambda \!dx\,
j^\tau(\tau s)
\cdot E(s)
\\
& -\tau \int_0^1\!ds
\int_{\partial \Lambda} \! d\sigma \, \lambda(s)
j^\tau (\tau s)
\cdot \hat n
\\
 & \; \; -\tau
 \int_0^1\!ds \int_\Lambda \!dx \,
j^\tau (\tau s)
\cdot
\chi\big(\rho^\tau(\tau s)\big)^{-1}
j^\tau (\tau s).
  \end{split}
\end{equation}
We can analyze this equation at the different orders in $1/\tau$,
obtaining an identity for each order. Direct computations yield that 
at order $\tau$ the right hand side of \eqref{lunga} vanishes, while
at order $1$ we get the first non trivial equation, 
\begin{equation}
\label{1t}
  \begin{split}
    &F\big(\bar \rho(1)\big)-F\big(\bar \rho(0)\big)\\
    &\; = \int_0^1\!ds \int_\Lambda\!dx \, E(s)\cdot g(s) -\int_0^1\!ds
    \int_{\partial \Lambda}\! d\sigma \,\lambda(s) g(s)\cdot \hat n
    \\
    &\;\; + \int_0^1\!ds \int_\Lambda \!dx\, r(s)
    J(s,\bar\rho(s))\cdot (\chi^{-1})'\big(\bar \rho(s)\big)
    J(s,\bar\rho(s)).
  \end{split}
\end{equation}
This is an interesting relationship as it
connects the variation of the free energy to the first order corrections
in a real transformation. It also carries relevant
information for transformations among equilibrium states, but it cannot be
derived within the framework of classical thermodynamics.
If we consider transformations among equilibrium states,
the second line on the right hand side of \eqref{1t}
vanishes when the intermediate states are also of equilibrium so that
$J(s,\bar\rho(s))=0$ for any $s$.  However the transformation can go
through non-equilibrium intermediate states.

\medskip
As a further application of the expansion \eqref{davide}, consider the
equation \eqref{Weff2} which expresses the energy balance in the time
interval $[0,\tau]$.  Recalling that we have already shown that the
last term vanishes in the quasi-static limit, we now estimate this
term when the transformation is given in term of a protocol and $\tau$
is large but finite.

We thus want to estimate, for large $\tau$
\begin{equation}\label{skype1}
\int_0^\tau\!dt\int_\Lambda \! dx \, 
J_S(t/\tau,\rho^\tau(t))\cdot\chi(\rho^\tau(t))^{-1}J_S(t/\tau,\rho^\tau(t)).
\end{equation}
Recalling \eqref{modello2}, the symmetric part of the current is
\begin{equation}\label{symm}
J_S(s,\rho)
=-\chi(\rho)\nabla \frac{\delta V_{\lambda(s),E(s)} (\rho)}{\delta \rho}
\end{equation}
where $V_{\lambda(s),E(s)}$ is the quasi-potential associated to
$(\lambda(s), E(s))$ (we regard $s$ here as a fixed parameter).

By \eqref{davide}, the symmetric current has the expansion
\begin{equation}\label{symm2}
J_S(s,\rho^\tau(\tau s) )
=-\frac 1 \tau
\chi(\bar\rho(s))\nabla\left(C^{-1}_s\star r(s)\right)
+ o\big(\tfrac 1{\tau} \big).
\end{equation}
where $\star$ denotes convolution and
$$
C^{-1}_s(x,y)=
\frac{\delta^2V_{\lambda(s),E(s)}(\bar\rho(s))}{\delta\rho(x)\delta \rho(y)}.
$$
Hence,
for slow transformations we get that \eqref{skype1} has the form
$\frac 1\tau \,B   + o \big(\tfrac 1{\tau}\big)$,
where
\begin{equation}
\label{1tau}
B = \int_0^1\!ds\!\int_\Lambda\!dx \, 
\nabla\big(  C_s^{-1}\star r(s) \big) \cdot 
\chi(\bar\rho(s))   \nabla\big(  C_s^{-1}\star r(s) \big).
\end{equation}

To illustrate the meaning of $B$, consider a transformation between
and through equilibrium states. Then, up to terms of order $1/\tau^2$, 
the work done in the (finite time) transformation is 
\begin{equation*}
  W_{[0,\tau]} =  \Delta F    + \frac 1\tau \, B
\end{equation*}
so that the inequality $B\ge 0$ is a restatement of the second
principle. In general, $B$ quantifies the additional energy
dissipated in the given transformation.
As we have shown, in the limit $\tau\to \infty$, all protocols
realize the equality $ W= \Delta F$. On the other hand, for finite
time $\tau$,  this identity cannot be achieved
and we can select the optimal protocol by minimizing $B$.

For transformations between and through equilibrium states, the
quasi-potential is given by \eqref{a1} so that $B$ has an explicit
expression.  In particular, we can  compute explicitly the
optimal protocol for transformations through homogeneous equilibrium states.  
Namely, we assume that the external
field vanishes and that the chemical potential does not depend on the
space variable. The protocol is thus defined by a real function
$\lambda(s)$, $s\in [0,1]$. The associated stationary solution
$\bar\rho(s)$ is also constant in space and solves $\lambda(s)=
f'(\bar\rho(s))$.  For simplicity we also assume that the diffusion
coefficient is a multiple of the identity; in this case the equation
\eqref{davide2bis} reduces the the classical Poisson equation
\begin{equation}
\begin{cases}
\partial_s \bar\rho(s) =
D(\bar\rho(s)) \Delta r(s) \\
r(s,x) =0, \quad x\in\partial \Lambda
\end{cases}
\end{equation}
whose solution is given by
\begin{equation}
  r(s,x) =  \frac{\partial_s \bar\rho(s)}{D(\bar\rho(s))}
  \int_\Lambda \!dy \, G_0(x,y)
\end{equation}
where $G_0$ is the Green function of the Dirichlet Laplacian on
$\Lambda$.
Since
\begin{equation}
\label{cov}
\frac{\delta^2V_{\lambda(s),E(s)}(\bar\rho(s))}{\delta\rho(x)\delta
  \rho(y)}
 = \frac{ D(\bar\rho(s))}{ \chi(\bar\rho(s))} \, \delta(x-y),
\end{equation}
then
\begin{equation}\label{B-1}
  B = \int_\Lambda \! dx \Big| \,\nabla\!\int_\Lambda \!dy \, G_0(x,y) \Big|^2
  \,
  \int_0^1\!ds \, \frac{[ \partial_s \bar\rho (s) ]^
    2}{\chi(\bar\rho(s))}.
\end{equation}
The dependence on space factorizes in the pre--factor which depends only
on the geometry of the domain. 
It is now straightforward to minimize $B$ with respect to
$\bar\rho(s)$ with the constraints $\bar\rho(0) =\bar\rho_0$,
$\bar\rho(1) =\bar\rho_1$. The minimizer is the unique function
satisfying the constraints such that
\begin{equation}
  \label{aff}
  \frac {\partial_s \bar\rho}{\sqrt{\chi(\bar\rho)}} = \mathrm{const.}
\end{equation}
The optimal protocol is then obtained by the relationship $\lambda =
f'(\bar\rho)$.
This  protocol does not correspond to a constant rate as one
could naively expect. In fact, \eqref{aff} shows that this rate has to be
adjusted to the response properties of the system.

\subsection{Dissipation}

The infinitesimal version of \eqref{04} gives the instantaneous energy
balance which reads
\begin{equation}
  \label{eb}
   \begin{split}
     \dot W &=
   \int_\Lambda \!dx\, \big[ f'(\rho) \dot \rho +  j \cdot
   \chi(\rho)^{-1} j \big]
   \end{split}
\end{equation}
where $\dot W$ is the power injected by the reservoirs
and external field in the system. Accordingly, $f'(\rho) \dot \rho$
represents the rate of change of the density of free energy while  $ j \cdot
\chi(\rho)^{-1} j$ is the dissipated power per unit volume.
For equilibrium states, the stationary density profile is
characterized by the vanishing of the current and therefore it minimizes the
dissipation. This is not the case for non-equilibrium stationary
states. Recalling \eqref{01}, the non-equilibrium stationary density
profile is characterized by the vanishing of the symmetric current and
this does not imply that the dissipation is minimal.
In view of the orthogonal decomposition \eqref{split},
\begin{equation*}
  \begin{split}
    \int_\Lambda \!dx\, J(\rho) \cdot \chi(\rho)^{-1} J(\rho)
    & = \int_\Lambda
    \!dx\, J_\mathrm{S} (\rho) \cdot \chi(\rho)^{-1} J_\mathrm{S}
    (\rho)
    \\
    &    + \int_\Lambda \!dx\, J_\mathrm{A} (\rho) \cdot
    \chi(\rho)^{-1} J_\mathrm{A} (\rho).
  \end{split}
\end{equation*}
The minimization of the left hand side is not achieved by making
the first term on the the right hand side equal to zero. Indeed, in the
simple case of a one dimensional rarefied gas
discussed in section~\ref{s:tr} the minimizer
of the left hand side with the prescribed boundary conditions
$\rho(0)=\rho_0$, $\rho(L)=\rho_1$ is
\begin{equation*}
  \hat\rho(x) = \big[ \sqrt{\rho_0} (1-x/L) + \sqrt{\rho_1} x/L\big]^2
\end{equation*}
while the stationary profile is $\bar\rho(x) = \rho_0 (1-x/L) +
\rho_1x/L$. Observe that, in accordance with the near to equilibrium
Prigogine principle \cite{Pri},
$\bar\rho- \hat\rho = O( [(\rho_1-\rho_0)/L]^2)$.

We remark that if we consider the renormalized work \eqref{Weff2},
the corresponding renormalized power is
\begin{equation}
  \label{eb2}
     {\dot{W}}^\mathrm{ren} =
   \int_\Lambda \!dx\, \big[ f'(\rho) \dot \rho +  J_\mathrm{S}(\rho) \cdot
   \chi(\rho)^{-1} J_\mathrm{S}(\rho) \big].
\end{equation}
Then, recalling \eqref{01}, the stationary density profile minimizes
the corresponding renormalized dissipation $\int_\Lambda \!dx
\,J_\mathrm{S}(\rho) \cdot \chi(\rho)^{-1} J_\mathrm{S}(\rho)$.
Once again, in terms of the renormalized
quantities, non--equilibrium stationary states behave as equilibrium states.

\subsection{Minimum dissipation principle}

Generalizing Onsager \cite{ons}, we introduce the
\emph{dissipation function}, \cite{MR2082195}
\begin{equation}
  \label{diss}
  \Phi (\rho,j) = \frac 12
  \int_\Lambda\!dx \big(j -J_\mathrm{A}(\rho)\big) \cdot
  \chi(\rho)^{-1} \big(j -J_\mathrm{A}(\rho)\big)
\end{equation}
and the functional
\begin{equation}
  \label{psi}
  \Psi(\rho,j) = -
  \int_\Lambda\!dx \, \frac {\delta V (\rho)}{\delta \rho}  \nabla \cdot j
  + \Phi(\rho,j).
\end{equation}
We can then reformulate the constitutive equation $j=J(\rho)$ in variational
terms,
\begin{equation}
  \label{ons}
    \Psi(\rho,j)
    = \mathrm{minimum},
\end{equation}
where the minimum is understood with respect to $j$ with
$\rho$ fixed. Indeed, by taking the variation of $\Psi$ with respect to $j$ we deduce
that the minimum is achieved for $j=J_\mathrm{S}(\rho) +J_\mathrm{A}(\rho)=J(\rho)$.

The difference with respect to the near equilibrium Onsager theory
(apart from the sign difference) is the insertion of the
antisymmetric current $J_A$ in the dissipation function \eqref{diss}
and the replacement of the entropy with the quasi-potential.
Accordingly, while in Onsager the minimum value of $\Psi$ is half of
the total dissipation, in our case
\begin{equation*}
  \min_j \Psi(\rho,j) =  \Psi \big(\rho, J(\rho)\big)
  = - \frac 12   \int_\Lambda\!dx \, J_\mathrm{S}(\rho)
  \cdot \chi(\rho)^{-1} J_\mathrm{S}(\rho)
\end{equation*}
is half the negative of the renormalized dissipation.

\subsection{Comments}

Within the scheme introduced, we considered only one conservation law
(the conservation of the mass) and accordingly we did not distinguish
between work and heat.  A model where heat is naturally introduced is
analyzed in \cite{olla}.

The splitting of the current \eqref{split} appears very interesting
conceptually.  However the two currents $J_S$ and $J_A$, apart some
special cases, are not easily accessible experimentally. In fact what
is directly measurable is the total current which coincides with $J_A$
in a stationary state while $J_S$ represents the total current in a
relaxation to an equilibrium state. In the general case their
computation require the knowledge of the quasi-potential. A
measurement of the quasi-potential, via rare fluctuations is hopeless
as very large times are involved. It can be either obtained from
calculations by solving a variational principle (see section
\ref{s:df}), or from simulations using algorithms like in
\cite{PhysRevLett.96.120603}. Otherwise it can be approximately
estimated from measurements of correlation functions in the stationary
state. In fact, as we shall see later, $V$ is the Legendre transform
of the generating functional of density correlations in the stationary
state.

These remarks imply that the renormalized work is not immediately
accessible. There are other possibilities to define a renormalized
work \cite{MR3162537}, which however have similar drawbacks. On the
other hand the approach developed in subsection \ref{FTT} allows, as
remarked in the introduction, a detailed analysis of quasi--static
transformations by relating explicitly, for example, the variation of
the free energy and the corrections to an infinitely slow
transformation \eqref{1t}.  Actually this approach provides an
infinity of relationships which should be further investigated.

Another benefit of finite time thermodynamics is related to the
possibility of optimizing the protocol of a transformation both in
equilibrium and non-equilibrium.

As in \cite{ons}, the dissipation function \eqref{diss} provides 
a variational characterization of the evolution equations. 
For the usefulness of the dissipation function in identifying the
various physical contributions, we refer to \cite{onsfuss} on the
irreversible processes in electrolytes.

\section{Statistics of density and current fluctuations}
\label{s:ft}

In this section we derive from the fundamental formula the large
deviations statistics separately for the density and the current.  We discuss
their singularities, that will be interpreted as non-equilibrium phase
transitions. The long range correlations of non-equilibrium states
will be connected to the non locality of the quasi-potential.

\subsection{Density fluctuations}
\label{s:df}
We start by deriving the probability of the  density trajectories.  
We fix a path $\rho=\rho(t,x)$, $ (t,x)\in[T_0,T_1]\times\Lambda$.
There are many possible trajectories $j=j(t,x)$, differing by
divergence free vector fields, satisfying the continuity 
equation associated to the given density
trajectory $\rho$. .  Optimizing over the possible currents we have
\begin{equation}\label{4.1}
I_{[T_0,T_1]}(\rho) =
\inf_{ \substack{ j \, : \\ \nabla \cdot j = -\partial_t \rho}}
\mathcal I_{[T_0,T_1]} (\rho , j).
\end{equation}
Then, the asymptotic probability of a density fluctuation is given by
\begin{equation*}
\begin{split}
& \mb P_{\rho_0} \big( \rho_\epsilon (t)
\approx \rho (t) \,,\, t\in [T_0,
 T_1] \, \big) \\
& \qquad \asymp \exp\big\{ - \epsilon^{-d} \,
I_{[{T_0},{T_1}]}(\rho) \big\}.
\end{split}
\end{equation*}
Due to the exponential character of such probability
estimates, only the minimum of the functional $\mc
I_{[T_0,T_1]}(\rho,j)$ over all possible currents $j$ is in fact relevant.

To find the optimal current in \eqref{4.1} we observe that, given any
trajectory $(\rho(t), j(t))$ satisfying the continuity equation, we
can introduce an external field $F$ defined by
\begin{equation}
\label{la20}
j(t)=J(\rho(t))+\chi(\rho(t))F
\end{equation}
so that
\begin{equation}\label{222}
\mc I_{[T_0,T_1]}(\rho,j) \,=\,
\frac 14 \int_{T_0}^{T_1}dt\int_\Lambda dx\,F\cdot \chi(\rho)F.
\end{equation}
The problem can be therefore formulated as follows.  Among all
possible external fields $F$, find the one that minimizes the right
hand side of \eqref{222} with the constraint 
\begin{equation*}
\nabla\cdot(J(\rho)+\chi(\rho)F)=-\partial_t\rho.    
\end{equation*}
We claim that the optimal $F$ is $F=-\nabla\pi$, where $\pi
:[T_0,T_1]\times\Lambda \to \mb R$ is the unique solution to the
Poisson equation
\begin{equation}
\label{la21}
 \nabla \cdot \big[ \chi(\rho) \nabla \pi \big]
= \partial_t \rho  + \nabla \cdot J(t, \rho)
\end{equation}
which vanishes at the boundary of $\Lambda$ for any $t\in[T_0,T_1]$.

Let $F= - \nabla \pi + \tilde F$, so that $\nabla \cdot \chi(\rho)
\tilde F=0$. Since $\pi$ vanishes at the boundary, an integration
by parts yields the orthogonality relationship
$\int_\Lambda\!dx\,\nabla\pi\cdot\chi(\rho)\tilde{F}=0$.  Whence,
\begin{equation}\label{111}
\int_\Lambda dx\,   F \cdot \chi(\rho)  F =
\int_\Lambda dx\, \Big\{ \nabla \pi \cdot \chi(\rho)  \nabla \pi
+ \tilde F \cdot \chi(\rho) \tilde F  \Big\}.
\end{equation}
By construction,
$\nabla\cdot (J(\rho)-\chi(\rho)\nabla\pi)=-\partial_t\rho$,
so that, by \eqref{111}, the choice of $F$ which minimizes \eqref{222} is obtained letting $\tilde{F}=0$.
We deduce
\begin{equation}
\label{I=}
\begin{split}
& I_{[T_0,T_1]} (\rho) = \\
& \frac 14 \int_{{T_0}}^{{T_1}}\!dt
\: \int_\Lambda dx\,
\big[ \partial_t \rho  + \nabla \cdot J(\rho)
\big] \, K(\rho)^{-1}
\big[ \partial_t \rho  + \nabla \cdot J(\rho) \big], \\
\end{split}
\end{equation}
where the positive operator $K(\rho)$ is defined on functions $\pi
:\Lambda\to \mb R$ vanishing at the boundary $\partial \Lambda$ by
\begin{equation*}
K(\rho) \pi = - \nabla \cdot\big( \chi(\rho) \nabla \pi \big).
\end{equation*}

The above argument shows that we can restrict to gradient external fields $F$
when we are looking for fluctuations only of the density $\rho$
(this corresponds to particular realizations of the noisy part of the current 
in the fluctuating hydrodynamics picture).
On the other hand, if we are looking for fluctuations of the current $j$,
then the corresponding external field $F$ is uniquely defined
by \eqref{la20} and will not be, in general, a gradient field.

\medskip
We now derive a variational formula for the quasi-potential.  From
equation \eqref{r28}, we deduce that
\begin{equation}
\label{r19}
V(\rho (T_0)) + I_{[T_0,T_1]}(\rho)=V(\rho(T_1)) +
I_{[-T_1,-T_0]}^*(\theta \rho)\;,
\end{equation}
where $I_{[-T_1,-T_0]}^*$ is the rate function of the adjoint
process. Let us consider the time interval taking $T_1=0$, $T_0=-T$.
Denoting by $\hat \rho$ a generic path satisfying
$\hat\rho(-T)=\bar\rho$, which implies
$V(\hat\rho(-T))=0$, and $\hat\rho(0)=\rho$, we obtain that
\begin{equation*}
I_{[-T,0]}(\hat \rho)=V(\rho) +
I_{[0,T]}^*(\theta \hat \rho).
\end{equation*}
Observing that $I_{[0,T]}^*(\theta \hat \rho) \ge 0 $ and that it is
equal to zero when $\theta \hat \rho$ solves the adjoint
hydrodynamics, we obtain that
\begin{equation}
\label{r07}
V (\rho) = \inf_{\hat\rho} I_{{(-\infty,0]}}(\hat\rho)\;,
\end{equation}
where the infimum is carried over all trajectories $\hat\rho$ such
that $\hat\rho({-\infty}) = \bar\rho$, $\hat\rho(0)
=\rho$.  
The optimal trajectory $\hat\rho$ satisfies
\begin{equation}
\label{r13}
I_{[0, \infty)}^*(\theta \hat\rho)=0. 
\end{equation}
Compare with \cite{MR2953753} for the finite dimensional case.

Optimal trajectories for non--reversible finite dimensional systems
have been seen in numerical simulations \cite{DMRH} and actually
experimentally observed in analogue electrical circuits with noise
modeling a two-dimensional diffusion process \cite{LMc}. We refer to
the bibliography in those articles for previous literature on the
topic.

We can summarize the previous analysis as follows.
While for equilibrium states the path leading to a fluctuation is the
time reversal of the relaxation path \cite{onsmach}, 
for non-equilibrium states the spontaneous emergence of a density
fluctuation takes place most likely following a trajectory which is the
time reversal of the relaxation path along the adjoint hydrodynamics. 
The optimal field to create the fluctuation is $2\nabla\frac{\delta
  V}{\delta \rho}$, that is minus twice the dissipative thermodynamic
force.  To understand the factor $2$ think of an electric circuit. To
invert the current one has to add minus twice the original electric
field.

From the identity \eqref{r07} we deduce that for profiles $\rho$
satisfying the boundary condition \eqref{2.3} $\delta V /\delta \rho$
vanishes at the boundary of $\Lambda$. It is in fact enough to
take the derivative of \eqref{r07} and notice that the optimal path
$\hat\rho$ has prescribed boundary values.

\medskip
The quasi-potential $V$ \eqref{r07} is a Lyapunov functional for the
hydrodynamic equations (H-theorem).  In fact, we can compute the rate
of decrease of $V(\rho(t))$ along the hydrodynamic equations.  In both
cases, using equation \eqref{r15-2}, we have
\begin{equation}
\label{HT}
\begin{split}
& \frac{d}{dt} V(\rho(t))  = \int_{\Lambda}dx\,
\frac{\delta V}{\delta \rho} (\rho(t)) \, \partial_t \rho(t) \\
&\qquad  = - \int_{\Lambda}dx\,
\nabla \frac {\delta V}{\delta \rho} (\rho(t)) \cdot
\chi (\rho(t)) \nabla \frac {\delta V}{\delta \rho}  (\rho(t))\;.
\end{split}
\end{equation}
Recalling \eqref{Ons-form} and \eqref{eb2}, we see that the rate of
decrease of $V$ is the renormalized dissipation. 
In particular, we have that $\frac{d}{dt} V(\rho(t))=0$ if and only if
$\frac {\delta V}{\delta \rho}(\rho(t))=0$. Since we assumed that
there exists a unique stationary profile $\bar\rho$, \eqref{HT} implies that
$\bar\rho$ is globally attractive.

\subsection{Hamiltonian structure}
\label{s:hs}

We regard the functional \eqref{I=} as an action function on the set
of density paths. The corresponding Lagrangian is
\begin{equation*}
\mc L (\rho, \partial_t \rho) =
\frac 14 \int_\Lambda dx\,
\big[ \partial_t \rho  + \nabla \cdot J(\rho)
\big] \, K(\rho)^{-1}
\big[ \partial_t \rho  + \nabla \cdot J(\rho) \big].
\end{equation*}

The associated Hamiltonian $\mathcal{H}(\rho,\pi)$ is obtained by the
Legendre transform of $\mathcal{L}(\rho,\partial_t \rho)$:
\begin{equation}
\label{lHJ}
\begin{split}
& \mathcal{H}(\rho, \pi) = \sup_{\xi}
\Big\{ \int_{\Lambda} dx\, \xi \, \pi
- \mathcal{L}( \rho, \xi ) \Big\} \\
&=\; \int_{\Lambda}dx\,  \Big\{ \nabla \pi \cdot \chi(\rho) \nabla  \pi
-   \pi \, \nabla \cdot J(\rho)\ \Big\}.
\end{split}
\end{equation}

The canonical equations associated to the Hamiltonian $\mc H$ are
\begin{equation}
\left\{
\begin{array}{lll}
\vphantom{\Big\{}
\partial_t \rho
&=&{\displaystyle \nabla \cdot \left( D(\rho )\nabla
\rho\right) -\nabla \cdot \chi(\rho) (E + 2\nabla \pi)}
\\
\vphantom{\Big\{}
\partial_t \pi &=&
{\displaystyle - \nabla \pi\cdot \chi'(\rho) (E+\nabla \pi)
- \text{\rm Tr} \{D(\rho) \, \text{\rm Hess} (\pi) \}}
\end{array}
\right.\label{ham-eq}
\end{equation}
where $\pi$ vanishes at the boundary of $\Lambda$, and $\rho$
satisfies \eqref{2.3}. In this formula Hess$(\pi)$ represents the
Hessian of $\pi$, Tr$A$ the trace of a matrix $A$ and $\chi'$ the
matrix with entries $\chi'_{i,j}(\rho)$.

Observe that $(\rho(t), 0)$ is a solution of the canonical equations if
$\rho(t)$ solves the hydrodynamic equation \eqref{2.1}-\eqref{2.3}. In
particular, $(\bar\rho,0)$ is an equilibrium point of
the canonical equations.

Within the Hamiltonian formalism, the variational problem \eqref{r07}
becomes a minimal action problem. Classical arguments in analytic
mechanics \cite{A} imply that the quasi-potential $V$ introduced in
\eqref{r07} solves the stationary Hamilton-Jacobi equation
\begin{equation}
\label{r14}
\mathcal{H}\Big(\rho, \frac{\delta V}{\delta \rho} \Big) =
\mathcal{H}\big(\bar\rho, 0 \big) =0.
\end{equation}
This is exactly the equation derived in \eqref{r15-2} by the time 
reversal argument.

We next discuss the time reversal within the Hamiltonian formalism.
Letting $\mathcal{L}^*(\rho, \partial_t \rho )$ be the Lagrangian
associated to the action function $I^*$, the time reversal
relationship \eqref{r19} implies the following relation between
Lagrangians:
\begin{equation}
\label{LL*}
\mathcal{L}(\rho, \partial_t \rho )=\mathcal{L}^*(\rho,-\partial_t\rho )
+\int_{\Lambda} dx\,  \frac{\delta V}{\delta \rho}\,\partial_t\rho.
\end{equation}
As a consequence, denoting by $\mc H^*$ the Hamiltonian associated
to $\mc L^*$,
\begin{equation}
\label{HH*}
\mathcal{H}(\rho, \pi)=
\mathcal{H}^*\Big(\rho, \,\frac{\delta V}{\delta\rho}- \pi\Big).
\end{equation}

Let us introduce the involution $\Theta$ on the phase space $(\rho,\pi)$
defined by
\begin{equation*}
\Theta (\rho,\pi) = \big( \rho,\, \frac{\delta V}{\delta \rho}(\rho)
- \pi \big).
\end{equation*}
Denoting by $\mc T_t$, $\mc T_t^*$ the Hamiltonian flow of $\mathcal
H$, $\mathcal{H}^*$, respectively, \eqref{HH*} yields that $\Theta$
acts as the time reversal in the sense that
\begin{equation}
\label{hamtr}
\Theta \circ \mc T_t = \mc T^*_{-t} \circ \Theta.
\end{equation}

The relationship \eqref{hamtr} is non trivial also for reversible
processes, i.e.\ when $\mathcal{H}=\mathcal{H}^*$, in such a case it
tells us how to change the momentum under time reversal.  This
definition of time reversal in a Hamiltonian context
agrees with the one given in \cite{MR0073464}.

\subsection{Path integral derivation of the Hamiltonian}
\label{s:pih}

In alternative to the previous argument, we provide here, following
e.g., \cite{DG2}, a derivation of the Hamiltonian \eqref{lHJ} from the
fundamental formula \eqref{r02} via a path integral calculation. 

When we are interested only in the fluctuations of the density, we can
use formally the fundamental formula as a probability distribution in
a path integral,
\begin{equation*}
\begin{split}
  &\mb P \big( (\rho_\epsilon, j_\epsilon) :\, \rho_\epsilon  \in A \big)
  \asymp \int_{A} \! \mc D \rho \! \int \! \mc D j \; 
  \delta \big( \partial_t \rho + \nabla \cdot j \big)  
  \\
  &
  \exp\Big\{ - \frac {\epsilon^{-d}}4 
  \int_{T_0}^{T_1}\!dt\!\int_\Lambda \! dx \, [j-J(\rho)]
  \cdot \chi(\rho)^{-1} [j-J(\rho)]
  \Big\}.
\end{split}
\end{equation*}
We can take into account the constraint of the $\delta$ function by
introducing an auxiliary field $\pi$. By Laplace asymptotics, 
\begin{equation*}
\begin{split}
  & \mb P \big( (\rho_\epsilon, j_\epsilon) :\, \rho_\epsilon  \in A \big)
  \asymp \int_{A} \! \mc D \rho \int \! \mc D j \int \! \mc D \pi \: 
  \\
  & 
  \exp\Big\{ - \epsilon^{-d} 
  \int_{T_0}^{T_1}\!dt\!\int_\Lambda \! dx \, \big( \partial_t \rho +
  \nabla \cdot j \big) \pi  
  \Big\}
  \\
  & \cdot 
  \exp\Big\{ - \frac {\epsilon^{-d}}4 
  \int_{T_0}^{T_1}\!dt\!\int_\Lambda \! dx \, [j-J(\rho)]
  \cdot \chi(\rho)^{-1} [j-J(\rho)]
  \Big\}.
\end{split}
\end{equation*}
Integrating by parts the term $(\nabla\cdot j)\pi $ and computing the
Gaussian integral over $j$ we get 
\begin{equation*}
\begin{split}
  &\mb P \big( (\rho_\epsilon, j_\epsilon) :\, \rho_\epsilon  \in A \big)
  \asymp \int_{A} \! \mc D \rho \! \int\! \mc D \pi 
  \\
  &\quad \exp\Big\{ - {\epsilon^{-d}} \int_{T_0}^{T_1}\!dt\,\Big[\int_\Lambda \! dx \, 
  \pi \, \partial_t \rho  -\mc H(\rho,\pi)\Big] \Big\}
\end{split}
\end{equation*}
where $\mc H$ is the Hamiltonian \eqref{lHJ}.
The exponential in the previous equation is the action corresponding
to the Hamiltonian $\mc H$. In the limit $\epsilon \to 0$ the
dominating contributions are thus in a neighborhood of solutions of  
the canonical equations \eqref{ham-eq}.

\subsection{Non differentiability of the quasi-potential: Lagrangian
  phase transitions}
\label{s:lpt}

The quasi-potential may exhibit singularities which can be interpreted
as non-equilibrium phase transitions.  
In a finite-dimensional setting, an analogous phenomenon is discussed
in \cite{473102246} where it is shown that the quasi-potential  
generically exhibits points of non differentiability. An interesting example of
this kind has been discussed in \cite{473102247}. We refer to
\cite{DMS} for further developments and examples. 

We discuss this phenomenon in the infinite dimensional context of the
MFT \cite{lpt}.
For equilibrium states, the quasi-potential $V$ is always convex and
the occurrence of first order phase transitions corresponds to the
presence of a flat part in the graph of $V$. In non-equilibrium states
$V$ is not necessarily convex and phase transitions without an
equilibrium analogue can occur. These phase transitions have a natural
geometric interpretation in the Hamiltonian formalism that we next
illustrate.

Recall that the Hamiltonian dynamics admits the equilibrium point
$(\bar{\rho}, 0)$.  The corresponding energy vanishes,
$\mc H (\bar\rho,0)=0$.  Consider the solution of the canonical
equations \eqref{ham-eq} with initial condition $(\rho, 0)$. As
$\bar{\rho}$ is globally attractive for the hydrodynamics, such a
solution of the canonical equations converges to the equilibrium
point $(\bar{\rho}, 0)$ as $t\to+\infty$. The set
$\{(\rho,\pi)\,:\,\pi=0\}$ is therefore the stable manifold $\mc
M_\mathrm{s}$ associated to the equilibrium point $(\bar{\rho},
0)$.  The unstable manifold $\mc M_\mathrm{u}$ is defined as the set
of points $(\rho,\pi)$ such that the solution of the canonical
equations starting from $(\rho,\pi)$ converges to $(\bar{\rho}, 0)$ as
$t\to - \infty$.  By the conservation of the energy, $\mc
M_\mathrm{u}$ is a subset of the manifold $\{(\rho,\pi)\,:\: \mathcal
H(\rho,\pi)=\mathcal H(\bar{\rho},0) =0\}$.

In the sequel we need the following result in Hamiltonian dynamics,
e.g., \cite{A}.  Given a closed curve $\gamma$ parametrized as
$\gamma(\alpha)=(\rho(\alpha),\pi(\alpha))$,  $\alpha\in [0,1]$, the integral
$\oint_\gamma \pi\,d\rho = \int_0^1\!d\alpha \int_\Lambda\!dx\,
\pi(\alpha,x)\, \partial_\alpha \rho(\alpha,x)$ is invariant under the
Hamiltonian evolution. 
This means that, by denoting with $\gamma_t$
the evolution of $\gamma$ under the Hamiltonian flow,
$\oint_{\gamma_t} \pi \,d\rho =\oint_\gamma \pi\,d\rho$.  
In view of this result, if $\gamma$ is a closed curve contained in the
unstable manifold $\mc M_\mathrm{u}$ then $ \oint_\gamma \pi\,d\rho =
\lim_{t\to -\infty} \oint_{\gamma_t} \pi\,d\rho =0$.  We can
therefore define the pre-potential $\mc V :\,\mc M_\mathrm{u}\to\mb R$
by
\begin{equation}
\label{f23}
\mc V (\rho, \pi) \;=\; \int_\gamma \hat\pi\,d\hat\rho ,
\end{equation}
where the integral is carried over an arbitrary curve
$\gamma(\alpha)$, $\alpha\in [0,1]$, in $\mc M_\mathrm{u}$
such that $\gamma (0) = (\bar{\rho}, 0)$ and 
$\gamma(1)=(\rho, \pi)$.
The possibility of defining such potential is usually referred to by
saying that $\mc M_\mathrm{u}$ is a Lagrangian manifold.

We now establish the relationship between the quasi-potential and the
pre-potential
\begin{equation}
\label{f13}
V(\rho) \;=\; \inf
\big\{ \mc V  (\rho, \pi)\,,\: \pi\,: \: (\rho,\pi) \in \mc M_\mathrm{u} \big\}.
\end{equation}
Indeed, fix $\rho$ and consider $\pi$ such that $(\rho,\pi)$ belongs
to $\mc M_\mathrm{u}$. Let $(\hat\rho(t), \hat\pi(t))$ be the solution
of the Hamilton equations starting from $(\rho, \pi)$ at $t=0$.  Since
$(\rho,\pi)\in \mc M_\mathrm{u}$, $(\hat\rho (t), \hat\pi(t))$
converges to $(\bar\rho, 0)$ as $t\to -\infty$.  Therefore, the path
$\hat\rho(t)$ is a solution of the Euler-Lagrange equations for the
action $I_{(-\infty, 0]}$, which means that it is a critical path for
\eqref{r07}.  Since $\mathcal L(\hat\rho,\partial_t \hat{\rho} ) =
\int_\Lambda \hat\pi \, \partial_t \hat{\rho} - \mathcal H
(\hat\rho,\hat\pi)$ and $\mathcal H (\hat\rho (t),\hat\pi(t)) =0$, the
action of such path $\hat\rho(t)$ is given by $I_{(-\infty,
  0]}(\hat\rho)=\mc V (\rho,\pi)$.  Hence the right hand side of
\eqref{f13} selects among all such paths the one with minimal action,
and the minimal action is, by definition, the quasi-potential
$V(\rho)$.

For equilibrium states, the quasi-potential is given by the expression
in \eqref{a1} and it is simple to check that the unstable manifold is
$\mc M_\mathrm{u} = \big\{ (\rho,\pi)\:: \, \pi =
f'(\rho)-f'(\bar\rho)\big\}$. In particular, $\mc M_{u}$ is globally a
graph which means that for every $\rho$ there exists an unique $\pi$
so that $(\rho,\pi)\in \mathcal M_{u}$.  On the other hand, for
non-equilibrium states the unstable manifold is not necessarily a
graph and it may happen, for special $\rho$, that the variational
problem \eqref{f13} admits more than a single minimizer (Figure 1.a).
The set of profiles $\rho$ for which the minimizer is not unique is a
caustic. In general, it is a codimension one submanifold of the
configuration space.  We call the occurrence of this situation a
Lagrangian phase transition. In this case, profiles arbitrarily close
to each other but lying on opposite sides of the caustic are reached
by optimal paths which are not close to each other. This implies that
on the caustics the first derivative of the quasi-potential is
discontinuous (Figure 1.b).

\medskip

\begin{figure}[h]
\label{f:lft}
\begin{picture}(200,165)(-60,-105)
{
\put(-60,55){(a)}
%
\thinlines
\put(-50,0){\line(1,0){160}}
\put(0,-30){\line(0,1){90}}
\put(-2.2,-2.2){$\bullet$}
\put(2,-7){$\bar{\rho}$}
\put(112,-2.5){$\rho$}
\put(-2,63){$\pi$}
%
\thicklines
\qbezier(-35,-15)(-10,-10)(0,0)
\qbezier(0,0)(30,30)(80,10)
\qbezier(80,10)(120,-5)(85,20)
\qbezier(85,20)(70,30)(40,30)
\qbezier(40,30)(0,30)(50,40)
\qbezier(50,40)(93,50)(95,53)
\put(-20,-11.5){\vector(-2,-1){0}}
\put(30,17){\vector(3,1){0}}
\put(70,45.5){\vector(3,1){0}}
\put(65,50){$\mc M_\mathrm{u}$}
%
\thinlines
\put(50,-10){\line(0,1){65}}
\put(52,-7){$\rho_\mathrm{c}$}
\put(47.7,15.5){$\bullet$}
\put(-10,15.5){$\pi_1$}
\put(47.7,27.5){$\bullet$}
\put(-10,27.5){$\pi_2$}
\put(47.7,37.7){$\bullet$}
\put(-10,37.7){$\pi_3$}
%
%
\put(-60,-55){(b)}
%
\thinlines
\put(-50,-100){\line(1,0){160}}
\put(0,-105){\line(0,1){60}}
\put(-2.2,-102.2){$\bullet$}
\put(2,-107){$\bar{\rho}$}
\put(112,-102.5){$\rho$}
%
\thicklines
\qbezier(-40,-60)(-30,-100)(0,-100)
\qbezier(0,-100)(40,-100)(50,-70)
\qbezier(50,-70)(80,-60)(95,-45)
\put(47.7,-72.5){$\bullet$}
\put(-35,-60){$V(\rho)$}
\put(47,-107){$\rho_\mathrm{c}$}
%
\thinlines
\put(41,-97){\line(1,3){14}}
\put(38,-74){\line(3,1){45}}
}
\end{picture}
\caption{
  (a) Picture of the unstable manifold.
  (b) Graph of the quasi-potential. $\rho_\mathrm{c}$ is a caustic
  point, e.g., $\mc V(\rho_\mathrm{c},\pi_1)=\mc V(\rho_\mathrm{c},\pi_3)$.
}
\end{figure}
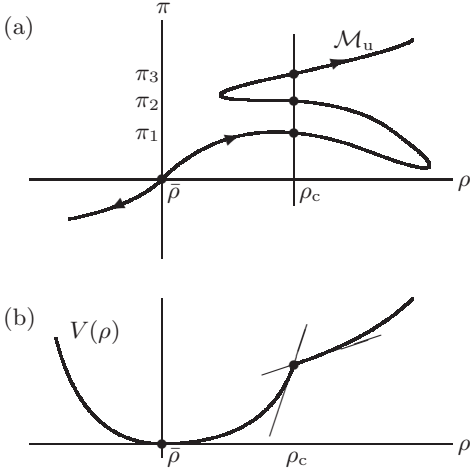

Recall the discussion in Section~\ref{s:df} showing that the optimal
field $F$ in \eqref{la20} to produce the profile $\rho$ is given by $F
= 2 \nabla \frac {\delta V}{\delta \rho}$. If $\rho$ is a caustic
point then the functional derivative of $V$ is not defined. However we
can take a profile $\rho + \tilde\rho$ close to caustic, compute the
derivative at $\rho+\tilde\rho$, and then take the limit as
$\tilde\rho \to 0$. However, since $V$ has a first order
discontinuity, we obtain different values for the
limiting derivative.  In this way, if $\rho$ is a caustic point, we
can construct different fields $F$ in \eqref{la20} such
that the corresponding action \eqref{222} is equal to $V(\rho)$. To
each field $F$ there corresponds an optimal trajectory for the
variational problem \eqref{r07}.

The previous geometrical considerations make plausible that, for a
non-equilibrium state, Lagrangian phase transitions do generically
occur.  The question whether a specific model, characterized by its
transport coefficients, exhibits such transitions is a completely
different story. 
It is remarkable, as we shall see in section \ref{s:exlpt},
that the weakly asymmetric simple exclusion can be proven analytically
to have such phase transitions.

\medskip
We conclude with some remarks on the possibility of observing
Lagrangian phase transitions.  Conceptually, they can be directly
detected from a detailed statistics of the stationary non-equilibrium
ensemble; Lagrangian phase transitions correspond to the presence of
corners in the graph of the probability distribution function in
logarithmic scale.  Alternatively, one could exploit the instability
of the exit path.  As mentioned, optimal exit paths
have been experimentally observed in noisy electronic devices with a
finite number of degrees of freedom \cite{LMc,PhysRevLett.100.130602}.
On the other hand, in thermodynamic systems the thermal fluctuations
are very small and the direct observation of Lagrangian phase
transitions appears quite difficult, as it requires an extremely long
time. 
The problem of large fluctuations admits an interpretation as a
control problem \cite{MR2082195}. This means that rather than
considering the optimal path, we look for the field driving the system
from the stationary state to a chosen profile with the minimal
energetic cost. The Lagrangian phase transition then corresponds to
the existence of two different optimal fields dissipating the same
energy.  In principle, these two fields can be theoretically
calculated and an experiment can be designed to check the predictions.

\subsection{Non locality of the quasi-potential
and long range correlations}

Long range correlations are a generic property of stationary non
equilibrium states which have been experimentally measured, see
\cite{DKS} for a review.  At the theoretical level, several approaches
have been developed around late 70's - early 80's, e.g.\
\cite{PhysRevA.26.950, PhysRevLett.42.287, MR732737}.  In the MFT,
long range correlations are a direct consequence of the non locality
of the quasi-potential. By perturbatively solving the Hamilton-Jacobi
equation, the equations for correlations of arbitrary order have been
obtained \cite{BDGJL-tow}.

We introduce the \emph{pressure} functional as the Legendre
transform of the quasi-potential $V$,
\begin{equation*}
V^\sharp (h) = \sup_\rho\Big\{\int_\Lambda \!dx\,  h \rho - V(\rho)\Big\}.
\end{equation*}
The large deviations
asymptotics \eqref{ld} implies
\begin{equation}
  \label{lgf}
  \lim_{\epsilon\to 0 } {\epsilon^{d}} \log E_P\left(
  \exp\Big\{  \epsilon^{-d} \int_\Lambda\! dx\,
  \rho_\epsilon \, h \Big\} \right) = V^\sharp (h),
\end{equation}
where we recall that $\rho_\epsilon$ denotes the empirical density
and $P$ is the stationary ensemble.
We point out that the above asymptotics does not imply in general
\eqref{ld}. Indeed, while the functional $V^\sharp $ is always convex, the
functional $V$ can be recovered as the Legendre transform of $V^\sharp $ only
when it is convex.  For example, for the KMP model, $V$ turns out to
be not convex, \cite{BGL}. On the other hand, \eqref{lgf} does suffice
to recover $V$ in a small neighborhood of the stationary profile
$\bar\rho$.

By taking derivatives, \eqref{lgf} yields the asymptotics of
truncated correlations of the empirical density,
\begin{equation}
  \label{acf}
  \lim_{\epsilon\to 0}   \big( \epsilon^{-d}\big)^{ n-1}  E_P\big(
  \rho_\epsilon(x_1) ; \,\cdots \, ;\rho_\epsilon(x_n)\big)
  = C_n(x_1,\dots,x_n)
\end{equation}
where
\begin{equation}
\label{alb5}
C_n(x_1,\dots,x_n)=
\frac{\delta^n V^\sharp }{\delta h(x_1)\cdots\delta h(x_n)}\,\Big|_{h=0}.
\end{equation}

By Legendre duality  we have the change of variable formula
$h=\frac{\delta V}{\delta\rho},\,\rho=\frac{\delta V^\sharp }{\delta h}$,
so that  the Hamilton-Jacobi equation  \eqref{r15-2}
can then be rewritten in terms of $V^\sharp $ as
\begin{equation}
\label{s2}
\int_\Lambda\!dx\, \nabla h \cdot
\chi\Big(\frac{\delta V^\sharp }{\delta h} \Big)\nabla h
+\int_\Lambda\!dx\,  \nabla h \cdot
 J \Big(\frac{\delta V^\sharp }{\delta h}\Big) = 0
\end{equation}
where $h$ vanishes at the boundary of $\Lambda$.
This an equation for the generating function $V^\sharp $, which by
Taylor expansion yields a recursive relationship for the macroscopic
correlations $C_n$.

Write the two-point correlation function in the form
$$
C_2(x,y)=C_{\mathrm{eq}}(x)\delta(x-y)+B(x,y)
$$
where
$$
C_{\mathrm{eq}}(x)= D^{-1}(\bar\rho(x))\chi(\bar\rho(x)).
$$
By expanding \eqref{s2} around the stationary profile $\bar\rho$
we  obtain the following equation for $B$
\begin{equation}\label{s5}
\mathcal L^\dagger B(x,y) = \alpha(x)\delta(x-y).
\end{equation}
The operator $\mathcal L^\dagger$ is the formal adjoint of the
differential operator $\mathcal L=L_x+L_y$, where
\begin{equation}\label{lx}
L_x = D_{ij}(\bar\rho(x))\partial_{x_i}\partial_{x_j} +
\chi^\prime_{ij}(\bar\rho(x)) E_j(x)\partial_{x_i}
\end{equation}
and, 
\begin{equation*}\label{alfa1}
\alpha(x) = \partial_{x_i}\big[
\chi^\prime_{ij}\big(\bar\rho(x)\big)\,
D^{-1}_{jk}\big(\bar\rho(x) \big) J_{k}(\bar \rho(x)) \big].
\end{equation*}
We are using the convention  that repeated indices are summed.
When $\alpha(x)=0$, due to the boundary conditions, the unique
solution to \eqref{s5} is $B=0$ and there are no
long range correlations.
In the case of equilibrium states $\alpha(x)=0$ since the
current vanishes. There are cases in which $\alpha(x)=0$ even
if $J(\bar \rho(x))\neq 0$.
This happens in the Ginzburg--Landau model \cite{GPV} where $\chi$ does not depend on $\rho$.
Another case is the zero range model discussed in Section~\ref{s:zr}.
If $\alpha(x)$ is non-vanishing the inhomogeneous equation \eqref{s5} has
a non-trivial solution and long range correlations are present.

Since $\mathcal L$ is an elliptic operator
(i.e. it has a
negative kernel), the sign of $B$ is determined by the sign of $\alpha$:
if $\alpha(x) \ge 0$, then $B(x,y) \le 0$,
while if $\alpha(x) \le 0$, then $B(x,y) \ge 0$.
For example, consider the following special case. The system is
one-dimensional, the diffusion coefficient is constant, 
$D(\rho)=D_0$,  the mobility $\chi(\rho)$ is a quadratic function of $\rho$, and there
is no external field, $E=0$. Then
\begin{equation}
\label{opp}
B(x,y) = - \frac 1{2D_0} \chi'' (\nabla \bar\rho)^2 \Delta^{-1}(x,y),
\end{equation}
where $\Delta^{-1}(x,y)$ is the Green function of the Dirichlet
Laplacian. 
Two well studied models, the symmetric exclusion process,
where $\chi(\rho)=\rho(1-\rho)$, and the KMP process,
where $\chi(\rho)=\rho^2$, meet the above conditions. Then \eqref{opp}
shows that their correlations have opposite signs.

By developing the arguments presented above, it is possible to deduce
recursive equations for the $n$-point correlations $C_n$; we refer to
\cite{BDGJL-tow} for the details of this analysis.

\medskip
The existence of long range correlations in stochastic lattice gases
and in particular in the symmetric simple exclusion process was first
established, using fluctuating hydrodynamics and a direct computation,
in \cite{MR732737}. We refer to \cite{gkr} for more recent microscopic
results.  Our derivation shows that long range correlations in
diffusive systems with a conservation law are a generic consequence of
inhomogeneous chemical potentials and external fields.  In real
systems couplings between different fluctuating quantities generate
non-equilibrium long range correlations as discussed in \cite{os}.
These authors consider the coupling of temperature fluctuations with
velocity fluctuations. The velocity fluctuating field formally appears
as an external field in the hydrodynamic equation for the temperature
fluctuations.  For the experimental situation the reader may consult
the review \cite{DKS}.

\subsection{Current fluctuations}
\label{s:cf}

Both from a theoretical and an experimental point of view, a natural
observable in non-equilibrium thermodynamics is the time averaged
current.
The corresponding fluctuations have been analyzed in \cite{BD04}. By
postulating an additivity principle, which relates the fluctuation of
the time averaged current in the whole system to the fluctuations in
subsystems, the corresponding asymptotic probability is deduced.
However, as pointed out in \cite{noiprlcurrent,MR2227084}, this
approach may underestimate the probability of fluctuations due to the
possible occurrence of a dynamical phase transition.

We show that the probability of fluctuations of the time averaged
current in the time window $[0,T]$ can be derived, without additional
assumptions, from the macroscopic fluctuation theory.  The probability
of observing a time averaged fluctuation $J$ can be described by a
functional $\Phi(J)$ which we characterize in terms of a variational
problem for the functional $\mathcal I_{[0,T]}$.

Recall that $j_\epsilon$ is the empirical current, described after
equation \eqref{r02}.  Given a vector field $J$, by the fundamental
formula \eqref{r02},
\begin{equation}
\label{LT}
\mb P_{\rho_0} \Big( \frac 1T \int_0^T \!dt \;
j_\epsilon(t) \approx J
\Big) \asymp  \exp \big\{-\epsilon^{-d}\, T \, \Phi_T (J)
\big\},
\end{equation}
where $\Phi_T$ is given by
\begin{equation}
\label{r23}
\Phi_T(J)\;=\; \frac 1T \, \inf_{(\rho,j)\in \mc A_{T}}
\mc I_{[0,T]} (\rho,j).
\end{equation}
In this formula, $\mc A_{T}$ is the set of paths $(\rho,j)$ whose
average current is $J$ and initial density is $\rho_0$,
\begin{equation*}
\begin{split}
& \mc A_{T}
= \Big\{ (\rho,j)  :\,
\frac1T \int_0^T \!\! {dt}  \, j(t) = J \,,\\
&\qquad\qquad 
\partial_t \rho=-\nabla\cdot j\,, 
\rho(0) = \rho_0
\Big\}.
\end{split}
\end{equation*}

By the local conservation of the mass, the asymptotic $T\to\infty$ of
the above probability is relevant only for divergence free vector
fields. Indeed, the case in which $J$ has not zero divergence leads
either to negative mass or to a mass condensation.

For a divergence free current $J$ the sequence $T\, \Phi_T(J)$ is
subadditive in $T$,
\begin{equation}
\label{r22}
(T+S) \Phi_{T+S}(J) \le T\, \Phi_T(J) + S\, \Phi_S(J)
\end{equation}
for $T$, $S\ge 0$. Indeed, let $(\rho_1,j_1) \in \mc A_{T}$ and
$( \rho_2,j_2) \in \mc A_{S}$. As $J=(1/T) \int_0^T \!dt \: j(t)$ is divergence free,
by the continuity equation
$\rho_1(0)=\rho_1(T) + \nabla \cdot \int_0^T \!dt \: j(t)=\rho_1(T)=\rho_2(0)$.
We may therefore glue the
trajectories $(\rho_1,j_1)$ and $(\rho_2,j_2)$, obtaining a trajectory
$(\rho,j)$ in $\mc A_{T+S}$ which satisfies
\begin{equation}
\label{r25}
\mathcal I_{[0,T+S]} (\rho,j)  = 
\mathcal I_{[0,T]} (\rho_1,j_1) +  \mathcal I_{[0,S]} (\rho_2, j_2)
\end{equation}
Therefore, optimizing over all the trajectories, we obtain \eqref{r22}.

Since for a divergence free current $J$ the sequence $T\, \Phi_T(J)$
is subadditive in $T$, $\Phi_T(J)$ converges to a limit denoted by
$\Phi (J)$, given by
\begin{equation}
\label{r24}
\begin{split}
\Phi (J)
 \; & = \; \lim_{T\to\infty} \; \inf_{(\rho,j)\in \mc A_{T}}
\frac 1T \; \mc I_{[0,T]} (\rho, j) \\
\; & = \;\inf_{T>0} \; \inf_{(\rho,j)\in \mc A_{T}}
\frac 1T \; \mc I_{[0,T]} (\rho, j)\;.
\end{split}
\end{equation}

The limit $\Phi(J)$ does not depend on the initial condition
$\rho_0$. Indeed given two
different initial conditions they can be connected by
a transient in a finite time that will be irrelevant for the limit.

We now prove that $\Phi$ is a convex functional. Let $0<p<1$ and
$J=pJ_1+(1-p)J_2$, we want to show that $\Phi(J)\leq
p\Phi(J_1)+(1-p)\Phi(J_2)$. Fix $T>0$ and an initial density profile
$\rho_0$.  Let $(\rho_1,j_1) \in \mc A_{pT}$, and $(\rho_2,j_2) \in
\mc A_{(1-p)T}$ be the optimal paths for the variational problem
\eqref{r23} associated to the currents $J_1$, $J_2$,
respectively. Therefore
\begin{equation*}
\begin{split}
& \Phi_{pT}(J_1) = \frac 1{pT} \ \mc I_{[0,pT]} (\rho_1, j_1) \;,\\
& \quad \Phi_{(1-p)T}(J_2) =
\frac 1{(1-p)T} \, \mc I_{[0,(1-p)T]} (\rho_2, j_2) \;.
\end{split}
\end{equation*}

By the same arguments used in \eqref{r25}, the path obtained
by gluing $j_1$ with $j_2$, denoted by $j$, is in the set $\mc
A_{T}$. Therefore,
$$
\Phi_T(J)  \le  \frac 1T \: \mc I_{[0,T]} (\rho, j)
 =   p \, \Phi_{pT}(J_1) + (1-p) \, \Phi_{(1-p)T}(J_2) .
$$
By taking the limit $T\to\infty$ and since the limiting function does
not depend on the initial condition, we conclude that $\Phi$ is
convex.  These arguments are standard in proving the existence and
the convexity of thermodynamic functions in equilibrium statistical
mechanics.

\medskip 
We introduce the functional $\mc U$ on the set of time independent profiles
$\rho=\rho(x)$ and $j=j(x)$ 
\begin{equation}
\label{la31}
\mc U (\rho,j) = \frac14
\int_{\Lambda} dx\,
[j - J(\rho)] \cdot \chi(\rho)^{-1}
[j - J(\rho)].
\end{equation}
We then  define $U$ on divergence free currents by
\begin{equation}
\label{r27}
U(J) = \inf_\rho\, \mc U(\rho, J),
\end{equation}
where the minimum is carried over all profiles $\rho$ satisfying the
boundary condition \eqref{2.3}.
We show that
\begin{equation}
\label{1ub}
\Phi(J) \le U(J)\;.
\end{equation}
To see this, since $\Phi(J)$ does not depend on the initial condition,
choose as initial condition the density profile $\rho_0$ which minimizes
\eqref{r27}.
Since $J$ is divergence free, the constant path $(\rho_0,J)$
lies in $\mc A_T$.
Hence, $\Phi_T(J)\leq\frac1T\mc I_{[0,T]}(\rho_0,J)=U(J)$.
The functional $U$ is in general non convex.

In one space dimension, the functional $U$ is the one introduced in
\cite{BD04}. Therefore, the additivity principle postulated there
provides the correct asymptotics when equality holds in \eqref{1ub}.
This is the case for some models and corresponds to the situation in
which the optimal path in \eqref{r24} does
not depend on time.  
On the other hand, as we shall see in Section \ref{s:dft}, for other
models the inequality in \eqref{1ub} is strict and this corresponds to
a spontaneous symmetry breaking of time translation invariance.

We now argue that for small deviations of the current, i.e.\
in a neighborhood of the stationary current $J(\bar\rho)$,
dynamical phase transitions do not occur, i.e.\ $\Phi=U$. 
Observe that for $J(\bar \rho)$ we have
$\Phi(J(\bar\rho))=U(J(\bar\rho))=0$ and this is uniquely realized by choosing
on the right hand side
of \eqref{r24} the time independent path $(\bar\rho,J(\bar\rho))$. For $J$
close to $J(\bar\rho)$ the optimal path for the right hand side of
\eqref{r24}, possibly time-dependent,  will be close to $(\bar\rho, J(\bar\rho))$. Since the path
$(\rho^*,J)$ where $\rho^*$ is the optimal profile in \eqref{r27} is a
stationary point for the right hand side of \eqref{r24}, by continuity it also will be
the global minimizer.

The asymptotics \eqref{LT} can be formulated in terms of the
moment generating function of the empirical current. For each time
independent, divergence-free vector field $v=v(x)$ we have
\begin{equation}
\label{fa1}
\lim_{T\to\infty}\lim_{\epsilon \to 0}
\frac{\epsilon^d}{T} \log
\mb E_{\rho_0} \Big[
e^{ \epsilon^{-d}\,  \int_0^Tdt \int_{\Lambda} dx\,
j_\epsilon (t) \cdot v }
\Big] = \Phi^{\sharp} (v)
\end{equation}
where $\mb E_{\rho_0}$ denotes the expectation with respect to the
probability distribution $\mb P_{\rho_0}$, and $\Phi^{\sharp} (v)$ is
the Legendre transform of $\Phi (J)$,
\begin{equation}
\label{r31}
\Phi^{\sharp} (v) = \sup_{J} \Big\{ \int_{\Lambda}dx\, v\cdot J
- \Phi (J)\Big\}.
\end{equation}
The supremum is carried over all the divergence free vector
fields $J$.

In connection with the functional $\Phi$, Varadhan suggested \cite{V} 
the possibility of the alternative variational representation
\begin{equation}
\label{r26}
\Phi(J) \;=\; \inf \big\<  \, \mc U (\rho(t),j(t)) \, \big \>.
\end{equation}
In this formula $\<\cdot\>$ represents the expectation with respect to
a stationary process $(\rho,j)$, and the infimum is carried over all
such stationary processes satisfying the continuity equation
$\partial_t\rho+\nabla\cdot j=0$ and the constraint $\< j(t)\> = J$.
Note that $\big\< \, \mc U (\rho(t),j(t)) \, \big \>$ does not depend
on $t$ by stationarity. 
The representation \eqref{r26} is not used in this paper.

\medskip
The fundamental formula \eqref{r02}--\eqref{r05} can be used to analyze
the fluctuation of the current flux across a surface. As shown in
\cite{vortices}, for models in two dimension the asymptotics for
closed or open curves are different due to the possible occurrence of
vortexes around the endpoints.

\subsection{Gallavotti--Cohen symmetry}

Denote by $\Phi^*$ the functional defined by the variational problem
\eqref{r24} with $\mathcal I^*$ in place of $\mathcal I$. By \eqref{r28} and since
$\theta j (t)= -j(-t)$,
\begin{equation*}
\Phi(J)=\Phi^*(-J)\;.
\end{equation*}
For equilibrium states this symmetry states that the functional
$\Phi$ is even.

Let us consider a path $j(t)$, $t\in [-T,T]$ such that $(2T)^{-1}
\int_{-T}^{T}\!dt \: j(t) = J$ for some divergence free vector field
$J$. Recalling the Einstein relation
$D(\rho)\chi(\rho)^{-1}=f''(\rho)$ we have that
$$
\chi(\rho)^{-1} J(\rho) = - \nabla f'(\rho) + E
$$
Recall \eqref{r05} and \eqref{r09}. Since $f'(\rho(x))=\lambda (x)$,
$x\in \partial \Lambda$, an integration by parts yields
\begin{equation}
\label{GC1}
\begin{split}
\frac{1}{2T}
\mc R_{[-T,T]}(\rho,j) & =
\frac{1}{2T} \mc R_{[-T,T]}(\theta\rho,\theta j) \\
& - \int_{\Lambda}dx\,  J\cdot E
+ \int_{\partial\Lambda} \!d\sigma \: \lambda \, J \cdot \hat{n}
\end{split}
\end{equation}
where $d\sigma$ is the surface measure on $\partial\Lambda$ and
$\hat{n}$ is the outward normal to $\Lambda$.  In particular this
relation implies that if $(\hat\rho,\hat\jmath) $ is an optimal path for
the variational problem defining $\Phi(J)$ then
$(\theta\hat\rho,\theta\hat\jmath)$ is an optimal path for the
variational problem defining $\Phi(-J)$.

By taking the limit $T\to \infty$ in (\ref{GC1}) we get
\begin{equation}
\label{r30}
\Phi(J)- \Phi(-J)=
- \int_{\Lambda}dx\,  J\cdot E
+ \int_{\partial\Lambda} \!d\sigma \: \lambda \, J \cdot \hat{n}  \;,
\end{equation}
which is a Gallavotti--Cohen type symmetry in our space time dependent
setup for macroscopic observables.
Note that the right hand side of \eqref{r30} is minus the energy given to
the system by the external field and the boundary reservoirs per unit
time \eqref{W=}.

When $\Phi=U$ the symmetry \eqref{r30} can be generalized as follows
\cite{PNAS}.
Consider $J$ and $J'$ two divergence free currents such that
$|J(x)|^2=|J'(x)|^2$ then it is immediately seen that
\begin{equation}
\begin{split}
U(J)- U(J')
&= \frac 12\int_{\partial\Lambda} \!d\sigma \:
\lambda \, \left(J-J'\right) \cdot \hat{n} \\
& - \frac 12\int_{\Lambda}dx\,  \left(J-J'\right)\cdot E
 \,.
\end{split}
\end{equation}
If $\Phi=U$, by taking $J'=-J$ we recover \eqref{r30}.

\subsection{Extended Hamiltonian structure}

In section \ref{s:hs} we discussed the Hamiltonian structure related to
the density fluctuations. Here we show that there is an underlying
(richer) Hamiltonian structure for the joint fluctuations of density
and current. 

To this end we write \eqref{r05} as an action associated to a
Lagrangian. This is possible using some simple changes of
variables. We consider the time interval $[0,T]$ and assume that the
external drivings do not depend on time. 
Let $A_0(x)$ be a vector field related to the initial condition
by $\nabla\cdot A_0(x)=\rho(x,0)$. For example we can fix $
A_0=-\nabla h$ where $h$ solves $\Delta h(x)=-\rho(x,0)$. We then define
the vector field
\begin{equation}\label{change-j-A}
A(t,x)=A_0(x)-\int_0^tj(s,x)\,ds\,,
\end{equation}
that, apart from the initial condition and a minus sign, is the time
integrated current.
Since $\rho$ and $j$ are related by the continuity equation we have
$j=-\partial_tA$ and $\rho=\nabla\cdot A$. 

We can then write  the rate functional \eqref{r05}
in terms of the vector field $A$ 
\begin{equation}
\begin{split}
\mc I_{[0,T]}(A)&=
\frac 14 \int_0^Tdt\int_\Lambda dx\,\Big(\partial_tA+J(\nabla
\cdot A)\Big)\\
&\quad \cdot \chi^{-1}(\nabla \cdot A)\Big(\partial_tA+J(\nabla\cdot
A)\Big)\,.
\end{split} 
\label{princ-A}
\end{equation}
Observe that, in this form, the constraint of the continuity equation is
automatically satisfied. 
Formula \eqref{princ-A} has the form of an action for the Lagrangian
\begin{equation}\label{lag-A}
\begin{split}
&\mathbb L\left(A,\partial_t A\right)=  \\
&\frac 14
\int_\Lambda \!dx \Big(\partial_tA+J(\nabla\cdot A)\Big)\cdot
\chi^{-1}(\nabla \cdot A)\Big(\partial_tA+J(\nabla\cdot A)\Big)\,. 
\end{split}
\end{equation}
The corresponding Hamiltonian is
\begin{equation}\label{ham-A}
\begin{split}
\mathbb H(A,B)=&\sup_\xi\left\{\int_\Lambda \!dx\, B(x)\cdot\xi(x)-\mathbb
  L\left(A,\xi\right)\right\} \\
=&\int_\Lambda dx\, \Big[B\cdot \chi(\nabla\cdot
A)B-B\cdot J(\nabla\cdot A)\Big] 
\end{split}
\end{equation}
and the canonical equations are
\begin{equation}\label{ham-eq-A}
\left\{
\begin{array}{ll}
\partial_tA =& 2 \chi(\nabla\cdot A)B-J(\nabla\cdot A)\,,\\ \\
\partial_tB=&-\nabla\Big[\textrm{Tr}\big(D(\nabla\cdot
A)\nabla^TB\big)\\
&+B\cdot \chi'(\nabla\cdot A)\left(E-B\right)\Big]\,, 
\end{array}
\right.
\end{equation}
where we denoted by $\nabla^TB$ the matrix having entries
$\left(\nabla^TB\right)_{i,j}=\partial_{x_i}B_j$ and recall that
$\textrm{Tr}(\cdot)$ denotes the trace. 

Given a solution $(\rho,\pi)$ of the canonical equations \eqref{ham-eq}
there corresponds a solution of \eqref{ham-eq-A} given by
\begin{equation}\label{up}
\left\{
\begin{split}
A(t)=& A_0-\int_0^t\!ds
\left[J(\rho(s))+2\chi(\rho(s))\nabla\pi(s)\right]\,,\\
B(t)=&-\nabla\pi(t)\,,
\end{split}
\right.
\end{equation} 
where $A_0$ satisfies the condition $\nabla\cdot A_0=\rho(0)$.

The momentum $B$ plays the role of the external field $F$ in
\eqref{la20}. When we look only at fluctuations of the density then
$B$ is a gradient vector field with potential $\pi$ as in \eqref{up}.
On the other hand when we study fluctuations of the current we need a
general vector field $B$. Correspondingly not all the solutions of
\eqref{ham-eq-A} are of the form \eqref{up}.

\section{Macroscopic models}
\label{sec3}

To illustrate the scope of the general theory developed so far, we
begin by discussing some cases where calculations can be made
explicitly.  From the point of view of the MFT, a system is defined by
the transport coefficient $D$ and $\chi$.  In this connection we
emphasize that many microscopic models can give rise to the same
macroscopic behavior encoded in such coefficients. Only in
special cases the microscopic models can be solved.  
Specific choices of the transport coefficients are named after the 
underlying microscopic models.
In Section \ref{s:8} we discuss how these coefficients can be
obtained from the microscopic dynamics.

\subsection{Equilibrium}
\label{s:eq}

We briefly look upon equilibrium states from the standpoint of
non-equilibrium.  Recall that we defined a system in the domain
$\Lambda$ to be in an equilibrium state when the current in the
stationary profile $\bar\rho$ vanishes, i.e.\ $J(\bar\rho) = 0$.  A
particular case is that of a homogeneous equilibrium state, obtained
by setting the external field $E=0$ and choosing a constant chemical
potential at the boundary, i.e., $\lambda(x) = \bar\lambda$.

For equilibrium states the quasi-potential, defined by the variational formula
\eqref{r07}, coincides with the functional $V$ in \eqref{a1}, that is
\begin{equation}
\label{Veq}
  V(\rho)
  = \int_{\Lambda} \!dx \:
  \big\{ f(\rho) - f(\bar\rho) -f'(\bar\rho)\big( \rho-\bar\rho\big)
  \big\}.
\end{equation}
We show that $V$ solves the Hamilton-Jacobi equation
\eqref{r15-2}. Its derivative is
\begin{equation}
\label{varf}
\frac{\delta V}{\delta\rho(x)} = f'(\rho(x))-f'(\bar\rho(x))
\end{equation}
so that, by an integration by parts,
\begin{equation}
\label{314}
  \begin{split}
    &\mc H \big(\rho, \tfrac{\delta V}{\delta\rho}\big) \\
   &  =
    \int_\Lambda\!dx\,
    \nabla\big[ f'(\rho) - f'(\bar\rho) \big] \cdot
    \chi(\rho) \nabla\big[ f'(\rho) - f'(\bar\rho) \big]
    \\
    &\quad + \int_\Lambda\!dx\,  \big[ f'(\rho) - f'(\bar\rho)
    \big] \nabla\cdot \Big[D(\rho)\nabla\rho - \chi(\rho)E \Big]
    \\
    & = \int_\Lambda\!dx\, \nabla\big[ f'(\rho) - f'(\bar\rho)
    \big] \cdot \chi(\rho) \Big[ \nabla f'(\bar\rho) - E \Big] =0
  \end{split}
\end{equation}
where we used \eqref{r29} and  $\nabla f'(\bar\rho) - E =
-\chi(\bar\rho)^{-1} J(\bar\rho)= 0$.

This statement is not sufficient to conclude that the functional in
\eqref{Veq} is the quasi-potential, observe for instance that $V=0$
always solves the Hamilton-Jacobi equation. In order to identify $V$
with the quasi-potential we need to verify that $V$ is the maximal
solution satisfying $V(\bar\rho)=0$. Clearly, $V$ is positive and zero
on  $\bar\rho$.
For checking that it is a maximal solution we refer to \cite{BDGJL-tow}.

\medskip
We next show that the condition $J(\bar\rho)=0$
is equivalent to either one of the following conditions.

\begin{itemize}
\item[--]{
There exists a function $\tilde\lambda: \Lambda \to \mb R$
such that
\begin{equation}
\label{cond_grad}
E(x)= \nabla\tilde\lambda (x)\,, \quad x\in \Lambda
\qquad \quad
\tilde\lambda(x) = \lambda (x) \,, \quad x\in \partial\Lambda\,,
\end{equation}
}
\item[--]{The system is \emph{macroscopically time reversal invariant}
    in the sense that for each profile $\rho$ we have 
    $J^*(\rho) =J(\rho)$.} 
\end{itemize}

We emphasize that the notion of macroscopic time reversal invariance 
does not imply that an underlying microscopic model satisfies the detailed
balance condition. Indeed, as it has been shown by explicit examples
\cite{MR1716811,MR1400378}, there are  microscopic
models not time reversal invariant for which $J^*(\rho) =J(\rho)$.

We start by showing that $J(\bar\rho) = 0$ if and only if
\eqref{cond_grad} holds.  From the local Einstein relation \eqref{r29}
and $J(\bar\rho) = 0$ we deduce
\begin{equation*}
E(x)= f''\big(\bar{\rho}(x) \big) \nabla\bar{\rho}(x)
= \nabla f'(\bar{\rho}(x))
\end{equation*}
hence \eqref{cond_grad}.
Conversely, let the external field $E$ be such
that \eqref{cond_grad} holds.  Since $f''$ is positive the function
$f'$ is invertible and we can define
$\bar{\rho}(x)=(f')^{-1}\big(\tilde\lambda(x) \big)$. The profile $\bar\rho$
satisfies \eqref{2.3} as well as $J(\bar{\rho})=0$.

We next show that $J(\bar\rho)=0$ if and only if $J(\rho)=J^*(\rho)$.
Suppose first that $J(\rho) = J^*(\rho)$.
By evaluating the Hamilton-Jacobi equation for $\rho=\bar\rho$ we
deduce $ \nabla \big[\delta  V (\bar\rho) / \delta \rho \big]
=0$. From the first equation in \eqref{r12} we then get $J(\bar\rho)=0$.
To show the converse implication, note that if $J(\bar\rho)=0$
then $V$ is given by \eqref{Veq}. We deduce that
\begin{equation*}
\chi(\rho) \nabla \frac {\delta V}{\delta \rho} =
D(\rho)\nabla\rho -  \chi(\rho) E = - J(\rho)
\end{equation*}
where we used \eqref{2.2}.
Recalling the first equation in \eqref{r12}
we get $J(\rho)=J^*(\rho)$.

\medskip
So far we have assumed the local Einstein relation and we have shown that
for equilibrium systems it implies \eqref{Veq}. Conversely,
we now show that macroscopic reversibility and \eqref{Veq}
imply the local Einstein relation \eqref{r29}.
If $J(\rho)=J^*(\rho)$ then \eqref{a2} holds, which reads, in view of
\eqref{Veq},  
\begin{equation}
\label{einbis}
\big[ \chi(\rho) f''(\rho)-D(\rho) \big] \nabla \rho
=
\chi(\rho) \big[ f''(\bar \rho)- \chi^{-1}(\bar \rho)D(\bar \rho)
\big]\nabla \bar \rho   
\end{equation}
where we used $J(\bar \rho)=0$ to eliminate $E$. 
Note that $J(\bar \rho)=0$  follows from the first equation in
\eqref{r12} and  $J(\rho)=J^*(\rho) $ without further assumptions.
Since $\rho$ and $\nabla \rho$ are arbitrary the local Einstein relation
$D = \chi \, f''$ follows from \eqref{einbis}.

\medskip
A peculiar feature of equilibrium states that allowed the explicit
derivation of the quasi-potential is that the optimal path for the
variational problem \eqref{r07} is the time reversal of the
hydrodynamic trajectory.
We emphasize that this can happen also if
the identity $J(\rho)=J^*(\rho)$ is violated but
$\nabla \cdot J(\rho)=\nabla
\cdot J^*(\rho)$ is satisfied.  Indeed, we next give an example of
a system not invariant under time reversal, i.e., with $J(\bar\rho)
\neq 0$, such that the optimal 
trajectory for the variational problem \eqref{r07} is the time
reversal of the solution to the relaxation trajectory.

Let $\Lambda = [0,1]$, $D(\rho)=\chi(\rho)=1$,
$\lambda(0)=\lambda(1)=\bar\lambda$, and a constant external field
$E\neq 0$. In this case the hydrodynamic evolution of the density is given
by the heat equation independently of the field $E$.
The stationary profile is $\bar\rho = \bar\lambda$, the associated
current is  $J(\bar\rho) = E \neq 0$.  By a computation analogous
to the one leading to \eqref{Veq}, we easily get that
\begin{equation*}
V(\rho) = \frac 12 \int_0^1\!dx \: \big[ \rho(x) - \bar\rho \big]^2
\end{equation*}
and the optimal trajectory for the variational problem \eqref{r07} is
the time reversal of the solution to the heat equation.
On the other hand $J(\rho)= - \nabla \rho +E$ while
$J^*(\rho)= -\nabla \rho -E$

\medskip
We remark that, even if $V$ is non local, the equality $J(\rho)=J^*(\rho)$ implies that the
thermodynamic force $- \nabla \left(\delta V / \delta \rho\right)$ is local.
Moreover, the first equation in \eqref{r12} reduces to the statement
\begin{equation}
\label{ein?}
J(\rho) =  - \chi(\rho) \nabla \frac{\delta V}{\delta \rho}   (\rho)
\end{equation}
so that $V$ can be obtained by integrating the above equation.
The identity \eqref{ein?} represents the general form, for equilibrium 
states, of the relationship between currents and thermodynamic
forces.  It holds both when the free energy is local and non
local. When \eqref{ein?} holds, the quasi-potential can be computed by an
integration.  An example of such a
situation with a non local free energy is provided by the ABC model on
a ring with equal densities \cite{EKKM}.

\subsection{Zero range}
\label{s:zr}

At the macroscopic level this model is specified by the choice
$\chi(\rho) = \phi(\rho)$ and $D(\rho) = \phi'(\rho)$, where $\phi$ is
an increasing function on $\mb R_+$. In particular, the local
Einstein relation \eqref{r29} holds with $f'= \varphi'/\varphi$.

This is a very special model in which the quasi-potential is a local
functional of the density that can be explicitly computed. It is given
similarly to the equilibrium case
\begin{equation}\label{zrqp}
V(\rho) \;=\; \int_{\Lambda} dx\, \big\{ f(\rho) -
f(\bar\rho) - f'(\bar\rho)
(\rho - \bar\rho) \big\},
\end{equation}
where $\bar\rho$ is the unique stationary solution of
\eqref{05}, which in the present case takes the form
\begin{equation*}
\left\{
\begin{split}
& \Delta \phi(\rho) = \nabla\cdot \phi(\rho) E \;, \quad
x\in\Lambda\;, \\
& \phi(\rho (x)) = e^{\lambda (x)}\;, \quad x\in\partial\Lambda\;.
\end{split}
\right.
\end{equation*}
The proof that the local functional \eqref{zrqp} solves the Hamilton-Jacobi equation \eqref{r15-2}
will be given in Section \ref{s:lqp}.

Assume that $d=1$, that the external field $E$ is constant, and that
$\Lambda = (0,1)$. We denote by $\lambda_0$, $\lambda_1$ the values of the
chemical potential at the endpoints. In this
context one can compute the functional $\Phi$ introduced in
\eqref{r24}. As we will show in section \ref{s:dft} for this model
there are no dynamic phase transitions and $\Phi=U$ where $U$ is the
functional introduced in \eqref{r27}.  Note that in one-dimension the
only vector fields with vanishing divergence are constant.
With the change of variable $\alpha(x)=\phi(\rho(x))$ the variational
problem \eqref{r27} reduces to
\begin{equation}\label{zrj}
  \inf_\alpha \frac 14\int_{0}^1dx\, \frac{\left(J+\nabla
      \alpha(x)-\alpha(x) E\right)^2}{\alpha(x)},
\end{equation}
where 
$\alpha(0)=e^{\lambda_0}=\varphi_0$, $\alpha(1)=e^{\lambda_1}=\varphi_1$.
This implies that $\Phi$ does not depend on the function $\phi(\rho)$
and in particular coincides with the one for a model of independent
particles, i.e.\, $\phi(\rho)=\rho$.

The optimal profile $\alpha$ of the variational problem
\eqref{zrj} is given by
\begin{equation*}
\alpha(x) =  C \Big( e^{E x} -a \Big)\Big( e^{- E x} - b \Big)
\end{equation*}
for suitable values of the constants $a,b,C$ to be determined by
the boundary conditions and the current $J$.
Using the explicit form of the minimizer we have
\begin{eqnarray*}
U(J) & = &J\log\left(\frac{\frac{J}{EA}+\sqrt{\frac{J^2}{E^2A^2}+4 
\frac BA}}{2}\right)\\
& -& E\left(A\sqrt{\frac{J^2}{E^2A^2}+4 \frac BA}-A-B \right),
\end{eqnarray*}
where $A=\frac{e^{\lambda_0}}{1-e^{-E}}$ and $B=\frac{e^{\lambda_1}}{e^{E}-1}$.
Its Legendre transform is
\begin{equation}
\label{fa2}
\begin{split}
\Phi^\sharp(v)
\;=\; E  \Big\{ A(e^{ v} -1)
+ B(e^{-v}-1) \Big\}.
\end{split}
\end{equation}
Notice that this solution converges, as $E\to 0$, to the solution
with no external field that can be easily obtained by the general formulas in
section \ref{s:cumulants}. For a microscopic counterpart see
\cite{1742-5468-2005-08-P08003}.

\subsection{Conditions for locality of the quasi-potential}
\label{s:lqp}

It is natural to ask under what conditions the quasi-potential
$V(\rho)$ is a local functional of the form \eqref{zrqp}, where
$\bar\rho=\bar{\rho}_{\lambda,E}$ is the stationary solution
associated to the boundary chemical potential $\lambda(x)$ and the
external driving field $E(x)$, and $f$ is the free energy density of
the model, related to the diffusion coefficient $D(\rho)$ and the
mobility $\chi(\rho)$ by the Einstein relation \eqref{r29}.

As a first observation, we show that $V(\rho)$ is local if and only if
\begin{equation}
\label{davide3}
\chi(\rho)^{-1} J_A(\rho)
\end{equation}
is independent of $\rho$.
Indeed, if $V(\rho)$ is as in \eqref{zrqp}, then
$J_S(\rho)=-\chi(\rho)\nabla\frac{\delta V}{\delta\rho}$ can be
computed explicitly, as well as $J_A(\rho)=J(\rho)-J_S(\rho)$.  The
result is
$$
J_A(\rho)=\chi(\rho)\big[E-\chi(\bar\rho)^{-1}D(\bar\rho)\nabla\bar\rho\big]
$$
Hence \eqref{davide3} is independent of $\rho$.
Conversely, if \eqref{davide3} is independent of $\rho$,
then the equation $\chi(\rho)^{-1}J_A(\rho)=\chi(\bar\rho)^{-1}J_A(\bar\rho)$,
can be rewritten, by \eqref{2.2}, \eqref{Ons-form}, and \eqref{r29}, as
$$
\nabla\frac{\delta V}{\delta\rho}(\rho) =\nabla\big[f'(\rho)-f'(\bar\rho)\big]
$$
This equation, together with the condition that $V(\rho)$ has a
minimum equal to $0$ for $\rho=\bar\rho$, gives \eqref{zrqp}.
For example, in equilibrium $J_A(\rho)=0$ for all $\rho$, and for the
(out of equilibrium) model of particles circulating on a ring driven
by a constant field $E$, described in Section \ref{s:tr}, we have
$\chi(\rho)^{-1}J_A(\rho)=E$, which is independent of $\rho$.

Next, we assume that the diffusion coefficient and the mobility are scalar matrices,
i.e. $D(\rho)_{ij}=D(\rho)\delta_{i,j}$ and $\chi(\rho)_{ij}=\chi(\rho)\delta_{i,j}$ ($i,j=1,\dots,d$).
We derive, in this case, an equivalent condition for the locality of the quasi-potential $V(\rho)$.
Assuming that $V(\rho)$ is local as in \eqref{zrqp}, we can use the
fact that \eqref{davide3} is independent of $\rho$ and the
orthogonality relation in \eqref{r18} to get
\begin{equation}
\label{davide2}
\int_{\Lambda}dx\,  J_S(\rho) \cdot\chi(\bar\rho)^{-1} J_A(\bar\rho)=0.
\end{equation}
We have $J_A(\bar\rho)=J(\bar\rho)$
and
\begin{equation}
\label{davide4}
\begin{array}{l}
\displaystyle{
\vphantom{\Big(}
J_S(\rho)
=-\chi(\rho)\nabla\frac{\delta V}{\delta\rho}
=-\chi(\rho)\big(f''(\rho)\nabla\rho-f''(\bar\rho)\nabla\bar\rho\big)
} \\
\displaystyle{
\vphantom{\Big(}
=
-\nabla\big(d(\rho)-d(\bar\rho)\big)
+\big(\chi(\rho)\chi(\bar\rho)^{-1}-1\big)\nabla d(\bar\rho)
},
\end{array}
\end{equation}
where $d(\rho)=\int^\rho d\alpha\, D(\alpha)$.
Using \eqref{davide4}, equation \eqref{davide2} can be rewritten as
\begin{equation}
\label{davide5}
\begin{array}{l}
\displaystyle{
\vphantom{\Big(}
\int_{\Lambda}dx\,
\frac{1}{\chi(\bar\rho)^{2}}
\Big(
-\big(d(\rho)-d(\bar\rho)\big)
\chi'(\bar\rho)
} \\
\displaystyle{
\vphantom{\Big(}
+\big(\chi(\rho)-\chi(\bar\rho)\big) D(\bar\rho)
\Big)J(\bar\rho)\cdot \nabla\bar\rho
=0
}.
\end{array}
\end{equation}
For this we used an integration by parts and the stationary equation
$\nabla\cdot J(\bar\rho)=0$.  Equation \eqref{davide5} is the desired
condition on the transport coefficients equivalent to the locality of
the quasi-potential $V(\rho)$. 
Indeed, if $V(\rho)$ is local we just proved that \eqref{davide5}
holds.  Conversely, if \eqref{davide5} holds, then the same
computation shows that the local functional $V(\rho)$ as in
\eqref{zrqp} solves the Hamilton-Jacobi equation \eqref{r15-2}.
In fact, such $V(\rho)$ is the quasi-potential.

For example, in equilibrium $J(\bar\rho)=0$, so \eqref{davide5} holds trivially.
In the model of particles circulating on a ring driven by a constant field $E$,
described in Section \ref{s:2.1}, we have $\nabla\bar\rho=0$,
so \eqref{davide5} still holds.
Furthermore, equation \eqref{davide5} holds for arbitrary choices of
external field $E$ and boundary chemical potential $\lambda$
provided that $D(\rho)$ and $\chi(\rho)$ are related by the following equation:
\begin{equation}
\label{davide6}
-\big(d(\rho)-d(\bar\rho)\big)
\chi'(\bar\rho)
+\big(\chi(\rho)-\chi(\bar\rho)\big) D(\bar\rho)
=0
\end{equation}
for arbitrary $\rho$ and $\bar\rho$.
This equation is an integral form of the following condition
\begin{equation}
\label{c=1}
D(\rho) \chi''(\rho) = D'(\rho) \chi'(\rho)
\end{equation}
It is easily seen that there are only two situations in which condition
\eqref{c=1} holds: for arbitrary $D(\rho)$ and $\chi(\rho)$ constant
in $\rho$, which corresponds to the called Ginzburg Landau model,
\cite{ippo,GPV}, and for $D(\rho)=c\chi'(\rho)$, for a constant $c$,
which corresponds to a ``generalized'' zero range model (the zero
range model is obtained for $c=1$).
We thus conclude, in particular, that in both
these cases the quasi-potential $V(\rho)$
is indeed local for arbitrary choices of
external field $E$ and boundary chemical potential $\lambda$.

Another situation in which \eqref{davide5} is satisfied
is when $ J(\bar \rho(x))\cdot \nabla \bar \rho(x)=0$ for any $x$.
This happens if $\Lambda$ is the $d$-dimensional torus and the external
field is of the form $-\nabla U+\tilde E$ with $\nabla \cdot \tilde E=0$
and $\nabla U(x)\cdot \tilde E(x)=0$ for any $x\in \Lambda$. This can be verified
with a simple calculation.

\subsection{Simple exclusion processes}
\label{sep-sec}

We consider here the boundary driven simple exclusion process in one
space dimension without external field. In particular we consider
$\Lambda=(-1,1)$ so that $\partial\Lambda=\pm 1$. The transport
coefficients in this case are $D(\rho)=1$ and
$\chi(\rho)=\rho(1-\rho)$ and the specific free energy is
$f(\rho)=\rho\log\rho +(1-\rho)\log(1-\rho)$.  We fix the chemical
potentials at left and right boundaries as $\lambda_\pm$,
correspondingly the macroscopic density will satisfy the boundary
conditions $\rho(\pm 1)=\rho_\pm$ as required by \eqref{2.3}.

By using a matrix representation of the microscopic invariant state
and combinatorial techniques, in 
\cite{DLS,DLS1} it is shown that the quasi-potential $V$ can be
expressed in terms of the solution of a non--linear ordinary
differential equation. We show how this result can be deduced
using the MFT.  
Namely, we consider the variational problem \eqref{r07} for the
one-dimensional simple exclusion process and show that the associated
Hamilton-Jacobi equation
\begin{equation}\label{HJsep}
\int_{\Lambda} \left(\nabla \frac{\delta V}{\delta\rho}
\rho (1-\rho)\nabla \frac{\delta V}{\delta\rho}
+\frac{\delta V}{\delta\rho} \Delta\rho\right)\, dx = 0
\end{equation}
can be reduced to the non-linear ordinary differential equation
obtained in \cite{DLS}.

We look for a solution of the Hamilton-Jacobi equation \eqref{HJsep}
by performing the change of variable
\begin{equation}\label{guess}
\frac{\delta V}{\delta\rho (x)} =
\log \frac{\rho(x)}{1-\rho(x)} - \phi(x;\rho)
\end{equation}
for some functional $\phi(x; \rho)$ to be determined satisfying the
boundary conditions $\phi(\pm 1 )= \log \rho(\pm 1 )/[1-\rho(\pm 1)]$.
Inserting \eqref{guess} into \eqref{HJsep}, we get that
\begin{equation*}
  \begin{split}
    0 &=
    \int_\Lambda\!dx\,
    \nabla \Big( \log \frac {\rho}{1-\rho}-\varphi\Big)
    \rho(1-\rho) \nabla \varphi
    \\
    &=  \int_\Lambda\!dx\,
    \Big[\nabla \rho \nabla \phi -
    \rho(1-\rho) (\nabla \varphi)^2 \Big].
  \end{split}
\end{equation*}
Adding and subtracting $e^\varphi/(1+ e^{\varphi})$, we may rewrite
the previous integral as
\begin{equation*}
  \begin{split}
    &\int_\Lambda\!dx\,
    \nabla \Big(\rho   - \frac {e^\varphi}{1+e^\varphi} \Big)
    \nabla \varphi\\
    &\quad
    \int_\Lambda\!dx\,
    \Big(\rho   - \frac {e^\varphi}{1+e^\varphi} \Big)
    \Big(\rho - \frac {1}{1+e^\varphi} \Big)
    (\nabla \varphi)^2.
  \end{split}
\end{equation*}
Since  $\rho - e^\phi/{(1+e^\phi)}$ vanishes at the boundary, an
integration by parts yields
\begin{equation}
  \label{shjsep}
  0= \int_\Lambda\!dx\,
    \Big(\rho   - \frac {e^\varphi}{1+e^\varphi} \Big)
    \Big(\Delta \varphi + \frac {(\nabla \varphi)^2}{1+e^\varphi}
    -\rho (\nabla \varphi)^2\Big).
\end{equation}
We thus obtain a solution of the Hamilton-Jacobi if we solve the following
ordinary differential equation which relates the functional
$\phi(x)=\phi(x; \rho)$ to $\rho$
\begin{equation}\label{dphi}
\left\{
\begin{array}{l}{\displaystyle
\frac {\Delta \phi(x)}{[\nabla\phi(x)]^2} +
\frac{1}{1+e^{\phi(x)}} = \rho (x)  \quad x\in (-1,1)\,, }
\\
\\
\phi(\pm 1) =  \log \rho(\pm 1 )/[1-\rho(\pm 1)]\,.
\end{array}
\right.
\end{equation}
As proven in \cite{DLS1} this equation admits a unique monotone
solution which is the relevant one for the quasi-potential.  Recalling
\eqref{10}, a computation shows that the derivative of the functional
\begin{equation}\label{Ssep}
V(\rho) =F(\rho)+ \int_{\Lambda}\!dx \, \left\{
 (1-\rho) \phi
 + \log\left[
\frac{\nabla\phi}{\nabla\bar\rho\left(1+e^{\phi}\right)}\right]
\right\}
\end{equation}
is given by \eqref{guess} when $\phi(x;\rho)$ solves \eqref{dphi}.
To prove that this is the maximal positive solution
we refer to \cite{BDGJL02}.

According to the general time reversal argument, see in particular
equation \eqref{r12}, the adjoint hydrodynamics can be written as
\begin{equation}
  \label{a.1}
  \partial_t \rho = \Delta \rho - 2 \nabla\cdot \big( \chi(\rho)
  \nabla \phi \big)
\end{equation}
where we used \eqref{guess} and $\phi$ has to be expressed in
function of $\rho$ by solving \eqref{dphi}.
As shown in \cite{BDGJL02}, it is remarkable that this non local
evolution can be directly related to the heat equation.
Let $\gamma=\gamma(t,x)$ be defined by
\begin{equation}
  \label{a.2}
  \gamma = \frac {e^{\phi}}{1 +e^{\phi}}
\end{equation}
where $\phi =\phi(t,x)$ is the solution to \eqref{dphi} when
$\rho=\rho(t,x)$ evolves according to \eqref{a.1}. Then $\gamma$
solves
\begin{equation}
  \label{a.3}
  \partial_t \gamma = \Delta \gamma
\end{equation}
with the appropriate initial and boundary conditions.

\medskip
One may be tempted to repeat the same computation in arbitrary
dimension; one would obtain a partial differential equation analogous
to \eqref{dphi}. However, in more than one dimension it does not
exist, in general, a functional $V$ whose derivative is given by
\eqref{guess} with $\phi$ and $\rho$ related by such partial
differential equation.

\medskip
For this model, as proved in \cite{DLS} for the one-dimensional case
and in \cite{BDGJL02} for higher dimensions, the quasi-potential
$V(\rho)$ is larger than the local functional \eqref{a1} with $\bar
\rho$ the non equilibrium stationary profile.  
For small fluctuations this follows from formula \eqref{opp}.

\medskip
An interesting result \cite{TKL-prl, TKL} is that this model and the
following Kipnis-Marchioro-Presutti model can be mapped into
equilibrium models. This result depends on the Hamiltonian structure
and the non local map \eqref{dphi}. We briefly outline the argument.
Recall the Hamiltonian \eqref{lHJ} that for the simple exclusion
process reads
\begin{equation}\label{Hamex}
\mathcal H(\rho,\pi)=\int_{-1}^1dx\,\left\{\rho(1-\rho)\left(\nabla \pi\right)^2+
\pi \Delta \rho \right\}
\end{equation}
with the boundary conditions $\rho(\pm 1)=\rho_{\pm}$ and $\pi(\pm 1)=0$.
Consider the symplectic
transformation $(\rho,\pi)\to (\varphi,\psi)$ given by
\begin{equation}\label{kur-eq}
\left\{
\begin{array}{l}
\nabla\left(\frac{1}{1-e^\psi}\right)=e^\pi-1-\rho(e^\pi+e^{-\pi}-2)\\
\nabla\left(\frac{\rho}{\rho+(1-\rho)e^\pi}\right)=e^\psi-1-\varphi(e^\psi+e^{-\psi}-2)\,.
\end{array}
\right.
\end{equation} 
The new
Hamiltonian has the same form of \eqref{Hamex}, that is
\begin{equation}
\tilde {\mathcal
H}(\varphi,\psi)=\int_{-1}^1dx\,\left\{\varphi(1-\varphi)\left(\nabla \psi\right)^2+
\psi \Delta \varphi \right\},
\end{equation}
but the boundary conditions are
$\nabla \varphi(\pm 1)=\nabla\psi(\pm 1)=0$ \cite{TKL-prl}. Since these
boundary conditions corresponds to an isolated exclusion process,
\eqref{kur-eq} realizes a map into an equilibrium system. 
In particular $\tilde {\mathcal H}$
satisfies \eqref{HH*} with $\tilde{\mathcal H}^*=\tilde {\mathcal H}$ and
the optimal exit trajectory is simply given by the time reversal of
the relaxation one. By mapping back this solution and computing the
corresponding action the expression \eqref{Ssep} for the
quasi--potential is recovered.

\medskip
From a physical point of view, besides the case of external
reservoirs, boundary conditions modeling a battery appear natural.
Namely, we can consider the system in a ring with an external field and 
take the limit in which the field becomes a delta function
localized at one point.
The application of the MFT to this case is discussed in \cite{MR2670735}.

\subsection{Kipnis-Marchioro-Presutti model}

We consider the one dimensional boundary driven
Kipnis-Marchioro-Presutti (KMP) model \cite{MR656869}. 
This is a diffusive system with transport coefficients given by
$D(\rho)=1$ and $\chi(\rho)=\rho^2$. It derives from a simple
stochastic model of heat conduction in a crystal. Like in the
exclusion process the computation of the quasi-potential can be
reduced to the solution of a non linear differential equation
\cite{BGL}.

The procedure is similar to the one for the simple exclusion
process. 
The Hamilton-Jacobi equation for the quasi-potential $V$
is
\begin{equation}\label{HJkmp}
\int_\Lambda \!dx\, \left(\nabla \frac{\delta V}{\delta\rho}
\rho^2\nabla \frac{\delta V}{\delta\rho}
+ \frac{\delta V}{\delta\rho} \Delta\rho\right)= 0.
\end{equation}
We assume that $\Lambda= (-1,1)$.
We shall also assume the macroscopic density
profile $\rho=\rho(x)$ satisfies the boundary conditions $\rho(\pm 1)=\rho_{\pm}$.
We emphasize that $\rho$ now represents an energy density.

We look for a solution of the Hamilton-Jacobi equation \eqref{HJkmp}
by performing the change of variable
\begin{equation}\label{guess-kmp}
\frac{\delta V}{\delta\rho (x)} =
\frac {1}{\alpha(x; \rho)} - \frac{1}{\rho(x)}
\end{equation}
for some functional $\alpha(x; \rho)$ to be determined satisfying the
boundary conditions $\alpha(\pm 1 )= \rho(\pm 1 )$.

With a calculation similar to the one in the previous section we find that
the quasi--potential is 
\begin{equation}\label{S-kmp}
V(\rho)=\int_{\Lambda}\!dx\,\Big(\frac{\rho}{\alpha}-1-\log \frac{\rho}{\alpha}-
\log \frac{\nabla \alpha}{\nabla \bar \rho}\Big),
\end{equation}
where $\alpha=\alpha(x; \rho)$ is the unique monotone solution to
\begin{equation}\label{1-d-kmp}
\left\{
\begin{array}{l}
\alpha^2\frac{\Delta \alpha}{\left(\nabla \alpha\right)^2}+\rho-\alpha=0 \\
\alpha(\pm 1)=\rho_{\pm}.
\end{array}
\right.
\end{equation}

By a direct computation it can be shown that $V(\rho)$ is not convex.
For this model, as proved in \cite{BGL}, the quasi-potential $V(\rho)$
is smaller than the local functional \eqref{a1} with $\bar \rho$ the
stationary profile.  For small fluctuations this follows from formula
\eqref{opp}.

\subsection{Exclusion process with external field}

The computation of the quasi-potential for the one-dimensional boundary
driven simple exclusion process reviewed in section \ref{sep-sec} can
be generalized to the case in which a constant external field is
applied to the system.  The first result has been obtained, using the
matrix approach, in \cite{MR2035624} and refers to the case in which
the driving due to the external reservoirs and the field are in the
same direction. In the same situation an approach based on the
macroscopic fluctuation theory is presented in \cite{MR2504852}. The
case when the field drives in the opposite direction with respect to
the boundary sources exhibits Lagrangian phase transitions and will be
discussed in the following section.

The weakly asymmetric boundary driven simple exclusion process is
defined, in appropriate units, by the following choices.  The
transport coefficient are $D=1$ and $\chi(\rho)=\rho(1-\rho)$ so that
the specific free energy is $f(\rho) = \rho\log \rho + (1-\rho) \log
(1-\rho)$.  Observe that the hydrodynamic equation is the viscous
Burgers equation.  We consider it on the space domain $\Lambda=(-1,1)$
with a constant external field $E$ and denote by $\lambda_\pm$ the
chemical potentials of the boundary reservoirs. We let $\rho_\pm
=e^{\lambda_\pm}/(1+e^{\lambda_\pm})$ be the boundary values of the
density.

The Hamilton-Jacobi equation for the quasi-potential \eqref{r15-2}
thus reads
\begin{equation}
  \label{HJsep-H}
  \begin{split}
      & \int_{\Lambda} \Big(\nabla \frac{\delta V}{\delta\rho}
    \rho (1-\rho)\nabla \frac{\delta V}{\delta\rho}
    \\
    &\qquad +\frac{\delta V}{\delta\rho} \big\{ \Delta\rho- E
    \nabla [\rho(1-\rho)] \big\} \Big)\, dx = 0.
  \end{split}
\end{equation}
As in the symmetric case we look for a solution $V$ whose derivative has the form
\begin{equation}
\label{phi-H}
  \frac{\delta V}{\delta\rho (x)} = f'(\rho(x))  - \phi(x;\rho)
\end{equation}
where, for density profiles $\rho$ satisfying the boundary conditions
$\rho(\pm 1)=\rho_\pm$,  we have $\phi(\pm 1, \rho)
=\lambda_\pm$.
Few integrations by parts similar to \eqref{shjsep} show that \eqref{HJsep-H}
is satisfied provided $\phi$ solves
\begin{equation}\label{dphi-H}
\left\{
\begin{array}{l}{\displaystyle
\frac {\Delta \phi(x)}{\nabla\phi(x) [\nabla \phi(x) -E]} +
\frac{1}{1+e^{\phi(x)}} = \rho (x)  \quad x\in (-1,1) }
\\
\\
\phi(\pm 1) =  \lambda_\pm\,.
\end{array}
\right.
\end{equation}
In order to identify the quasi-potential we need show that $\phi$ is
properly defined, namely that \eqref{dphi-H} has a unique solution,
and that there exists a functional $V$ with derivative given by
\eqref{phi-H}.

Fix $\rho_-<\rho_+$ and observe that when $E=E_0 \equiv
[\lambda_+-\lambda_-]/2$ the model describes an inhomogeneous
equilibrium state as in  Section~\ref{s:eq}.
We here consider the case in which $E<E_0$ that corresponds to a
negative stationary current.
Recalling $F(\rho) = \int_{\Lambda} dx\, f(\rho)$,  we introduce the auxiliary functional of
two variables
\begin{equation}
  \label{auxG}
  \begin{split}
    & \mc G (\rho,\phi) =  F(\rho) + \int_{\Lambda}\!dx \, \Big\{
    (1-\rho)\varphi -\log\big(1+e^{\varphi}\big)
    \\
    & \;\; + \frac 1E \Big[ \nabla\varphi \log \nabla \varphi
    - (\nabla \varphi - E) \log (\nabla \varphi  - E) \Big] \Big\}
  ,\end{split}
\end{equation}
which has the property that \eqref{dphi-H} is the stationarity
condition $\delta \mc G / \delta \phi =0$ while
$\delta \mc G/\delta \rho = f'(\rho) - \phi$ is the right
hand side of \eqref{phi-H}.

The functional $\mc G$ is well defined
provided $\phi$ is increasing and $\nabla \phi \ge E$.
In \cite{MR2504852} is shown that
\eqref{dphi-H} has a unique solution $\phi$ satisfying these requirements.
The quasi-potential, up to an additive constant that is fixed by the
normalization $V(\bar\rho)=0$, can thus be expressed in terms of the
auxiliary functional $\mc G$ as
\begin{equation*}
  V(\rho) = \sup_{\phi} \mc G(\rho,\phi) = \mc G\big(\rho,\phi(\rho)\big)
\end{equation*}
where $\phi(\rho)$ is the solution to \eqref{dphi-H}.
Indeed, if $\phi(\rho)$ solves \eqref{dphi-H} then  by chain rule
\begin{equation*}
  \begin{split}
    \frac {\delta V}{\delta \rho} (\rho) & =  \frac {\delta \mc G}{\delta \rho} (\rho,\phi(\rho) )
    + \frac {\delta \mc G}{\delta \phi}
    (\rho,\phi(\rho) ) \, \frac {\delta \phi }{\delta \rho} (\rho)
\\  & =
    f'(\rho) -\phi(\rho)
  \end{split}
\end{equation*}
The fact that $\phi$ solving \eqref{dphi-H} corresponds to a maximum
of $\mc G$ follows from the concavity with respect to $\phi$ of $\mc
G$.

We mention that the computation reducing the (infinite dimensional)
Hamilton-Jacobi equation  \eqref{HJsep-H} to the (one dimensional)
problem \eqref{dphi-H} can be extended to the models with constant
diffusion coefficient, quadratic mobility, and constant external
field \cite{BGL,DG2}.

\subsection{An example of Lagrangian phase transition}
\label{s:exlpt}

As in the previous Section, we consider the one-dimensional boundary
driven weakly asymmetric exclusion process on the interval $(-1,1)$
with $\lambda_- <\lambda_+$.  We consider here the case in which the
driving from the field is in the opposite direction with respect to
the one from the boundary reservoirs and the the stationary current
$J(\bar\rho)$ is positive, that is $E> E_0=[\lambda_+-\lambda_-]/2$.
We shall show that for $E\gg E_0$ this model provides an example of a
Lagrangian phase transition, see Section~\ref{s:lpt}. This appears to
be the first concrete example where this can be rigorously proven \cite{BDGJL-cpam}.

As a first step, we discuss the change of variable \eqref{phi-H} in
the framework of the underlying Hamiltonian structure.  Recalling that
the Hamiltonian is given in \eqref{lHJ}, we perform the symplectic
change of variables
\begin{equation}
\label{sympl}
  \begin{cases}
    \varphi = f'(\rho) - \pi \\
     \psi = \rho
  \end{cases}
\end{equation}
where we recall that $f(\rho)=\rho\log \rho + (1-\rho)\log(1-\rho)$ is
the specific free energy.

In the new variables $(\varphi,\psi)$ the Hamiltonian ${\widetilde{\mc
    H}} (\varphi,\psi) = \mc H (\psi, f'(\psi) -\varphi)$ reads
\begin{equation*}
  \begin{split}
    {\widetilde{\mc H}} (\varphi,\psi)
= &\int_{-1}^1\!dx\, \big\{ \psi(1-\psi)\, (\nabla \varphi)^2
\\
&- [\nabla \psi + E \, \psi(1-\psi)] \, \nabla \varphi   + E\, (\rho_+ -\rho_-) \big\}
  \end{split}
\end{equation*}
where we used that $\rho(\pm 1)=\rho_\pm$.
The corresponding canonical equations are
\begin{equation}
\label{nhf}
\left\{
\begin{array}{l}
\partial_t \phi  = \Delta \phi  - (1- 2\psi)  \, \nabla \phi
\, (E - \nabla \phi) \\
\partial_t \psi = -  \Delta \psi  -  E\, \nabla [
\psi(1-\psi )] +  2\, \nabla \big[\psi(1-\psi) \, \nabla \phi \big]
\end{array}
\right.
\end{equation}
with the boundary conditions inherited from \eqref{sympl}.

In the new variables the equilibrium position $(\bar\rho,0)$ becomes
$(f'(\bar\rho),\bar\rho)$. The associated
stable manifold is $\mc M_\mathrm{s} = \{ (\varphi,\psi):\, \varphi=f'(\psi)\}$.
As shown in \cite{lpt} the unstable manifold is given by
\begin{equation}
\label{Sig}
\begin{split}
  \mc M_\mathrm{u} = \Big\{ & (\varphi,\psi) \, :\:
 0< \nabla \varphi < E, \\
& \psi = \frac{1}{1+e^\varphi} - \frac{\Delta \varphi}
{\nabla \varphi (E-\nabla \varphi)}
\Big\}.
\end{split}
\end{equation}
Note that in the variables $(\varphi,\psi)$ the unstable manifold
$\mc M_\mathrm{u}$ can be described as the graph of a single-valued function
while this is not the case in the original variables $(\rho,\pi)$,
recall Fig. 1 (a) of Section \ref{s:lpt}.

In view of the expression \eqref{Sig} of the unstable manifold, the
pre--potential $\mc V $ in \eqref{f23} can be obtained by direct
computations using the new variables $(\phi,\psi)$. 
Let $\mc G$ be the functional (compare with \eqref{auxG})
\begin{equation}
\label{Gbis}
  \begin{split}
  & \mc G(\rho,\phi) 
  = \int_{-1}^1 \!dx \, \Big\{ f(\rho) +(1-\rho)\phi
  -\log\big(1+e^{\phi}\big)
  \\
  &\; + \frac {1}{E} \Big[ \nabla \phi \log {\nabla \phi} +
  (E -\nabla \phi)\log \big( E -{\nabla \phi} \big) \Big]
  \Big\}
  \end{split}
\end{equation}
up to an additive constant fixed by the normalization 
$\mc G (\bar\rho,f'(\bar\rho))=0$. Then the pre-potential (in the
original variables) is 
\begin{equation}
     \mc V (\rho,\pi) =   \mc G(\rho,f'(\rho) -\pi).   
\end{equation}
We deduce that the quasi-potential is given, up to an additive constant, by
\begin{equation}
  \label{Veps}
  {V}(\rho) = \inf \big\{
  \mc G  (\rho, \varphi) \,,\: \varphi \,: \: (\varphi, \rho) \in
  \mc M_\mathrm{u} \big\}.
\end{equation}
According to the general arguments in Section \ref{s:lpt} the
pre-potential is defined on the unstable manifold $\mc
M_\mathrm{u}$. On the other hand, the right hand side of \eqref{Gbis}
extends to a function defined for all $\phi$ satisfying $0< \nabla \phi <
E$.  By denoting
still with $\mc G $ this extension we realize that the condition
$(\phi,\rho) \in \mc M_\mathrm{u}$ is equivalent to
$\delta \mc G  (\rho,\phi) / \delta \phi =0$. We conclude that \eqref{Veps}
still holds if  the constraint between $\rho$ and $\phi$ is dropped.

When the external field $E$ is large enough, the weakly asymmetric
exclusion process exhibits Lagrangian phase transitions. Namely the
variational problem in \eqref{Veps} admits more than a single critical
point or equivalently the equation on the second line of \eqref{Sig}
has multiple solutions.  We argue as follows.  Consider first the
limiting case $E=\infty$ in which the hydrodynamic equation becomes
the inviscid Burgers equation and corresponds to the asymmetric simple
exclusion process examined in \cite{MR1964689}.  The
functional $\mc G $ becomes
\begin{equation}
\label{Gin}
\mc G _\infty  (\rho,\varphi)  = \int_{-1}^1 \!dx\,
\Big[ f(\rho) +(1-\rho)\varphi  -\log\big(1+e^{\varphi}\big)
\Big].
\end{equation}
In this limit the variational problem \eqref{Veps} becomes a
one-dimensional problem and it is possible to exhibit explicitly
density profiles such that uniqueness fails.
For instance this is the case if $\rho$ is of the form drawn in Fig.~2.
By a continuity (topological) argument one shows that this phase
transition persists also for $E$ finite and large.

\begin{figure}[h]
\label{f:rc}
\begin{picture}(200,65)(0,-5)
{
%
\thinlines
\put(0,0){\vector(1,0){170}}
\put(0,0){\vector(0,1){60}}
\put(-1.2,-2.2){$\cdot$}
\put(-2.2,-8){$-1$}
\put(158.6,-2.2){$\cdot$}
\put(157.5,-8){$1$}
\put(160,0){\line(0,1){60}}
\put(173,-2){$x$}
%
\put(-1.3,7.6){$\cdot$}
\put(-10,7.6){$\rho_-$}
\put(158.7,47.6){$\cdot$}
\put(162,47.6){$\rho_+$}
\put(0,30){\line(1,0){160}}
%
\thicklines
\qbezier(0,10)(50,50)(80,30)
\qbezier(80,30)(110,10)(160,50)
\put(29,27.7){$\bullet$}
\put(78,27.7){$\bullet$}
\put(127,27.7){$\bullet$}
\put(100,40){$\rho(x)$}
%
\thinlines
\multiput(37,29)(3,0){13}{$\cdot$}
\multiput(40,31)(3,0){11}{$\cdot$}
\multiput(49,33)(3,0){7}{$\cdot$}

\multiput(84,26)(3,0){13}{$\cdot$}
\multiput(87,24)(3,0){11}{$\cdot$}
\multiput(93,22)(3,0){7}{$\cdot$}
}
\end{picture}
\caption{
  Graph of a caustic density profile for $E=\infty$.
  The shaded regions have equal area.
}
\end{figure}
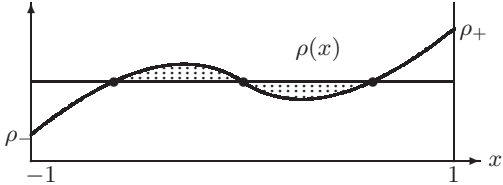

Comparing the formula \eqref{auxG} of the previous section with
\eqref{Gbis} we obtain easily, by inserting the absolute value inside
the argument of the logarithm, an expression for $\mc G$ that covers
both cases. On the other hand, if $E< E_0$ the function $\phi
\mapsto \mc G (\rho, \phi)$ has a unique critical point which
corresponds to a maximum, while for $ E> E_0$ it may have more
critical points and the quasi-potential is obtained in correspondence to
the global minimum.

The analysis of the weakly asymmetric exclusion process has been further
developed, by considering density profiles $\rho$ with more critical
points, in \cite{abk}, see also references therein.

\subsection{Reaction-diffusion dynamics}

In this Section we discuss the case in which the macroscopic
dynamic is not a conservation law but there is a reaction term
allowing creation/destruction of particles in the bulk.
This class of models, with added random forces, has been
investigated in the literature, e.g.\ \cite{tau} for a recent reference. 
Here we just show, in a specific example, how the basic principles of the
MFT need to be modified to cover these processes.

The macroscopic evolution has the form
\begin{equation}
\label{eqn6}
\partial_t \rho = \Delta \rho + b(\rho)-d(\rho)=  \Delta \rho + K(\rho),
\end{equation}
where $b$ and $d$ are respectively the creation and destruction
rates. For simplicity we restrict to the case $\Lambda=(-1,1)^d$ with 
periodic boundary conditions.

This evolution can be derived as the typical behavior of some
underlying stochastic microscopic dynamics in which particles can jump
on the lattice and be created or destroyed. For instance, as shown in
\cite{DFL}, it can be derived from the so-called \emph{Glauber+Kawasaki} process that
we describe in Section \ref{sec:g+k}.

The associated large deviation functional for the density trajectories was first
calculated in \cite{JLV}
\begin{eqnarray}\label{eqn14}
I_{[0,T]}(\rho) =
\int_0^{T}  \!dt \int_\Lambda \!dx \, \Big\{
\frac{1}{4}\nabla H\cdot \rho (1-\rho) \nabla H
\quad \quad \quad\nonumber
\\
+ b(\rho) \big(1-e^{H}+ H e^{H}\big)
+ d(\rho) \big(1- e^{-H}-H e^{-H} \big)
\Big\},\nonumber\\
\end{eqnarray}
where the external potential $H$ is connected to the fluctuation
$\rho$ by
\begin{equation}
\label{eqn10}
\partial_t \rho = \Delta\rho
-\nabla \cdot \big( \rho(1-\rho) \nabla H \big)
+b(\rho)e^H-d(\rho)e^{-H}.
\end{equation}

The structure of the functional $I$ reflects the Poissonian nature of
the underlying microscopic dynamics.  
The Hamiltonian associated to the large deviation functional
\eqref{eqn14}--\eqref{eqn10} for this model is 
\begin{eqnarray}
{\cal H}(\rho, \pi)=\int_{\Lambda} \!dx \, \bigg\{
 \pi\Delta\rho  +
(\nabla \pi)^2 \rho(1-\rho)
\nonumber\\
- b(\rho) \big(1-e^{\pi}\big) - d \big(\rho)(1-e^{-\pi}\big) \bigg\},
 \label{HA}
\end{eqnarray}
where $\pi$ is the conjugate momentum.
Observe that while $I$ has an implicit expression, since $H$ has to
be expressed in terms of $\rho$ by solving \eqref{eqn10},  
the Hamiltonian $\mc H$ has a closed form.

As $\mc H$ is not quadratic, the  Hamilton-Jacobi equation
\begin{equation}
{\cal H}\left(\rho, \frac {\delta V}{\delta \rho}\right)=0
\label{HJ}
\end{equation}
is very complicated but
can be solved in some special cases.
This happens when $b(\rho)=c_1(1-\rho)h(\rho)$ and $d(\rho)=c_2\rho
h(\rho)$ where $c_i$ are positive constants and $h(\rho)$ is a
positive function. In this case \cite{GJLV}
\begin{equation}\label{QP-GK}
V(\rho)=\int_\Lambda \!dx \, \Big\{ \rho \log \frac{\rho}{\bar c}+(1-\rho)
\log\frac{(1-\rho)}{(1-\bar c)}\Big\}, 
\end{equation}
where $\bar c=c_1/(c_1+c_2)$. This corresponds to the situation that we call
macroscopic reversibility of which the validity of microscopic
detailed balance (see \eqref{inv-glaub}) is a special case. 

\medskip
In the general case equation \eqref{HJ} can be solved by successive
approximations using as an expansion parameter $\rho - \bar \rho$
where $\bar \rho$ is a solution of $B(\rho)=D(\rho)$ that is a
stationary solution of hydrodynamics.  More precisely we look for an
approximate solution of (\ref{HJ}) of the form
\begin{equation}
\label{apprS}
V(\rho) = \frac {1}{2} \int_{\Lambda\times \Lambda} \! dx \, dy \, 
[ \rho(x) - \bar \rho]
k(x,y)[\rho(y) - \bar \rho] + o(\rho - \bar \rho)^2.
\end{equation}
By inserting (\ref{apprS}) in (\ref{HJ}) one can show that  $k(x,y)$ satisfies
the following equation
\begin{equation}
\label{ke}
\begin{split}
& \bar{\rho}(1 - \bar{\rho}) \Delta_x k(x,y)
-b_0 k(x,y) 
\\
&- \Delta_x \delta(x-y)
+ (d_1 - b_1)\delta(x - y)=0,
\end{split}
\end{equation}
where 
\begin{equation*}
  b_1=b'(\bar\rho), \quad d_1=d'(\bar\rho), \qquad b_0=b(\bar{\rho})=d(\bar{\rho}).
\end{equation*}

If $V$ is a local functional of the density, $k(x,y)$ must be of  the
form  $k(x,y)=g(\bar{\rho})\delta(x-y)$ which inserted in (\ref{ke})
gives
\begin{equation}
g(\bar{\rho})=[\bar{\rho}(1-\bar{\rho})]^{-1}
\end{equation}
and
\begin{equation}\label{cond-1}
b_0[\bar{\rho}(1-\bar{\rho})]^{-1}-(d_1-b_1)=0.
\end{equation}
Condition \eqref{cond-1} is satisfied in the cases when \eqref{QP-GK}
is the quasi-potential.
On the other hand if $b_0,b_1,d_1$ do not satisfy the last equation
the quasi-potential cannot be a local functional of the density.

For this model it is possible to prove \cite{BL} an analogue of the
fundamental formula \eqref{r02},\eqref{r05}.
The hydrodynamic equation has a local source term $K$ and we are
interested in the joint fluctuations of $\rho$, $J(\rho)=-\nabla \rho$,
$K(\rho)= b(\rho)-d(\rho)$. The large deviation functional is
\begin{equation}
\label{eq:functional}
{\mathcal I}_{[0,T]} ( \rho,j, k)
= \int_0^T  \!dt \int_\Lambda \!dx\,
\Big\{ {\frac 14}\frac{\big|j-J(\rho) \big|^2}{ \rho(1-\rho)}
+ \Psi ( \rho , k ) \Big\},
\end{equation}
with
\begin{eqnarray}
\label{eq:dissipative}
\Psi (\rho, k)& =& b(\rho) + d(\rho) - \sqrt{k^2 + 4  d(\rho) b(\rho)}\nonumber \\
&+& k \log \left( \frac{\sqrt{k^2 + 4  d(\rho) b(\rho)} + k}
{2 b(\rho)} \right).
\end{eqnarray}
Here $\rho$, $j$ and $k$ are connected by the equation
\begin{equation}
\partial_t \rho=-\nabla j + k.
\end{equation}
The rate function \eqref{eqn14} can be recovered from \eqref{eq:functional} 
by optimizing with respect to $j$ and $k$.

For driven diffusive systems, we have shown that long range
correlations of the density are a generic feature of non-equilibrium
states. If $b(\bar\rho)=d(\bar\rho)$ the reaction diffusion dynamics
does not exhibit a macroscopic current and, in this respect, may be
regarded as an equilibrium state. On the other hand, the previous
discussion implies that long range correlations do appear if
\eqref{cond-1} is violated. From the point of view of the MFT,
violation of \eqref{cond-1} corresponds to a breaking of macroscopic
reversibility. We refer to \cite{BJ,BDGJL-sips} for more details.

\subsection{Mean field models}

The macroscopic fluctuation theory can be applied to diffusion
processes coupled via a mean field interaction, \cite{Gaw}. A prototype of such
systems is the Kuramoto model with noise. This is a system of $N$
coupled planar rotators described by the phases $\theta_i$ in a
rotating magnetic field with amplitude $H$ and frequency $F$. In the
frame comoving with the rotators, the evolution is given by the
Langevin equations
\begin{equation*}
  \dot \theta_i = F - H \sin \theta_i - \frac JN \sum_{j=1}^{N} \sin
  (\theta_i -\theta_j) + \sqrt{2 \kappa T} \alpha_i
\end{equation*}
where $J$ is the coupling constant, $\kappa$ the Boltzmann constant
$T$ the temperature, and $\alpha_i$ independent white noises.

If the frequency $F$ vanishes this is an equilibrium model and the
stationary ensemble has a Gibbsian description with a mean field
interaction that undergoes a phase transition. On the other hand for
$F\neq 0$ it is a non-equilibrium model. With the proper definition of
the current $J(\rho)$,  the fundamental formula of the macroscopic
fluctuation theory holds and thus allows an analysis of the asymptotic
properties of this model. In particular, the quasi-potential can be
computed perturbatively. Moreover, the current fluctuations exhibit
rich and interesting phenomena of the type of the dynamical phase
transition that will be discussed in Section \ref{s:dft}.

\subsection{Models with several conservation laws }

So far we have considered for simplicity
conservative models with only one conservation law.
The theory however is not limited by this restriction  and
models with more than one thermodynamic variable have
been considered.

We mention in particular the work \cite{B}.
It deals with a stochastic heat conduction model for solids.
The system is in contact with two heat baths at different temperatures.
There are two conserved quantities: the energy and the deformation between atoms.
The author establishes the hydrodynamic limit for the two conserved quantities
and calculates a large deviation functional
analogous to \eqref{I=} for the joint fluctuations of
the energy and the deformation. From this formula he obtains the quasi-potential for
temperature fluctuations which is the same as for the KMP model \eqref{S-kmp}.

Another interesting case is the ABC model \cite{EKKM,CDE}.  In this case
there are three conserved quantities but only two are independent.
The hydrodynamic equations are not of the standard form \eqref{2.2}
but the quasi-potential can be calculated exactly when the 
total densities of the the three species are equal. 
It is non local but this is not in contradiction with our previous
statements due to the non standard form of the hydrodynamics.  It
satisfies the Hamilton-Jacobi equation which in this case is
equivalent to \eqref{ein?} due to reversibility. 
If the total densities are not equal the MFT has been used in 
\cite{BDLvW} to compute perturbatively the quasi-potential.

\section{Thermodynamics of currents}
\label{s:6}

The study of current fluctuations is one of the most interesting topic
that can be developed within the MFT and has received a considerable
attention in the literature.  In this section we first discuss a
striking prediction of the theory on the possibility of dynamical
phase transitions in current fluctuations leading to a state of the
system spontaneously breaking time translation invariance
\cite{noiprlcurrent,MR2227084}. We then show that universal properties
of the cumulants of the time averaged current can be obtained both in
stationary and non stationary states.

\subsection{Examples of dynamical phase transition}
\label{s:dft}

Recalling the discussion in Section~\ref{s:cf}, we first show that,
under some structural conditions on the transport coefficients, the
identity $\Phi= U$ holds.  In this case the additivity principle in
\cite{BD04} is satisfied and there are no dynamical phase transitions.
The computation of $\Phi$ is simpler as we have to solve a time
independent variational problem.

We assume that the matrices $D(\rho)$ and $\chi(\rho)$ are multiples of
the identity. In the case with no external field, $E=0$, if
\begin{equation}
\label{c<}
D(\rho) \chi''(\rho) \le D'(\rho) \chi'(\rho)
\quad \textrm{ for any } \rho
\end{equation}
then $\Phi=U$, which implies also that $U$ is convex.
Moreover if
\begin{equation}
\label{c=bis}
D(\rho) \chi''(\rho) = D'(\rho) \chi'(\rho)
\quad \textrm{ for any } \rho
\end{equation}
then $\Phi=U$ for any external field $E$.

For the proof of these statements we refer to \cite{MR2227084} where
we also discuss the case with periodic boundary conditions which
requires the further restriction that $D$ is constant.  Condition
\eqref{c<} is satisfied e.g., for the symmetric simple exclusion
process, where $D=1$ and $\chi(\rho)=\rho(1-\rho)$, $\rho\in[0,1]$.
We recall that, as shown in Section~\ref{s:lqp}, condition
\eqref{c=bis} implies the locality of the quasi potential and is
satisfied by the zero range and the Ginzburg Landau processes.

\smallskip
To exemplify situations in which $\Phi < U$, that is the presence of a
dynamical phase transition, consider the fluctuations of the time
averaged current in the one dimensional case with periodic boundary
conditions.
Two models have been discussed so far, the KMP model and the exclusion
process with an external field.

In \cite{MR2227084} by simple arguments (Jensen inequality and
convexity properties of the transport coefficients), we find
sufficient conditions on $D$, $\chi$, $E$ and $J$ implying that the
optimal profile for the variational problem \eqref{r27} defining the
functional $U$ is the constant one. More precisely we show that if $D$
is constant and $J^2/\chi(\rho)+E^2\chi(\rho)$ is a convex function in
$\rho$ then
\begin{equation}\label{U-m}
U(J)=\frac 14
\frac{\left(J-E\chi(\bar\rho)\right)^2}{\chi(\bar\rho)}. 
\end{equation}

Under suitable conditions, we shall exhibit a time-dependent path for
which $(1/T) \mc I_{[0,T]} (\rho,j)$ is strictly less then $U$. This
implies the inequality $\Phi< U$.
Let $\Lambda =(0,1)$ and $(\rho(t),j(t))$ be a periodic trajectory, with time averaged
current $J$, in the form of a traveling
wave of velocity $v$, 
\begin{equation}\label{tv}
\left\{
\begin{array}{l}
\rho(t,x)=\rho_0(x-vt)\\
j(t,x)=J+v\left[\rho_0(x-vt)-\bar\rho\right]\,,
\end{array}
\right.
\end{equation}
where $\rho_0$ is an arbitrary periodic function with period one  such that
$\int_0^1\!dx\,  \rho_0(x)=\bar\rho$. As functions of $t$, $\rho$ and
$j$ are periodic with period $1/v$.
It is easy to verify that the continuity equation holds and that the
time average of $j$ over the time interval $v^{-1}$ is equal to $J$. For this choice
we have
\begin{eqnarray}\label{tv-u}
& &\Phi(J)\leq v \, \mathcal I_{[0,v^{-1}]}(\rho,
j)=\frac{v}{4}\int_0^{v^{-1}}\! dt\,  \mathcal U(\rho(t),j(t)) \nonumber \\ 
& &=\frac{1}{4}\int_0^1\! dx\,
\frac{\left\{J+v\left[\rho_0-\bar\rho\right]-J(\rho_0)\right\}^2}{\chi(\rho_0)} 
\end{eqnarray}
As shown in \cite{MR2227084} under the condition
\begin{equation}\label{c-torus}
\left[1-\frac{E^2\chi^2(\bar\rho)}{J^2}\right]\chi''(\bar\rho)>0,
\end{equation}
for $J$ large enough it is possible to
find $\rho_0$ and $v$ such that the right hand side
of \eqref{tv-u} is less then \eqref{U-m}.

\medskip
Consider the KMP model.
Since $\chi''>0$ condition \eqref{c-torus} is satisfied when $E$ is
small enough, in particular in the case of no external field, that is
for an equilibrium state.
The above argument thus provides a complete analytic proof of
the strict inequality $\Phi<U$.  
The existence of this dynamical phase transition has been also
observed in simulations in \cite{HG1}.  An open problem is whether the
phase transition exists in the case of a boundary driven model. At the
numerical level so far the answer has been negative \cite{HG0}.

In the case of the exclusion process, since $\chi''<0$, in order to
have a dynamical phase transition we need an external field.
This case has been discussed in \cite{bdcurrpt}. 
When $E$ and $J$ are small,  $\Phi=U$ and  the optimal
density profile for the variational problem \eqref{r27} defining $U$
is  constant.  
These authors perform a linear stability analysis 
showing, in particular, that 
the constant profile becomes
unstable for sufficiently large external fields and currents and
conclude the existence of a dynamical phase transition.  By a
numerical computation, they also show that the traveling wave path is
the optimal one for the variational problem \eqref{r24} defining
$\Phi$.

\subsection{Cumulants of the current and their universality properties}

\label{s:cumulants}

We define the average total current as
\begin{equation}
  \label{jvsq}
  \mc Q_{\epsilon, T}
  = \frac 1T \int_0^T\!dt\int_\Lambda \!dx \, j_\epsilon(t,x)\,,
\end{equation}
whose relationship with the microscopic dynamics
will be detailed in section \ref{emp-dc}.
In the limit $\epsilon\to 0$ and $T\to \infty$
$\mc Q_{\epsilon,T}$ converges to $\int_{\Lambda}\!dx\, J(\bar\rho)$,
where $J(\bar\rho)$ is the hydrodynamic current corresponding to the
stationary density profile $\bar\rho$.
The MFT allows to describe the asymptotic
behavior of the cumulants of $\mc Q_{\epsilon, T}$. We present
in this section 
some results obtained in 
\cite{BD04,MR2054160, ADLW,ABDS13}.

We assume throughout this section that there is no external field,
$E=0$. We start with the case of a one-dimensional boundary driven
system and choose $\Lambda=(0,1)$. Since in one space dimension
the only divergence-free vector fields are the constant fields, the
analysis of the asymptotic behavior of $\mc Q_{\epsilon, T} $ is
equivalent to \eqref{LT}.
Assume that $\rho_0 < \rho_1$ so that the stationary current is
negative.
The asymptotics of the cumulants of $Q_{\varepsilon, T}$ can be
deduced from the general formulas \eqref{fa1}, \eqref{r31} by computing the
derivatives of $\Phi^\sharp$ at $0$. Note that the behavior of
$\Phi^\sharp$ in a neighborhood of $0$ corresponds to the behavior of
$\Phi$ in a neighborhood of the stationary current $J (\bar\rho)$.
In view of the continuity argument given in the paragraph
before \eqref{fa1}, we can compute the cumulants analyzing the
time-independent variational problem \eqref{r27}.  The same continuity
argument implies that, in a neighborhood of $J(\bar\rho)$, the optimal
$\rho$ for \eqref{r27} is increasing.

As shown in \cite{BD04}, we then obtain
\begin{equation}
\label{la32}
\Phi(J) = U(J) = \frac J4 \int_{\rho_0}^{\rho_1} \frac{D(\rho)}{\chi(\rho)}\,
\Big( 2 \,-\, \frac{2 + A(J) \chi(\rho)}
{\sqrt{ 1 + A(J) \chi(\rho)}} \Big)\, d\rho
\end{equation}
where $A(J)$ is related to $J$ by
\begin{equation}
\label{la15}
J=  -\, \int_{\rho_0}^{\rho_1}
\frac{D(\rho)} {\sqrt{ 1 + A(J) \chi(\rho)}}
\, d\rho.
\end{equation}
By taking the Legendre transform \eqref{r31} we deduce that for $\theta$ small
\begin{equation}
\label{la13}
\Phi^{\sharp}(\theta) \,=\, -\, \frac {B(\theta)}4 \, \Big[
\int_{\rho_0}^{\rho_1} \frac{D(\rho)}{\sqrt{ 1 + B(\theta) \chi(\rho)}}\,
d\rho \Big]^2 ,
\end{equation}
where $B$ is a related to $\theta$ by
\begin{equation}
\label{la14}
\theta\,=\, \frac 12
\, \int_{\rho_0}^{\rho_1} \frac{D(\rho)}{\chi(\rho)}\,
\Big( 1\, -\, \frac{1} {\sqrt{ 1 + B(\theta) \chi(\rho)}} \Big)\, d\rho.
\end{equation}

Denote by $(\Phi^\sharp)^{(k)}$ the $k$-th derivative of $\Phi^\sharp$
and by $C_k$ the $k$-th cumulant of $\mc Q_{\epsilon,T}$
From \eqref{fa1} we deduce that 
\begin{equation}
\label{la02}
C_k \approx \Big( \frac {\epsilon}{T}\Big)^{k-1}
(\Phi^\sharp)^{(k)} (0), \qquad k\ge 1,
\end{equation}
where the approximation becomes exact as $\epsilon \to 0$ and $T\to \infty$.
We point out that the cumulants calculated in \cite{BD04}
are related to a random variable which differs from $\mc Q_{\epsilon,T}$
by the scaling factor $\epsilon/T$.

By expanding \eqref{la13} and \eqref{la14} in powers series we can compute
the derivatives of $\Phi^\sharp$ .The first three are
$(\Phi^\sharp)^{(1)} (0) = - I_1$, $(\Phi^\sharp)^{(2)} (0) =
I_2/I_1$, $(\Phi^\sharp)^{(3)} (0) = - 3( I_3 I_1 - I^2_2)/I^3_1$,
where
\begin{equation*}
I_n \;=\; \int_{\rho_0}^{\rho_1} D(\rho)\,
[2 \chi(\rho)]^{n-1} d\rho \;, \quad n=1,2,3.
\end{equation*}

In the case where $D$ is constant and $\chi(\rho) = \rho(1-\rho)$,
which corresponds to the case of the simple exclusion process,
condition \eqref{c<} holds. By the results of section \ref{s:dft}
we get that $\Phi(J)=U(J)$ for all $J$.
The optimal solution $\rho$ of the variational problem \eqref{r27} for $U$
has been computed in \cite{BD04}. For any value of $\theta$ one then gets the closed form
\begin{equation*}
\Phi^\sharp(\theta)=\left(\mathrm{arcsinh} \sqrt{\omega}\right)^2\,.
\end{equation*}
where
\begin{equation}
\label{omega}
\omega \,=\, \rho_0 (e^\theta -1) \,+\, \rho_1 (e^{-\theta} -1) \,+\,
\rho_0 \rho_1  (e^\theta -1)  (e^{-\theta} -1)\,.
\end{equation}

The computation of the Legendre transform has been extended to higher
dimensions in \cite{ABDS13}.
Consider a domain $\Lambda$ in dimension
$d>1$ and assume that there are two external reservoirs,
at densities $\rho_A$ and $\rho_B$,
in the regions $A,B\subset \Lambda$.
For $J$ close to the stationary value or globally under the
assumption \eqref{c<} we have that $\Phi$ is equal to $U$.
The fluctuations of the net flow between $A$ and $B$
are analyzed in \cite{ABDS13} where it is shown that
\begin{equation}
\label{cap}
  \Phi^\sharp (\theta) = {\mathrm{Cap}}_\Lambda \big( A, B\big)
   \, \Phi^\sharp_1 (\theta)
\end{equation}
where $\Phi^\sharp_1$ is computed for a one-dimensional system on the
interval $(0,1)$ with boundary densities
$\rho_A$, $\rho_B$, and ${\mathrm{Cap}}_\Lambda \big( A, B\big)$ is
the capacity, that depends only on the geometry, of a condenser formed
by $A$ and $B$ in $\Lambda$.
From \eqref{cap} it follows in particular that  the ratio between
any pair of cumulants is the same as in one dimension.

\medskip
We now turn to the one-dimensional ring. Under the assumption that
$D(\rho)$ is constant and that $\chi(\rho)$ is concave, 
in \cite{MR2227084}  it is proven  that $\Phi(J)=U(J)$. 
Moreover, if $1/\chi(\rho)$ is a convex function then 
$U(J) \,=\, (1/4) (J^2/\chi(\bar\rho))$. 
Therefore, under the two previous conditions, the
Legendre transform $\Phi^{\sharp} $ of $\Phi$ is simply given by
\begin{equation*}
\Phi^{\sharp}(\theta) \,=\, \theta^2\, \chi(\bar\rho).
\end{equation*}
As $\Phi^{\sharp}$ is quadratic, in view of \eqref{la02},
the limiting variance of $\varepsilon^{-1}T\mc Q_{\epsilon, T}$
is equal to $2\chi(\bar\rho)$ while the remaining cumulants vanish
as $\epsilon\to 0$ and $T\to \infty$.
The finite size corrections to this Gaussian behavior are studied in
\cite{ADLW}.
The relationship between the variable $Q_t$ used in this reference  and \eqref{jvsq} is
\begin{equation*}
  \mc Q_{\epsilon,T}=\frac{\epsilon^2}{T}Q_{\epsilon^{-2}T}.
\end{equation*}
In our notation, the finite size correction to
the function $\Phi(J)$ is
\begin{equation*}
\Phi_\epsilon (J) \;=\; \ \Phi (J)
-\, \epsilon \Big\{  \frac{J^2}{4\chi}
+ D \, \mc F\Big( \frac{J^2 \chi''}{16 D^2 \chi}\Big) \Big\}
\,+\, o(\epsilon).
\end{equation*}
In this formula, $D=D(\bar\rho)$, $\chi = \chi(\bar\rho)$, $\chi'' =
\chi''(\bar\rho)$, and
\begin{equation*}
\mc F(u) \,=\, \sum_{k\ge 2} \frac{B_{2k-2}}{(k-1)! k!} (-2u)^k\,,
\end{equation*}
where $B_n$ are the Bernoulli numbers, the coefficients of the
expansion $x(e^x-1)^{-1} = \sum_{n\ge 0} B_n x^n/n!$.
Accordingly, the finite size correction to  $\Phi^\sharp$ up to
first order in $\epsilon$ is
\begin{equation*}
\Phi^\sharp_\epsilon (\theta) \,=\,
\Phi^\sharp (\theta)
+ \epsilon \Big\{  \chi \theta^2 +
D \, \mc F\Big( \frac{\chi \chi''}{4 D^2} \theta^2 \Big) \Big\}.
\end{equation*}

From this expansion we derive the asymptotic for the cumulants of the
integrated current. More precisely, recalling \eqref{jvsq} the
variance of $\mc Q_{\epsilon, T}$ (including the first order
correction) is  \begin{equation*}
  C_2 \approx   \frac{\epsilon}{T} (1+\epsilon)   \, 2 \chi,
\end{equation*}
while the cumulant of order $2k$, $k\ge 2$, is
\begin{equation*}
  C_{2k} \approx   \frac{\epsilon^{2k}}{T^{2k-1}}
  \,
  B_{2k-2}\, \frac{(2k)!}{(k-1)! k!}
  D\, \Big( \frac{ - \chi \chi''}{2 D^2} \Big)^k.
\end{equation*}

\subsection{Current fluctuations for non stationary  infinite systems}

The MFT has been applied also to study current fluctuations for
diffusive infinite systems in non stationary states.  More precisely,
in \cite{DG2} the authors consider a diffusive stochastic lattice gas
on the infinite lattice with step initial condition. This
means that at the initial time the particles are distributed 
in a non steady state having density $\rho_a$ at the left of the
origin and density $\rho_b$ at the right. Let $Q_\tau$ be the net flow of
particles across the origin  up to time $\tau$ and
let
\begin{equation}\label{DGfor1}
  \Phi^\sharp(\theta)=\lim_{\tau\to +\infty}\frac{1}{\sqrt \tau}\log
  \mathbb E\left(e^{\theta Q_\tau}\right)
\end{equation}
be the corresponding generating function of the cumulants.
The appearance of the $\sqrt\tau$ in this formula is due to the fact
that a law of large numbers holds for $Q_\tau/\sqrt\tau$ for large
$\tau$.  In \eqref{DGfor1} the expected value can be interpreted in
two different ways depending on whether we consider fluctuations of
the initial condition (annealed case) or not (quenched case).
In the annealed case $\Phi^\sharp(\theta)$ satisfies a
relationship reminiscent of the Gallavotti-Cohen symmetry.
In \cite{DG2} the authors argue that $\Phi^\sharp(\theta)$ in
\eqref{DGfor1} can be computed using MFT.
The correct asymptotic behavior is obtained considering the scaling
parameter $\epsilon=\left(\sqrt{\tau}\right)^{-1}$ and letting the
macroscopic time $T$ vary on the finite window $[0,1]$. Since there is
conservation of the mass and the system is one dimensional the net
flow $Q_\tau$ in this approximation will coincide with
\begin{equation}\label{DGfor2}
\sqrt{\tau}\int_0^{+\infty}\left(\rho_\epsilon(x,1)-\rho_\epsilon(x,0)\right)\,dx\,.
\end{equation}
By \eqref{r02} and \eqref{r05}, in the annealed regime
$\Phi^\sharp(\theta)$ can be obtained as 
\begin{equation}
\label{DGfor3}
\begin{split}
& \Phi^\sharp(\theta)=
\inf \bigg\{-V_\mathrm{in}(\rho(0))+\theta \int_0^{+\infty}\!dx\,
\left[\rho (x,1)-\rho (x,0)\right]\
\\
&\qquad -\int_0^1\!dt\int_{-\infty}^{+\infty}\!dx\,
\frac{\left[j+D(\rho) \nabla \rho \right]^2}{\chi(\rho)}\bigg\}\,,
\end{split}
\end{equation}
where the infimum is carried out over all $(\rho,j)$ satisfying the
continuity equation.
The term $V_\mathrm{in}$ is due
to fluctuations of the initial condition.
This is a product of Bernoulli distributions
of parameter $\rho_a$ in the negative axis and $\rho_b$
in the positive one. The functional $V_\mathrm{in}$
coincides with \eqref{a1} with $\bar \rho(x)$
substituted by
\begin{equation}\label{DGfor}
\rho_a\left(1-\theta(x)\right)+\rho_b\theta(x)\,,
\end{equation}
where $\theta(x)$ is the Heaviside function.
In the quenched case there is an expression similar to \eqref{DGfor3}
but without the term $V_\mathrm{in}$ and the minimization has to be done
over all the $(\rho,j)$ such that $\rho(x,0)$ coincides with
\eqref{DGfor}.
The variational problem \eqref{DGfor3} and the corresponding
one for the quenched case
cannot be solved explicitly in general.
An exact solution is possible
for free particles \cite{DG2} and in some cases
for the symmetric exclusion process
\cite{PhysRevE.89.010101}.

In the annealed case for the symmetric
exclusion process it is possible
to apply some symmetry argument to \eqref{DGfor3}
showing that the dependence of $\Phi^\sharp(\theta)$
on the parameters $\rho_a$, $\rho_b$ and $\theta$
is only through their combination
$\omega$ as in  \eqref{omega} (with $\rho_0$ and $\rho_1$ replaced by
$\rho_a$ and $\rho_b$).
This means that $\Phi^\sharp(\theta)=F(\omega)$
for a suitable function $F$ whose
explicit expression has been obtained in \cite{DG1}
using microscopic combinatorial
arguments. An open problem is to recover such an
expression using instead \eqref{DGfor3}.
Still a symmetry argument for \eqref{DGfor3} shows that, in the annealed case,
from the exact  expression for the symmetric simple exclusion it is
possible to obtain the expression of
$\Phi^\sharp(\theta)$ for other models,
like the KMP model,
having constant diffusion matrix and quadratic mobility.

Another result that can be deduced from \eqref{DGfor3}
is a non-Gaussian decay of the distribution
of the net flow $Q_\tau$. This holds under some conditions on the
transport coefficients both in the annealed and in the quenched regime.
More precisely, under some conditions that hold e.g., for the
exclusion process, for large  $\tau$ and large $q$ 
the net flow $Q_\tau$ has the super-Gaussian statistics
\begin{equation}\label{DGfor5}
\mathbb P\left(\frac{Q_\tau}{\sqrt\tau}\approx q\right)
\asymp e^{-\alpha \sqrt\tau \, q^3}\,,
\end{equation}
for a suitable positive constant $\alpha$.
On the other hand,  in \cite{1742-5468-2013-12-P12011} it is shown that
the KMP model in the quenched regime exhibits instead a sub-Gaussian
statistics.

\section{Hyperbolic conservation laws}
\label{s:7}

The MFT for hyperbolic conservation laws is less
developed than the case of driven diffusive systems. In this section
we show however how some results can be obtained by taking the formal
limit of vanishing viscosity. We restrict the discussion to the
one-dimensional inviscid Burgers equation \cite{burgers}, which
is a simple model for a compressible fluid.  It can be obtained as
hydrodynamic limit of the asymmetric exclusion process under Eulerian
rescaling of space-time, that is keeping $x/t$ fixed.

\subsection{Hydrodynamics}
The hydrodynamic equation is
\begin{equation}
  \label{h.1}
  \partial_t \rho +\nabla \chi(\rho) =0
\end{equation}
where $\chi(\rho)=\rho(1-\rho)$ is the mobility of the exclusion
process (called \emph{flow} in the context of hyperbolic conservation laws) and
we consider an external field toward the right with unit strength.
The standard inviscid Burgers equation, that is usually written in the form
$\partial_t u + \nabla u^2 =0$, can be obtained from \eqref{h.1}
by a simple change of variables. According to the interpretation in
terms of the exclusion process, we shall however consider $0\le
\rho\le 1$.

An important difference between the evolution \eqref{h.1} and the
driven diffusive (parabolic) equations considered before is that,
even if the initial condition is smooth, the solution to \eqref{h.1}
may develop singularities, called \emph{shocks}, after a finite time.
This is easily seen by the method of characteristics. Indeed, in
the Lagrangian coordinates, an element of the ``fluid'' at local
density $\rho$ has a velocity $v_\rho=\chi'(\rho)=1-2 \rho$.
In particular, low density regions $\rho\ll 1$ will overtake the
regions of intermediate density $\rho\approx 1/2$ resulting in the
formation of a singularity.

Let us discuss these shock solutions to \eqref{h.1} in more
detail. Consider the function
\begin{equation}
  \label{h.2}
  \phi(x)=\phi_{\rho_-,\rho_+}(x)=
  \begin{cases}
    \rho_- & x < 0 \\
    \rho_+ & x>0
  \end{cases}
\end{equation}
describing a shock from $\rho_-$ to $\rho_+$. If we set
\begin{equation}
  \label{h.3}
  v= v_{\rho_-,\rho_+} = \frac{ \chi(\rho_+)
  -\chi(\rho_-)}{\rho_+-\rho_-} = 1 - (\rho_+ + \rho_-)
\end{equation}
then it is not difficult to check that $\phi(x-vt)$ solves \eqref{h.1}
in the sense of distributions. Observe that as $\rho_+ -\rho_-\to 0$
the shock velocity $v_{\rho_-,\rho_+}$ approaches the velocity of the
characteristics.  As far as the hydrodynamic equation \eqref{h.1} is
concerned, both $\rho_-<\rho_+$ and $\rho_->\rho_+$ are
allowed. These cases correspond to quite different situations from a
physical point of view. Recalling that we have chosen an external
field toward the right, the case $\rho_- <\rho_+$ corresponds to a low
density region at the left blocked by a high density region (a pile of
particles in the microscopic picture) at the right and appears a
natural feature of the system. On the other hand, the case
$\rho_->\rho_+$ does not have a natural interpretation and should be
regarded as unphysical.

\medskip
The hyperbolic evolution \eqref{h.1} can be obtained from the driven
diffusive equation in the limit of vanishing viscosity.
Namely, by considering
\begin{equation}
  \label{h.3b}
  \partial_t \rho + \nabla \chi(\rho) =
  \nu \nabla \big( D(\rho) \nabla \rho\big)
\end{equation}
and taking the formal limit $\nu \to 0$.
For the exclusion process $D$ is constant but for a while we consider
arbitrary diffusion coefficient. By setting $\nu=0$ we recover the
evolution \eqref{h.1}, but as we next show there is another condition
from \eqref{h.3b} that survives in the limit $\nu\to 0$ and rules
out the unphysical decreasing shocks.  Let $h(\rho)$ be a convex
function (an \emph{entropy} in the terminology of hyperbolic
conservation laws)
and let $g(\rho)$ be the function defined by
\begin{equation}
  \label{h.4}
  h'(\rho) \chi'(\rho) = g'(\rho).
\end{equation}
In the terminology of hyperbolic conservation laws $g$ is called the
\emph{entropy flow}  associated to $h$.
Multiplying \eqref{h.3b} by $h'(\rho)$ we deduce
\begin{equation*}
  \begin{split}
   & \partial_t h(\rho) + \nabla g(\rho)  =
  \nu h'(\rho) \nabla\big(D(\rho) \nabla \rho\big)
  \\
  &\quad  = - \nu h''(\rho) D(\rho)  \big(\nabla \rho\big)^2
  + \nu  \nabla\big( h'(\rho) D(\rho) \nabla \rho\big).
  \end{split}
\end{equation*}
Since the last term is a total derivative and $h''(\rho)\ge 0$,
by taking the limit $\nu\to 0$, we deduce the inequality
$\partial_t h(\rho) + \nabla g(\rho) \le 0$.
We conclude that the appropriate formulation of \eqref{h.3b} in the
vanishing viscosity limit is
\begin{equation}
  \label{h.5}
  \begin{cases}
    \partial_t \rho + \nabla \chi(\rho) = 0 \\
    \partial_t h(\rho) + \nabla g(\rho) \le 0
  \end{cases}
\end{equation}
where $h$ is an arbitrary convex function and $g$ is defined by
\eqref{h.4}. In view of the specific form of the flow $\chi(\rho)$
(more precisely in view of its concavity), it is simple to check that
increasing shocks, i.e.\ $\phi_{\rho_-,\rho_+}(x-v_{\rho_-,\rho_+}t)$
with $\rho_-<\rho_+$ solve \eqref{h.5} while decreasing shocks do not.

Observe that while \eqref{h.1} in invariant under time and space
reflection the entropy condition in \eqref{h.5} is not and implies a
time arrow. The initial value problem corresponding to
\eqref{h.5} on the whole line is well posed 
\cite{serre}, while uniqueness fails for \eqref{h.1}.

\medskip
As we would like to include boundary reservoirs in the model, we
need to discuss the role of boundary conditions when the hyperbolic
evolution \eqref{h.5} is considered on the interval $\Lambda=(0,1)$.
More precisely, we consider boundary reservoirs with chemical
potentials $\lambda_0,\lambda_1$ at the endpoints of $\Lambda$ and
denote by $\rho_0$, $\rho_1$ the corresponding values of the density,
i.e., $\lambda_i= f'(\rho_i)$.
While for driven diffusive systems the effect of the boundary
reservoirs is to fix the value of the density, for hyperbolic
conservation laws
the situation is more subtle. As we have discussed above, the
hyperbolic evolution develops shocks which may occur also at the
boundary. In this case the value of the density at the boundary will not be
fixed by the reservoirs but rather constrained by the admissibility of
the shock.  The boundary conditions will
thus be given in terms of inequalities and not of identities.

Referring to 
\cite{serre} for the general theory of boundary
conditions for hyperbolic conservation laws, we only discuss the case
of the Burgers equation.  At the left endpoint $x=0$ the reservoir's
density is $\rho_0$ and the appropriate boundary condition is the
following. If $\rho_0\le 1/2$ then $1-\rho_0\le \rho(t,0) \le 1$ while
if $\rho_0\ge 1/2$ then $\rho_0 \le \rho(t,0)\le 1$. Likewise, at the
right endpoint $x=1$ the reservoir's density is $\rho_1$ and the
boundary condition is the following. If $\rho_1\le 1/2$ then $0\le
\rho(t,1) \le 1/2$ while if $\rho_1\ge 1/2$ then $0\le \rho(t,1)\le
1-\rho_1$.

\subsection{Large fluctuations}

We discuss first the case of periodic boundary conditions.
As for the case of driven diffusive system, we want to compute the
probability of a space-time fluctuation of the density and
current. Due to the singular behavior of the hyperbolic evolution,
there are two different large deviations regimes. In order to violate
the continuity equation in \eqref{h.5} we need to apply an external
field over a macroscopic part of the system. On the other hand if we
consider a solution to \eqref{h.1} with shocks, we can violate the
entropy condition in \eqref{h.5} (which allows only increasing shocks)
by applying a  field localized on the shocks.
In terms of the microscopic dynamics,
consider a high density region We describe only the probability of fluctuations violating the entropy
condition in \eqref{h.5} which are, so to speak, much less improbable
and the relevant ones for the computation of the quasi-potential.
For such fluctuations the density and current are directly related.
Since we do not violate the continuity equation \eqref{h.1},
once we specify the fluctuation $\rho$ of the density the current will
be given by $\chi(\rho)$.
The corresponding action functional has been derived in \cite{jensen,
MR2073328}. The answer is amazingly simple: in order to violate the
entropy condition we need only to pay the corresponding entropy cost.
The subtle point is to decide which is the correct entropy to
use. Note in fact that the entropy condition in \eqref{h.5} does not
depend on the function $h$: if it holds for some convex $h$
($g$ is then given by \eqref{h.4}) then it holds for all convex $h$.
In order to find the correct choice of $h$ we need to go back to the
small viscosity approximation \eqref{h.3b}. At this level the
physical  entropy $h$ is selected by the Einstein condition $h''(\rho) =
D(\rho)/\chi(\rho)$. For the exclusion process $D=1$ so that $h$ is
the equilibrium free energy $f$, i.e.,
\begin{equation}
  \label{berne}
  h(\rho)= f(\rho) = \rho \log \rho +(1-\rho) \log (1-\rho).
\end{equation}
The Jensen-Varadhan large deviation formula for this inviscid Burgers
equation then reads
\begin{equation*}
  \mathbb{P}\big(\rho_\epsilon \approx \rho, t\in[T_0,T_1] \big)
  \asymp \exp\big\{ -\epsilon^{-1} I_{[T_0,T_1]} (\rho)\big\}
\end{equation*}
with $I(\rho)$ finite only for $\rho$ satisfying \eqref{h.1} and for
such $\rho$ given by
\begin{equation}
\label{Ihyp}
  I_{[T_0,T_1]}(\rho)=
  \int_{T_0}^{T_1}\!dt\int_0^1\! dx\,
  \big[ \partial_t f (\rho) + \nabla g (\rho) \big]_+
\end{equation}
where $[a]_+ = \max \{ 0,a\}$ is the positive part of $a$, $f$ as in
\eqref{berne}, and $g$ satisfies \eqref{h.4}.
As discussed in \cite{MR2227085} the expression \eqref{Ihyp} can be derived
from \eqref{I=} by considering the limit of vanishing viscosity.

While the structure of the functional $I$ in \eqref{Ihyp} is very
different from the case of driven diffusive systems of Section~\ref{s:ft}, the
time reversal symmetry of Section~\ref{s:tr} holds also in this case.
Since we are considering  periodic boundary conditions, the total mass
$m=\int_0^1\!dx \rho(x)$ is conserved. The quasi-potential is then
\begin{equation*}
  V(\rho) = \int_0^1\!dx\, \big[ f(\rho) - f(m) \big].
\end{equation*}
Since the time reversed dynamics can be realized
by inverting the external field, the adjoint hydrodynamics is
obtained by replacing $\chi(\rho)$ with $-\chi(\rho)$ so that
\begin{equation*}
  I_{[T_0,T_1]}^*(\rho) =
  \int_{T_0}^{T_1}\!\int_0^1\! dx\,
  \big[ \partial_t f (\rho) - \nabla g (\rho) \big]_+.
\end{equation*}
It is now simple to check that \eqref{r28} holds also in the
hyperbolic regime, i.e.,
\begin{equation*}
  V(\rho(T_0)) + I_{[T_0,T_1]}(\rho) =
  V (\rho(T_1)) + I_{[-T_1,-T_0]}^*(\theta \rho).
\end{equation*}

\medskip
We now discuss the large deviations asymptotics in the presence of
boundary reservoirs.  Since the boundary condition for the hyperbolic
evolution \eqref{h.5} discussed in the previous Section can be in
formulated as entropic conditions at the boundary,
we need to add to the Jensen-Varadhan functional \eqref{Ihyp} the
boundary terms that take into account the total entropy production at
the boundary. For the exclusion process, these terms have been
computed in \cite{MR2227085} by considering the limit of vanishing
viscosity. They have the form
\begin{equation*}
  \begin{split}
  I^{(0)}_{[T_0,T_1]} (\rho) = \int_{T_0}^{T_1}\!dt\,
  s^{(0)}(\rho(t,0), \rho_0)
  \\
  I^{(1)}_{[T_0,T_1]} (\rho) = \int_{T_0}^{T_1}\!dt\,
  s^{(1)}(\rho(t,1), \rho_1)
  \end{split}
\end{equation*}
where $\rho_0$, $\rho_1$ are the densities of the boundary reservoirs
and the functions $s^{(0)}$, $s^{(1)}$ are explicitly given in
\cite{MR2227085}.  Accordingly, the full rate function is
\begin{equation}
\label{h.10}
  I_{[T_0,T_1]} (\rho)=
  I^{\mathrm{bulk}}_{[T_0,T_1]}(\rho)+
  I^{(0)}_{[T_0,T_1]} (\rho) +I^{(1)}_{[T_0,T_1]} (\rho)
\end{equation}
with $I^{\mathrm{bulk}}$ given by \eqref{Ihyp}.

By considering the variational problem \eqref{r07}, i.e.,
$V(\rho)=\inf I_{(-\infty,0]}(\hat\rho)$, with the constraint
$\hat\rho(0)=\rho$, for the action functional \eqref{h.10}, the
formulas for the quasi-potential of the boundary driven asymmetric
exclusion process derived in \cite{DLS} by exact computations on the
microscopic ensembles can be obtained within the MFT formalism. We
refer to \cite{Bah10}, for the details of such computations that, as
there discussed, can be generalized to higher space dimensions and to
the models satisfying the symmetry $\chi(\rho) = \chi(\psi(\rho))$ for
some decreasing $\psi$.

\section{Microscopic models}
\label{s:8}

Models have played a fundamental role in equilibrium statistical mechanics.
The Ising model provided the first proof that statistical mechanics can explain
the existence of phase transitions and was a main guide in the study of
critical behavior. A reason for this effectiveness is the circumstance
that the macroscopic behavior is, to a considerable extent, independent of
the microscopic details. Hence different systems exhibit qualitatively the
same phenomenology at large scales.
This section requires some basic notions on probability theory and
Markov processes, see e.g., \cite{MR1689633}.

Stochastic lattice gases are a collection of particles performing
random walks on a lattice in continuous time and interacting
with each other. These particles are to be considered
indistinguishable. Accordingly, the microscopic state is specified by
giving the occupation number in each site of the lattice.  The effect
of the interaction is that the \emph{jump rates} depend on the local
configuration of the particles, i.e., on the occupation numbers of the
nearby sites.  For non-isolated systems we model the effect of the
reservoirs by adding creation/annihilation of particles at the
boundary. The effect of an external field is modeled by perturbing the
rates and giving a net drift toward a specified direction.

As basic microscopic model we consider a stochastic lattice gas in a
finite domain, with an external field, and either with periodic
boundary conditions or with particle reservoirs at the boundary.  The
dynamics can be informally described as follows.  Associated to each
lattice site there is an independent Poisson clock of parameter
depending on the local configuration.  When the clock rings, a
particle jumps from this site to a neighboring site.  In the case of
particle reservoirs, superimposed to this dynamics, at the boundary
particles are created and annihilated at exponential times.

Fix $\Lambda \subset \mb R^d$ and, given $\epsilon>0$, let
$\Lambda_\epsilon = \Lambda\cap\epsilon\mb Z^d$ its discrete
approximation. The microscopic configuration
is given by the collection of occupation variables $\eta(i)$, $i\in \Lambda_\epsilon$,
representing the number of particles at the site $i$.
We denote by $\Omega_\epsilon$ the space of all possible configurations.
The microscopic dynamics $\{\eta_t\}_{t\in\mb R}$
of the configuration of the system is formally specified in terms
of its infinitesimal \emph{generator} $L$, defined
as follows. Let $f:\Omega_\epsilon\to \mb R$ be an observable, then
\begin{equation}
\mb E \big( f (\eta_{t+h}) \big| \eta_t \big) - f(\eta_t) = (L f)(\eta_t) \, h
+ o(h)
\,,
\end{equation}
so that the expected infinitesimal increment of $f(\eta_t)$ is
$(Lf)(\eta_t)\, dt$.
Recall that $\mb E$ denotes the expectation over trajectories on the configuration space.
The transition probability of the Markov process
$\eta_t$ is then given by the kernel of the
semi--group generated by $L$, i.e.\
\begin{equation}
p_t(\eta,\eta') = e^{t L} (\eta,\eta')
\,.
\end{equation}

We can rewrite the full generator $L$ as follows
\begin{equation}
\label{genmc}
\begin{array}{l}
\displaystyle{
L f(\eta)
=  \sum_{i,j\in\Lambda_\epsilon}
c_{ij}(\eta) \big[ f(\sigma^{ij}\eta) - f(\eta) \big]
} \\
\displaystyle{
+ \sum_{\pm,i\in\Lambda_\epsilon}
c^\pm_{i}(\eta) \big[ f(\sigma_\pm^{i}\eta) - f(\eta) \big]
\,,}
\end{array}
\end{equation}
where
$\sigma^{ij}\eta$ is the configuration obtained from $\eta$ letting one particle jump from $i$ to $j$,
$\sigma_\pm^{i}\eta$ are the configurations associated to the creation or annihilation of a particle in site $i$,
and $c_{i,j}(\eta)$, $c^\pm_i(\eta)$ are the corresponding jump rates.
We denote by $\partial\Lambda_\epsilon$ the interior boundary of $\Lambda_\epsilon$,
i.e. the collection of sites $i\in\Lambda_\epsilon$
at distance $\epsilon$ from $\epsilon\mb Z^d\backslash\Lambda_\epsilon$.
The cases when the rates $c^\pm_i(\eta)$ are zero except for
$i\in\partial\Lambda_\epsilon$ correspond to conservative bulk
dynamics, with a hydrodynamic equation as in \eqref{2.1}.  In these
cases, creation and annihilation of particles at the boundary describe
the interaction with the external reservoirs.  Models with non-zero
creation/annihilation rates $c^\pm_i(\eta)$ also in the bulk
correspond to reaction-diffusion equations, an example being
\eqref{eqn6}.

A physical state of the system corresponds to a probability distribution
$P$ (ensemble) on the configuration space $\Omega_\epsilon$.  A
state is \emph{invariant} (stationary) under the dynamics if
\begin{equation}
\label{muinvpt}
\sum_{\eta\in \Omega_\epsilon} P(\eta) \, e^{t L } (\eta,\eta') =
P (\eta').
\end{equation}
Namely, if we distribute the initial condition $\eta$ according to
$P$, then the distribution of $\eta_t$, at any later time $t\geq0$, is again $P$.
A necessary and sufficient condition for a state $P $ to be
invariant is
\begin{equation}
\label{muinv}
E_{P } \big( L f  \big) =0
\quad \textrm{for all observables } f,
\end{equation}
where $E_P$ denotes the expectation with respect to $P$.

All the models that we consider are \emph{irreducible}, i.e.\ there is
a strictly positive probability to go from any configuration to any
other.  In this case, according to general results on Markov
processes, the invariant state is unique and it coincides with the
limiting distribution of the system when $t\to\infty$.

If the generator $L$ satisfies the detailed balance condition with
respect to some distribution $P$, namely
\begin{equation}\label{rev}
E_{P}(gL f) = E_{P}(f L g)
\,,
\end{equation}
for all observables $f,g$,
then $P$ is necessarily an invariant state. In such a case the
process is said to be \emph{time reversal invariant}.
This terminology is due to the following fact.
Let ${\mb P}_{\eta}$ be the probability distribution on the space of paths
$\{\eta_t\}_{t\geq0}$ with initial condition $\eta_0=\eta$, and let
${\mb P}$ be the stationary process, i.e., the distribution on the
space of paths with initial configuration $\eta_0$ distributed
according to the invariant state $P$.
Since $P$ is invariant, the distribution $\mb P$ is invariant with respect
to time shifts.  We can thus regard $\mb P$ as a distribution on paths
defined also for $t\le 0$.  This probability distribution is invariant under time
reversal if and only if the detailed balance condition \eqref{rev}
holds.  Indeed, if $\vartheta$ is the time reversal, i.e.,
$(\vartheta \eta)_t :=\eta_{-t}$, we have that $\mb P \circ \vartheta$
is the stationary Markov process with generator $L^*$, the adjoint of
$L$ with respect to $P$, and condition \eqref{rev} is precisely the
condition that $L^*=L$.

An equivalent form of the detailed balance condition \eqref{rev} is as follows
\begin{equation}\label{rev2}
P(\eta)c(\eta,\eta')
=P(\eta')c(\eta',\eta)
\,,
\end{equation}
for all configurations $\eta,\eta'\in\Omega_\epsilon$.
In this equation $c(\eta,\eta')$ is the transition rate  from the
configuration $\eta$ to $\eta'$, which can be either a jump rate
$c_{ij}(\eta)$, if $\eta'=\sigma^{ij}\eta$, or a creation/annihilation
rate $c_i^\pm(\eta)$, if $\eta'=\sigma^i_\pm\eta$.

When the unique invariant state does not satisfy the detailed balance
condition \eqref{rev2}, the corresponding process is not time reversal
invariant.  Time reversal invariant processes
correspond to equilibrium thermodynamic states.  The converse is not
necessarily true: there can be microscopic models not invariant under time reversal 
corresponding to equilibrium macroscopic states \cite{MR1400378,MR1716811,BJ,GJLV}.  
This is not surprising:
going from the microscopic to the macroscopic description there is
loss of information.

Next, we describe in some detail some of the most studied microscopic models,
which allow a detailed mathematical analysis.
They are microscopic counterparts of the macroscopic models discussed in Section \ref{sec3}.

\subsection{The simple exclusion process}

The boundary driven simple exclusion process, on a domain
$\Lambda\subset\mb R^d$, is defined letting particles move according
to independent simple random walks, with the exclusion rule that there
cannot be more than one particle in a single lattice site (hard core
interaction).  This gives a kind of classical Pauli principle. It is
appropriate to remark that the simple exclusion process is a special
case of the Kawasaki spin dynamics \cite{PhysRev.145.224}.  This is a
conservative dynamics that satisfies detailed balance with respect
to a Gibbs distribution. The simple exclusion process corresponds to
the case of a constant Hamiltonian.

According to the exclusion rule, the space of all possible
configurations of the system is
$\Omega_\epsilon=\{0,1\}^{\Lambda_\epsilon}$.  In terms of the
generator \eqref{genmc} this corresponds to the following choice of
the bulk jump rates:
\begin{equation}\label{sep-rates}
\left\{\begin{array}{l}
\displaystyle{
\vphantom{\Big(}
c_{ij}(\eta)=\eta(i)(1-\eta(j))
\,\,\text{ for }\,\, |j-i|=\epsilon
\,,
} \\
\displaystyle{
\vphantom{\Big(}
c_{ij}(\eta)=0
\,\,\text{ otherwise }
\,.
}
\end{array}\right.
\end{equation}
The interaction with the boundary reservoirs is described by
creation and annihilation rates $c_i^\pm(\eta)$ for $i\in\partial\Lambda_\epsilon$.
Let $\lambda(x)$ be the chemical potential of the boundary reservoirs
(it is a continuous function on a neighborhood of $\partial\Lambda$).
The corresponding creation/annihilation rates are as follows
\begin{equation}\label{sep-rates-bound}
\left\{\begin{array}{l}
\displaystyle{
\vphantom{\Big(}
c_i^-(\eta)=\eta(i)
\sum_{j\in\partial\Lambda^o_\epsilon(i)}
\frac{1}{1+e^{\lambda(j)}}
\,\text{ for }\,\, i\in\partial\Lambda_\epsilon,
} \\
\displaystyle{
\vphantom{\Big(}
c_i^+(\eta)=(1-\eta(i))
\sum_{j\in\partial\Lambda^o_\epsilon(i)}
\frac{e^{\lambda(j)}}{1+e^{\lambda(j)}}
\,\,\text{ for }\,\, i\in\partial\Lambda_\epsilon,
} \\
\displaystyle{
\vphantom{\Big(}
c_i^\pm(\eta)=0
\,\,\text{ otherwise,}
}
\end{array}\right.
\end{equation}
where $\partial\Lambda^o_\epsilon(i)$ denotes the set of all sites at
distance $\epsilon$ from $i$ outside of $\Lambda_\epsilon$. Observe
that the rates at the corners of $\Lambda_\epsilon$ differ as there are
more  neighbors.

The model is clearly irreducible,
hence, as explained above,
there is a unique invariant state $P$ satisfying \eqref{muinv},
corresponding to the limiting distribution of the system.
When the chemical potential of the boundary reservoirs is constant,
$\lambda(i)=\lambda$ for all $i\in\Lambda_\epsilon$,
the detailed balance condition \eqref{rev2} holds.
The corresponding stationary state is given by the product 
distribution 
\begin{equation}\label{sep-eq}
P(\eta)
=\prod_{i\in\Lambda_\epsilon}
\frac{e^{\lambda\eta(i)}}{1+e^{\lambda}}
\,.
\end{equation}
On the other hand, when the chemical potential $\lambda(i)$ at the
boundary is not constant, the model is not time reversal invariant and
the stationary ensemble is not product.

\subsection{The zero range model}

In the zero range model there is no bound on the number of particles
which can occupy the same site, hence the space of all possible
configurations of the system is $\Omega_\epsilon=\mb
N^{\Lambda_\epsilon}$.  The dynamics is defined letting a particle
interact only with the other particles present in the same lattice
site.  The interaction can be either attractive or
repulsive. 
The bulk jump rates are:
\begin{equation}\label{zr-rates}
\left\{\begin{array}{l}
\displaystyle{
\vphantom{\Big(}
c_{ij}(\eta)=g(\eta(i))
\,\,\text{ for }\,\, |j-i|=\epsilon
\,,
} \\
\displaystyle{
\vphantom{\Big(}
c_{ij}(\eta)=0
\,\,\text{ otherwise.}
}
\end{array}\right.
\end{equation}
where $g:\,\mb N\to\mb R^+$ is a function such that $g(0)=0$ and
$g(k)>0$, $k\ge 1$, describing the type of interaction.  In particular, the
choice of linear function $g(k)=\alpha k$ corresponds to the ideal gas
(independent random walks).
Also in this model the boundary creation and annihilation rates
are associated to the chemical potential $\lambda(x)$ of the reservoirs,
which, as before, is a continuous function on a neighborhood of $\partial\Lambda$.
The boundary rates are 
\begin{equation}\label{zr-rates-bound}
\left\{\begin{array}{l}
\displaystyle{
\vphantom{\Big(}
c_i^-(\eta)=g(\eta(i))  \, \big| \partial  \Lambda^o_\epsilon(i) \big|
\,\,\text{ for }\,\, i\in\partial\Lambda_\epsilon
\,,} \\
\displaystyle{
\vphantom{\Big(}
c_i^+(\eta)=
\sum_{j\in\partial\Lambda^o_\epsilon(i)}
e^{\lambda(j)}
\,\,\text{ for }\,\, i\in\partial\Lambda_\epsilon
\,,} \\
\displaystyle{
\vphantom{\Big(}
c_i^\pm(\eta)=0
\,\,\text{ otherwise,}}
\end{array}\right.
\end{equation}
where $|\partial\Lambda^o_\epsilon(i)|$ denotes the cardinality of the
set $\partial\Lambda^o_\epsilon(i)$, recall \eqref{sep-rates-bound}.

If the function $g$ grows fast enough,
there is a unique invariant state $P$ satisfying \eqref{muinv}.
The peculiarity of this model is that,
for arbitrary chemical potential $\lambda(x)$,
the invariant distribution is a product distribution.
It has the following form:
\begin{equation}\label{zr-eq}
P(\eta)
=\prod_{i\in\Lambda_\epsilon}
\frac1{Z(\varphi(i))}
\frac{\varphi(i)^{\eta(i)}}{g(\eta(i))!}
\,,
\end{equation}
where $g(k)!=g(k)g(k-1)\dots g(1)$,
and $Z(\varphi)=\sum_{k\in\mb N}\frac{\varphi^k}{g(k)!}$.
The function $\varphi:\,\Lambda_\epsilon\to\mb R^+$ solves
the discrete Laplace equation
\begin{equation}\label{zr-phi}
\Delta_\epsilon\varphi(i)
=
\sum_{|j-i|=\epsilon}(\varphi(j)-\varphi(i))=0
\,,
\end{equation}
with boundary condition $\varphi(i)=e^{\lambda(i)}$ for lattice sites
$i$ immediately outside of the boundary. Also in this case,
when the boundary chemical potential is constant $\lambda(i)=\lambda$
for all $i\in\partial\Lambda_\epsilon^o$, the detailed balance condition \eqref{rev2} holds,
the solution to \eqref{zr-phi} is constant $\varphi=e^\lambda$,
and \eqref{zr-eq} describes an equilibrium state.

\subsection{The Glauber+Kawasaki model}\label{sec:g+k}

We consider here a Glauber+Kawasaki model
for which the conservative part of the dynamics is given by the same rates
\eqref{sep-rates} as in the exclusion process,
while the non conservative part of the dynamics \eqref{genmc},
associated to the creation and annihilation rates $c_i^\pm(\eta)$,
extends over all sites $i$ of the domain $\Lambda_\epsilon$.
In general, the creation and annihilation rates $c_i^\pm(\eta):\,\Omega_\epsilon\to\mb R^+$
are functions, translation invariant,
depending only on the value of the configuration $\eta$
on sites $j\neq i$ at distance at most $k\epsilon$ form $i$
($k$ is a fixed positive integer).
When the site $i$ is near the boundary,
the rates $c_i^\pm(\eta)$ will also depend on the value of the chemical potential
of the reservoirs.

For simplicity, we write an explicit formula for the creation and annihilation rates
only on the torus, i.e. when $\Lambda=[0,1]^d$ with periodic boundary conditions.
Let $\tau_i$ be the shift operator on the configuration space $\Omega_\epsilon$,
defined by $\left[\tau_i\eta\right](j)=\eta(j-i)$.
Then
\begin{equation}
\left\{
\begin{array}{l}
c^+_i(\eta)=(1-\eta(i))b(\tau_{-i}\eta)\\
c^-_i(\eta)=\eta(i)d(\tau_{-i}\eta)
\end{array}
\right.
\end{equation}
where the functions $b(\eta)$ and $d(\eta)$,
associated to the ``birth'' and ``death'' of particles,
only depend on the occupation numbers $\eta(j)$
for sites $j$ at distance at most $k\epsilon$ from the origin.

Recall that any product Bernoulli distribution
\begin{equation}\label{bernoulli}
P^p(\eta)=\prod_{i\in\Lambda_\epsilon}p^{\eta(i)}(1-p)^{1-\eta(i)}
\end{equation}
(cf. \eqref{sep-eq}) is time reversal invariant for the conservative
part of the dynamics.
Hence, $P^p$ will be time reversal invariant
with respect to the full dynamics provided that
\begin{equation}\label{inv-glaub}
\frac{d(\eta)}{b(\eta)}=\frac{1-p}{p}
\end{equation}
for all $\eta\in\Omega_\epsilon$.
Indeed, \eqref{inv-glaub} guarantees that the detailed balance condition \eqref{rev2}
holds also for the non conservative part of the dynamics.
Therefore, if ${d(\eta)}/{b(\eta)}$ is constant in $\eta$,
the stationary state $P$ is as in \eqref{bernoulli},
where $p$ is uniquely determined by \eqref{inv-glaub}.
When $d(\eta)/ b(\eta)$ is not constant,
the corresponding stationary state $P$ is, in general, not invariant
under time reversal and not product.

\subsection{The Kipnis-Marchioro-Presutti model}

The Kipnis-Marchioro-Presutti (KMP) model \cite{MR656869}, originally
proposed as a simple solvable model of heat conduction, does not fit
exactly in the general framework outlined above and we need to modify
the notation accordingly. We discuss this model only in the
one-dimensional case.

This model describes a linear chain of harmonic oscillators with a
random exchanges of energy between nearest neighbors and possibly
heat baths at the boundary sites. As usual, let $\Lambda = (0,1)$ be
the macroscopic domain and $\Lambda_\epsilon = \Lambda \cap \epsilon
\mathbb Z$ be the corresponding discrete chain.  In each lattice site $i\in
\Lambda_\epsilon$ there is an harmonic oscillator and we call
$(q(i),p(i))$ its canonical coordinates so that its energy is 
$H_i(q(i),p(i)) = q(i)^2+ p(i)^2$. The oscillators are mechanically
uncoupled, i.e.\ the total energy is $H =\sum_i H_i$, but the dynamics
has a stochastic term which induces an interaction. More precisely, on
the bonds $(i, i+\epsilon)$ there are independent Poissonian
clocks. When the clock across the bond $(i,i+\epsilon)$ rings,
we compute the energy $E= H_i (q(i),p(i))+
H_{i+\epsilon}(q(i+\epsilon),p(i+\epsilon))$ and redistribute the
canonical coordinates of the two oscillators uniformly to new values
$(q'(i),p'(i)), (q'(i+\epsilon),p'(i+\epsilon))$ chosen uniformly on
the surface $H_i (q'(i),p'(i))+
H_{i+\epsilon}(q'(i+\epsilon),p'(i+\epsilon))=E$.
On the boundary sites there are other two independent Poissonian
clocks. When a clock rings at a boundary site $i\in \partial
\Lambda_\epsilon$, choose the new value of the coordinates
$(q'(i),p'(i))$ according to the following rules forgetting the old
configuration  $(q(i),p(i))$.
Sample a value of the energy $E$ according to an exponential distribution
of parameter $\lambda(i)$ and let $(q'(i),p'(i))$ be uniformly
distributed on the surface $H_i(q'(i),p'(i))=E$.

A peculiar feature of this model is that the local energies $H_i$ have
a closed Markovian evolution. In the sequel, we shall denote by
$\eta(i)\in \mathbb R^+$ the energy of the oscillator at site
$i\in\Lambda_\epsilon$ and describe formally their evolution. Observe
that from a statistical mechanics viewpoint these are indeed the
relevant quantities.
Let us define for $p\in [0,1]$
$$
\left[\sigma_p^{i,j}\eta\right](k)=\left\{
\begin{array}{ll}
\eta(k) & \textrm{if}\ k\neq i,j
\,, \\
p(\eta(i)+\eta(j)) & \textrm{if}\ k=i
\,, \\
(1-p)(\eta(i)+\eta(j)) & \textrm{if}\ k=j
\,.
\end{array}
\right.
$$
and for $s \in \mathbb R^+$
$$
\left[\sigma_s^{i}\eta\right](k)=\left\{
\begin{array}{ll}
\eta(k) & \textrm{if}\ k\neq i
\,, \\
s & \textrm{if}\ k=i
\,.
\end{array}
\right.
$$
For this model the general formula \eqref{genmc} has to be substituted by
\begin{equation}
\label{genmc-kmp}
\begin{array}{l}
\displaystyle{
L f(\eta)
=  \sum_{i,j\in\Lambda_\epsilon\,,
|i-j|=\epsilon}
\int_0^1 \!dp \, \big[ f(\sigma_p^{ij}\eta) - f(\eta) \big]
} \\
\displaystyle{
+ \sum_{i\in\partial\Lambda_\epsilon}
\int_0^{+\infty}\!ds\, \lambda(i) e^{-\lambda(i)s}
\big[ f(\sigma_s^{i}\eta) - f(\eta) \big]
},
\end{array}
\end{equation}
where $\lambda(i)$ are the temperatures of the boundary thermostats.
The generator in \eqref{genmc-kmp} describes a stochastic evolution in which
every pair of nearest neighbor sites after an exponential time
redistribute the sum of their energies between the two sites in a uniform way.
This mechanism preserves the total energy of the system. At a boundary
site $i$ after an exponential time  the energy is replaced
by the value of an exponential random variable of parameter $\lambda(i)$.

If $\lambda(i)=\lambda$ for both boundary sites, then
the model is time reversal invariant.
The corresponding equilibrium state $P$ is given by the following
product distribution on $\left(\mathbb R^+\right)^{\Lambda_\epsilon}$
\begin{equation}\label{kmp-eq}
dP(\eta)
=\prod_{i\in\Lambda_\epsilon}
\lambda e^{-\lambda\eta(i)}d\eta(i).
\end{equation}
On the other hand when $\lambda(i)$ is not constant the model is not
time reversal invariant, the invariant state is not product, and an
explicit representation is not known (except for the case of a single
oscillator \cite{BDGJL-sips}).

We refer to  \cite{PhysRevE.88.022110} for a variant of this model in
which part of the energy is dissipated.

\subsection{Weakly asymmetric models}

We now show how to modify the stochastic models described above in
order to take into account the action of an external vector field. Let
$F:\Lambda \to \mathbb R^d$ be a vector field, describing the force
acting on the particles of the system.  When the system goes from the
configuration $\eta$ to the configuration $\sigma^{i,j}\eta$, the work
done by the force field $F$ is
\begin{equation}\label{disc-lavoro}
F_{i,j}=\int_{[i,j]} F\cdot dl
\,,
\end{equation}
where $[i,j]$ is the oriented segment from $i$ to $j$
(which has length of order $\epsilon$).
The perturbed rates are defined by
\begin{equation}\label{pert-rates}
c^F_{i,j}(\eta)=c_{i,j}(\eta)e^{F_{i,j}/2}
\,.
\end{equation}
When $|i-j|$ is of order $\epsilon$,
then the work \eqref{disc-lavoro} is of order $\epsilon$ and we have
$$
c^F_{i,j}(\eta)=c_{i,j}(\eta)\big(1+  \tfrac {F_{i,j}}{2} \big)+o(\epsilon)
\,.
$$
If $F=-\nabla H$ is a gradient vector field, then
$F_{i,j}=H(i)-H(j)$,
and equation \eqref{pert-rates}
becomes
\begin{equation}\label{pert-rates-gr}
c^F_{i,j}(\eta)=c_{i,j}(\eta)e^{ [H(i)-H(j)]/2}
\,.
\end{equation}
For the KMP model the net amount of energy flown across the bond
$(i,j)$ when the configuration $\eta$ is transformed into
$\sigma^{i,j}_p\eta$ is given by $(1-p)\eta(i)-p\eta(j)$. The
perturbed dynamics on the bond $(i,j)$ is defined by
\begin{equation}\label{pert-kmp}
\int_0^1 dp e^{\left[(1-p)\eta(i)-p\eta(j)\right]
  F_{i,j}/2 } \left[f(\sigma^{i,j}_p\eta)-f(\eta)\right] \,,
\end{equation}
where in this case $F_{i,j}$ is the work done per unit energy.

Observe that $F_{i,j}$ in \eqref{disc-lavoro} is of order
$\epsilon$. Namely, on the microscopic scale the external field is
small with the scaling parameter. This is the reason for the name
weakly asymmetric. The case in which $F_{i,j}$ in \eqref{pert-rates}
is of order one corresponds to asymmetric models. In this case the
hydrodynamics is given by hyperbolic conservation laws and not by
driven diffusive equations. We refer to \cite{MR1707314} for periodic
boundary conditions and to
\cite{MR2885612} for the case of models with reservoirs.

\subsection{Empirical density and current}
\label{emp-dc}

In order to pass from a microscopic model
to the corresponding macroscopic system,
it is convenient to introduce some intermediate quantities,
called the empirical density and the empirical current.

The empirical density associated to the configuration $\eta\in\Omega_\epsilon$
is defined as
\begin{equation}\label{emp-meas}
\rho_\epsilon(\eta;x)
=
\epsilon^d\sum_{i\in\Lambda_\epsilon}\eta(i)\delta(x-i)
\,,
\end{equation}
where $\delta(x-i)$ is the delta distribution concentrated at site $i$.
It gives a positive distribution on the domain $\Lambda$,
describing the local densities of particles.
It is equivalently defined by
\begin{equation}\label{emp-meas2}
\int_\Lambda dx\, \rho_\epsilon(\eta;x) f(x)
=
\epsilon^d\sum_{i\in\Lambda_\epsilon} \eta(i)f(i)
\,,
\end{equation}
for a continuous function $f:\,\Lambda\to\mb R$.

The empirical current is associated to a trajectory $\eta_t$, $t\in[0,T]$,
of the particle system on the configuration space.
Denote by $N^{i,j}_T$ the number of particles that jump from $i$ to
$j$ in the time interval $[0,T]$.
At the boundary, for $i\in\Lambda_\epsilon$ and
$j\in\partial\Lambda_\epsilon^o(i)$, $N^{i,j}_T$ is the number of
particles leaving the system at $i$ by jumping to $j$ (annihilation),
while $N^{j,i}_T$ is the number of particles entering the system in
$i$ jumping from the reservoir site $j$ (creation).
The difference $Q^{i,j}_T = N^{i,j}_T - N^{j,i}_T$ is the
net number of particles flowing across the oriented bond $(i,j)$ in
the time interval $[0,T]$.
The instantaneous
current $d Q^{i,j}_{t}/ dt $ is thus a sum of $\delta$--functions localized
at the jump times across the unoriented bond $\{i,j\}$ with weight $+1$,
respectively $-1$, if a particle jumps from $i$ to $j$, respectively
from $j$ to $i$.
The empirical current is defined as
\begin{equation}\label{emp-curr}
j_\epsilon(\eta;t,x)
=
\epsilon^d\sum_{\{i,j\}}
(j-i) \delta(x-i) \frac{d Q^{i,j}_t}{dt}
\,,
\end{equation}
where the sum is over unoriented bonds $\{i,j\}$
such that $|i-j|=\epsilon$.
Note indeed that the product
$(j-i) \frac{d Q^{i,j}_t}{dt}$
is symmetric with respect to the exchange of $i$ and $j$.
The empirical currents $j_\epsilon$ is a distribution on 
$\Lambda\times[0,T]$ with values in $\mb R^d$, describing the local flux of particles.
It is equivalently defined by
\begin{equation}\label{emp-curr2}
\begin{array}{l}
\displaystyle{
\int_0^T\!dt
\int_\Lambda \!dx\,
j_\epsilon(\eta;t,x)\cdot F(x,t)
} \\
\displaystyle{
=
\epsilon^d
\sum_{(i,j)}
\sum_{k=1}^{N^{i,j}_T} (j-i)\cdot F(i,\tau^{i,j}_k)
\,, }
\end{array}
\end{equation}
for a continuous vector field $F:\,\Lambda\times[0,T]\to\mb R^d$.
In the above equation $\tau^{i,j}_k$, $k=1,\dots,N^{i,j}_T$, denote the times at which particles jump from site $i$
to site $j$.
In \eqref{emp-curr2} the sum is over oriented bonds $(i,j)$ such that $|i-j|=\epsilon$.

In the one-dimensional case, recalling the definition \eqref{jvsq} of
the average total current $\mc Q_{\epsilon, T}$, by choosing $F=1$, we
get
$$
\mc Q_{\epsilon, T}
=
\frac{\epsilon^2}{T}
\sum_{i}
Q^{i,i+\epsilon}_T
\,.
$$
From the previous formula one can deduce the relationship between
$\mc Q_{\epsilon, T}$ and analogous quantities considered in
\cite{BD04,MR2054160, ADLW,ABDS13}.

\subsection{Hydrodynamic limits }
The models introduced in the previous subsections have a non trivial scaling
limit under a diffusive rescaling. Since the lattice size is $\epsilon$
this corresponds to speed up the dynamics multiplying the transition rates
by $\epsilon^{-2}$.
The basic formula we will use is
\begin{equation}\label{basic-hydro}
N^{i,j}_t=\epsilon^{-2}\int_0^tc^{i,j}(\eta_s)ds +M^{i,j}_t
\,.
\end{equation}
Formula \eqref{basic-hydro} is derived by classic
arguments in the theory of Markov processes, see e.g. \cite{MR636252}.
Given the configuration $\eta_t$ at time $t$,
the expected value of the increment $N^{i,j}_{t+dt}-N^{i,j}_t$ is $\epsilon^{-2}c^{i,j}(\eta_t)dt$.
The last term $M^{i,j}_t$ thus describes the microscopic fluctuation
(in probabilistic language, it is a martingale).
From \eqref{basic-hydro} we get
\begin{equation}\label{basic-hydro2}
Q^{i,j}_t=\epsilon^{-2}\int_0^t q^{i,j}(\eta_s)ds +\tilde M^{i,j}_t
\,,
\end{equation}
where
\begin{equation}\label{curr-ist}
q^{i,j}(\eta)=c^{i,j}(\eta)-c^{j,i}(\eta)
\end{equation}
is the mean instantaneous current across the bond $(i,j)$,
and $\tilde M^{i,j}_t$ is a fluctuation term, which plays the same role as $M^{i,j}_t$ does in \eqref{basic-hydro}.

The models we introduced in the previous sections are of gradient
type. This means that there exists a function $h(\eta)$, depending on
the configuration $\eta$ only through a finite number of lattice
sites, such that
\begin{equation}\label{grad-cond}
q^{i,j}(\eta)=h(\tau_i \eta)-h(\tau_j \eta),
\end{equation}
where, as before, $\tau_i$ denotes the shift on $\Omega_\epsilon$.

For the simple exclusion process we have $h(\eta)=\eta(0)$, while for
the zero range model we have $h(\eta)=g(\eta(0))$. The construction
for the KMP model is slightly different and $Q^{i,j}_t$ represents the
net amount of energy flowing  across the bond  $(i,j)$  in the time
interval $[0,t]$. 
The mean instantaneous current appearing in formula
\eqref{basic-hydro2} in this case becomes
\begin{equation}\label{kmp-curr}
  q^{i,j}(\eta)=\int_0^1 dp\Big(\left[\sigma^{i,j}_p\eta\right](j)+ 
  \left[\sigma^{j,i}_p\eta\right](j)-2\eta(j)\Big)
  \,.
\end{equation}
From \eqref{kmp-curr} we deduce that \eqref{grad-cond} still holds with $h(\eta)=\eta(0)$.

In order to discuss the hydrodynamic behavior, we first observe that
the definition of $Q^{i,j}_t$ implies the discrete continuity equation
\begin{equation}\label{disc-cont}
\eta_t(i)-\eta_0(i)=-\sum_{j\,:\, |j-i|=\epsilon}Q^{i,j}_t.
\end{equation}
In view of \eqref{basic-hydro2}, we can rewrite this equation as
\begin{equation}
\label{disc-cont-ist}
\begin{split}
\eta_t(i)-\eta_0(i)
=& -\epsilon^{-2}\sum_{j\,:\, |j-i|=\epsilon}\int_0^t
ds\,q^{i,j}(\eta_s)\\
& + \mathrm{ fluctuation}
\end{split}
\end{equation}
Consider now a test function $\psi\colon \Lambda \to \mathbb R$.
By integrating \eqref{disc-cont-ist} in space we deduce
\begin{equation}
\label{step-idro}
\begin{split}
&\int_\Lambda\!dx\,  \psi(x)  \rho_\epsilon(\eta_t;x)
-\int_\Lambda\!dx\,  \psi(x)  \rho_\epsilon(\eta_0;x)
\\
&=\epsilon^d \int_0^t\!ds \sum_i h(\tau_i\eta_s)\left[\epsilon^{-2}\sum_{j\,:\,|j-i|=
\epsilon}\left(\psi(j)-\psi(i)\right)\right]
\\
&\; +o\left(1\right)
\end{split}
\end{equation}
where we used \eqref{grad-cond} and a discrete integration by parts.
The term $o(1)$ represent the space integral of  the fluctuation in
\eqref{disc-cont-ist}.
It vanishes as $\epsilon \to 0$ as the random variables $M^{i,j}_t$
have mean zero and are almost independent for different bonds.
Observe that the term inside square brackets in \eqref{step-idro} is a discrete
version of the Laplacian of $\psi$; namely,
\begin{equation}\label{appr-lap}
\epsilon^{-2}\sum_{j\,:\,|j-i|=
\epsilon}\left(\psi(j)-\psi(i)\right)=\Delta \psi(i)+o\left(1\right).
\end{equation}

As already mentioned, both for the exclusion and the KMP processes
condition \eqref{grad-cond} holds with $h(\eta) = \eta(0)$. In these cases,
by taking the limit $\epsilon \to 0$ and denoting by $\rho(t,x)$ the
limit of $\rho_\epsilon(\eta_t,x)$, \eqref{step-idro} yields
directly
\begin{equation*}
  \begin{split}
  &\int_\Lambda\!dx\, \rho(t,x)\,\psi(x)
  -  \int_\Lambda\!dx\, \rho(0,x)\,\psi(x)
  \\
  &\quad =\int_0^t \!ds\int_\Lambda\!dx\,
  \rho(s,x)\,\Delta\psi(x)
  \end{split}
\end{equation*}
which is the weak formulation of the heat equation
\begin{equation*}
  \partial_t \rho = \Delta \rho.
\end{equation*}
This is the hydrodynamic equation for both the simple exclusion
and the KMP processes. Namely, for these models $D(\rho)=1$.

For the zero range process we have instead $h(\eta)=g(\eta(0))$.
Therefore, \eqref{step-idro} yields directly a closed equation for the density
only in the case $g(k)=k$, corresponding to independent particles.
In order to derive the hydrodynamic equation we need a 
mathematical formulation of the local equilibrium assumption.
The basic idea is the following.
Fix a point $i\in\Lambda_\epsilon$ and consider a macroscopically small,
but microscopically large, neighborhood $B(i)$ of $i$.
Since the total number of particles is locally conserved,
on the macroscopic time scale, the system in $B(i)$
is essentially in the homogeneous equilibrium state corresponding to the
average density in $B(i)$.
Therefore, we can replace $h(\eta)$ with the corresponding ensemble average.

In order to compute this average, we describe the equilibrium states
$P^\rho$ of the zero range process.
These are product distributions of the form \eqref{zr-eq} with
$\phi(i)$ constant and equal to the solution of
\begin{equation}\label{zr-equation}
\rho=\phi\frac{Z'(\phi)}{Z(\phi)}
\,.
\end{equation}
Let
\begin{equation*}
  \Phi(\rho) =
E_{P^\rho}(g(\eta(0)))
\,.
\end{equation*}

In view of the previous discussion,
in \eqref{step-idro} we can replace
\begin{equation}\label{idro-somma}
\epsilon^d\sum_i h(\tau_i\eta_t)\Delta \psi(i)
\,,
\end{equation}
by
\begin{equation}\label{idro-somma2}
\epsilon^d\sum_i \Phi
\left(
\frac1{|B(i)|}\int_{B(i)}\rho_\epsilon(\eta_t;x)dx
\right)
\Delta \psi(i)
\,,
\end{equation}
where $|B(i)|$ denotes the volume of $B(i)$.
We refer to \cite{MR1707314,ippo} for the (quite technical) proof of this statement.

Since $B(i)$ is macroscopically infinitesimal, by taking the limit
$\epsilon\to0$ in \eqref{step-idro}, we derive the weak formulation of
the non-linear diffusion equation
\begin{equation}\label{idro-limite-fin}
\partial_t \rho=\Delta \Phi(\rho)
\,,
\end{equation}
which is the hydrodynamic equation for the zero range process.
We conclude that for the zero range process $D(\rho)=\Phi'(\rho)$.

\medskip
We discuss now the hydrodynamic scaling limit of the empirical
current.
In order to obtain a microscopic expression for the mobility we
consider the case of weakly asymmetric models.
Recalling \eqref{disc-lavoro}, for such models
\eqref{basic-hydro2} holds with \eqref{curr-ist} replaced by
\begin{equation}
\label{curr-ist-w}
\begin{split}
  & q^{i,j}(\eta) = c^F_{i,j}(\eta)-c^F_{j,i}(\eta)
\\
& \quad =  c_{i,j}(\eta)-c_{j,i}(\eta)
+ \frac 12 \big[ c_{i,j}(\eta)+c_{j,i}(\eta)\big]F_{i,j}+
 o\left(\epsilon\right)
\\
& \quad =  h(\tau_i \eta) -h(\tau_j \eta)
+ \frac 12 \big[ c_{i,j}(\eta)+c_{j,i}(\eta)\big]F_{i,j}+
 o\left(\epsilon\right)
\end{split}
\end{equation}
where we used the gradient condition \eqref{grad-cond} for the rates
without external field.

For the KMP model, using \eqref{pert-kmp} and \eqref{kmp-curr}, we
instead get
\begin{equation}\label{per-kmp-curr}
q^{i,j}(\eta)=\eta(i)-\eta(j)+\frac
13\Big(\eta^2(i)+\eta^2(j)-\eta(i)\eta(j)\Big)F_{i,j}+o\left(\epsilon\right)
\end{equation}

Let $G\colon [0,T]\times\Lambda  \to \mathbb R^d$ be a test vector field.
Recalling the definition of the empirical current \eqref{emp-curr}, by
using \eqref{curr-ist-w} and writing the sum over unoriented bonds as
$1/2$ the sum over oriented bonds we get
\begin{equation}
\label{idro-curr}
\begin{split}
  &\int_0^t\!ds \int_\Lambda \!dx\, j_\epsilon(\eta;s,x)\cdot G(s,x) \\
  &\;
   = \epsilon^d \int_0^t\!\! ds\, \sum_i h(\tau_i\eta_s)
 \Big[ \frac {\epsilon^{-1}} {2}
 \sum_{j\,:\,|j-i|=\epsilon}
 \left( G (s,i)- G (s,j) \right) \Big]\\
&\;
+ \frac {\epsilon^{d-1}}{2}  \int_0^t\!\! ds\, \sum_i
\sum_{j\,:\,|j-i|=\epsilon}
\frac 12 \big[ c_{i,j} (\eta)+c_{j,i}(\eta)\big]
F_{i,j} G(s,i)
\\
&\; + o(1)
\end{split}
\end{equation}
where the term $o(1)$ is due to the fluctuation in \eqref{basic-hydro2}.

Observe that
\begin{equation}\label{disc-1}
\frac {\epsilon^{-1}}{2}
\sum_{j\,:\,|j-i|=\epsilon}\left(G (i)-G (j) \right)=\nabla \cdot G(i)+o(1)
\end{equation}
and, since $F_{i,j}$ is of order $\epsilon$,
\begin{equation}\label{disc-2}
\frac {\epsilon^{-1}}{2}
\sum_{j\,:\, |j-i|=\epsilon} F_{i,j}  G(i)
=  F(i)\cdot G(i)+o(1).
\end{equation}
We now define in general
\begin{equation}
  \label{trancoeff}
  \Phi(\rho) = E_{P^{\rho}} (h) ,\qquad
  \chi(\rho) = \frac 12  E_{P^\rho}\left[ c_{i,j} +c_{j,i} \right]
\end{equation}
where $P^\rho$ denotes the homogeneous equilibrium state with
density $\rho$. By the same local equilibrium argument discussed
before, we get that the right hand side of \eqref{idro-curr} converges
as $\epsilon \to 0$ to
\begin{equation}\label{corrente-i}
\int_0^t\!ds \int_\Lambda\!dx\,  \Phi(\rho)\nabla \cdot G
+\int_0^t\!ds\int_\Lambda\!dx\, \chi(\rho) F\cdot G
\end{equation}
which is the weak form of
\begin{equation*}
  J(\rho) = -\nabla \Phi (\rho) + \chi(\rho) F
  = -\Phi' (\rho)\nabla \rho  + \chi(\rho) F
\end{equation*}
that is \eqref{2.2} with $D(\rho)=\Phi'(\rho)$ and $\chi (\rho)$  as
in \eqref{trancoeff}.

In the case of the exclusion process,
$c_{i,j}(\eta)+c_{j,i}(\eta)=\eta(i)(1-\eta(j))+\eta(j)(1-\eta(i))$ so
that $\chi(\rho)=\rho(1-\rho)$. In the case of the zero range we have
$c_{i,j}(\eta)+c_{j,i}(\eta)=g(\eta(i))+g(\eta(j))$ so that
$\chi(\rho)=\Phi(\rho)$.  For the KMP model the equilibrium state is a
product of exponential distribution so that, using \eqref{per-kmp-curr},
we can deduce $\chi(\rho)=\rho^2$.

When the condition \eqref{grad-cond} does not hold, the model is
called \emph{non gradient}. In this case the deduction of the
hydrodynamic equation is more complicated. Referring to
\cite{MR1707314}
for the detail of this derivation, we mention
that in the general case the diffusion coefficient is linked to the
microscopic dynamics by a Green-Kubo formula, see
\cite[II.2.2]{ippo}.

\smallskip
We discussed the hydrodynamic limit without considering the boundary
terms.  At the boundary there is a Glauber dynamics speeded up by a
factor $\epsilon^{-2}$ that keeps fixed the density at a value
determined by the local chemical potential of the external reservoirs
\cite{ELS}.

\subsection{Large fluctuations}
\label{ld-slg}

In this section we derive, for models satisfying the gradient
condition \eqref{grad-cond}, the fundamental formula \eqref{r02}.
For simplicity, we restrict to models without external field.

We need an expression for the relative distribution of two stochastic
particle systems.  Since these processes can be constructed using
independent Poisson processes, we start by giving the
relative distribution of two Poisson processes.
More precisely, consider two Poisson processes with parameters
depending on the value $N_t$ of the process. The first one has
parameter $c(N_t)$ and the second one is obtained from the first with
a time dependent perturbation and has parameter $c(N_t)e^{F(t)/2}$. Then
the ratio between the two ensembles on the time window $[0,t]$ is,
see \cite[App.~A]{BDGJL02},
\begin{equation}\label{RN-poi}
  \begin{split}
   \frac{d\mathbb P}{d\mathbb P^F}\Big|_{[0,t]}
   =&
 \exp\Big\{\int_0^t\left[c(N_s)e^{F(s)/2}-c(N_s)\right]
     \\
     &\quad - \frac 12 \sum_kF(\tau_k)\Big\},
  \end{split}
\end{equation}
where the $\tau_k$ are the jump times.

Consider a macroscopic fluctuation $(\rho(s),j(s))$ in the time window
$[0,t]$ of the empirical density and current satisfying the continuity
equation.  In order to estimate the probability of this fluctuation we
introduce an external field $F$ such that $(\rho(s), j(s))$ becomes
typical, that is its probability is close to one as $\epsilon\to 0$.
The external field $F$ that we need to introduce is obtained solving
the equation
\begin{equation}\label{eq-per-F}
-D(\rho)\nabla \rho+\chi(\rho)F=j.
\end{equation}
The ratio between the distributions of the original particle system and the one
obtained with the perturbation $F$ in the time window $[0,t]$
can be computed by using \eqref{RN-poi}.  Recalling
\eqref{disc-lavoro}, we get
\begin{equation}\label{RN-part}
  \begin{split}
    & \frac{d\mathbb P }{d\mathbb P^F }\Big|_{[0,t]} =
    \exp\Big\{ \int_0^t\!ds \sum_{|i-j|=\epsilon}\epsilon^{-2}c_{i,j}(\eta_s)
        \left(e^{F_{i,j}(s)/2 }-1\right)
        \\
  &\qquad \qquad
  - \frac 12
  \sum_{|i-j|=\epsilon} \sum_{k=1}^{N^{i,j}_t}F_{i,j}(\tau^{i,j}_k)
  \Big\}.
  \end{split}
\end{equation}
Recalling \eqref{emp-curr2}, the second term at the exponent above is
equal to $\epsilon^{-d} \frac 12 \int_0^t\!ds
\int_{\Lambda}\!dx\,  j_\epsilon \cdot F$. By expanding up to
second order $e^{F_{i,j}(s)/2}$, using the antisymmetry of $F_{i,j}(s)$ with
respect to $i,j$ and  the gradient condition
\eqref{grad-cond}, we rewrite the first term as
\begin{equation*}
  \begin{split}
    & \epsilon^{-d} \int_0^t\!ds\, \epsilon^{d-2} \sum_i
    \sum_{j\,:\,|j-i|=\epsilon}
    c_{i,j}(\eta_s)
    \Big[ \frac 12 F_{i,j}(s) +  \frac 18  F_{i,j}(s)^2 \Big]
    \\
    & = \int_0^t\!ds\, \sum_i
    \Big\{ h(\tau_i \eta_s )
    \frac {\epsilon^{-2}}{4}  \sum_{j\,:\,|j-i|=\epsilon}
    \big[ F_{i,j}(s) -F_{j,i}(s) \big]
    \\
    & +  {\epsilon^{-2}}  \sum_{j\,:\,|j-i|=\epsilon}
    \frac 12 \big[c_{i,j} (\eta_s) +c_{j,i} (\eta_s) \big]
    \frac 18 F_{i,j}(s)^2 \Big\}
    \\
    & \approx
    \epsilon^{-d} \int_0^t\!ds\int_\Lambda\!dx \, \Big\{
    \frac 12 \Phi(\rho_\epsilon ) \nabla \cdot F
    + \frac 14 F \cdot \chi(\rho_\epsilon) F \Big\}
 \end{split}
\end{equation*}
where we used local equilibrium as in the previous section, see in
particular \eqref{trancoeff} for the microscopic definition of the
transport coefficients. Since $F$ satisfies \eqref{eq-per-F} we
finally deduce that
\begin{equation}
\label{rncirca}
   \begin{split}
    \frac{d\mathbb P|_{[0,t]}}{d\mathbb P^F|_{[0,t]}}
    &\approx  \exp\Big\{
    \epsilon^{-d}
    \int_0^t\!ds \int_\Lambda\!dx \, \Big[
    \frac 12 (j-j_\epsilon) \cdot F
    \\
    &\qquad
        - \frac 14 F \cdot \chi(\rho_\epsilon) F \Big] \Big\}.
\end{split}
\end{equation}

We now estimate the probability of the fluctuation $(\rho(s),j(s))$,
$s\in [0,t]$. We write
\begin{equation}
\label{change-meas}
\begin{split}
 \mathbb P\Big(\left(\rho_\epsilon,j_\epsilon\right)\sim(\rho,j)\Big)
= \mathbb E^F\left(\frac{d\mathbb P|_{[0,t]}}{d\mathbb P^F|_{[0,t]}}
  {\boldmath 1}_{ \left(\rho_\epsilon,j_\epsilon\right)\sim (\rho,j)}  \right),
\end{split}
\end{equation}
where $\boldmath 1_A$ denotes the indicator of the set $A$.
By using \eqref{rncirca} and the fact that under the perturbed
distribution $(\rho_\epsilon,j_\epsilon)\approx (\rho,j)$ we finally get
\begin{equation}
  \label{kovlike}
  \begin{split}
   & \mathbb P\Big(\left(\rho_\epsilon,j_\epsilon\right)\sim(\rho,j)\Big)
   \\
&\qquad \approx \exp\Big\{
   - \epsilon^{-d} \frac 14
    \int_0^t\!ds \int_\Lambda\!dx \,
    F \cdot \chi(\rho) F \Big] \Big\}
  \end{split}
\end{equation}
which, in view of \eqref{eq-per-F}, concludes the proof of the
fundamental formula.

In the case in which one considers only the fluctuations of the
density, as in Section~\ref{s:df}, formula \eqref{kovlike} has been
first obtained in \cite{MR978701} for the exclusion process.

As for the hydrodynamic limit, we discussed the large deviation
asymptotic without considering the boundary terms.  At the boundary
there are independent Glauber dynamics speeded up by $\epsilon^{-2}$
so that the asymptotic probability to observe a density fluctuation on
a region $\Gamma\subseteq \partial \Lambda$ of the boundary is of the
order $e^{-\epsilon^{-(d+1)}\left|\Gamma\right|}$ which is much
smaller than $e^{-\epsilon^{-d}}$.  The fluctuations whose
probability is exponentially small in $\epsilon^{-d}$ have therefore the values
of the density at the boundary fixed by the reservoirs.

\subsection{Quasi-potential and relative entropy}

We consider two states of a system and establish a connection between
the quasi-potential $V$ and the relative entropy between the
corresponding ensembles.

This connection is readily established in equilibrium.  For
simplicity, consider the case of lattice gases without external field
and constant chemical potential, i.e.\ the case of homogeneous
equilibrium states. The Gibbs distribution on the volume $\Lambda$ is
\begin{equation}
\label{gibbs}
P^\lambda_\Lambda (\eta) = \frac {1}{Z_\Lambda (\lambda)}
\exp\Big\{ -  H_\Lambda (\eta) +
  \lambda \sum_{i\in\Lambda} \eta(i) \Big\},
\end{equation}
where $H_\Lambda (\eta)$ is the energy of the
configuration $\eta$, $\lambda$ is the chemical potential,
and $Z_\Lambda (\lambda)$ is the grand-canonical
partition function. Recall that we have included the dependence on the
temperature in the Hamiltonian. According to the standard postulates of
statistical mechanics, the pressure $p$ is given by
\begin{equation}
  \label{press}
p(\lambda) = \lim_{\Lambda \uparrow {\boldmath Z}^d}
\frac 1{|\Lambda|}
\log Z_\Lambda(\lambda),
\end{equation}
and the free energy per unit volume $f$ is obtained as the Legendre
transform of $p$,
\begin{equation*}
f(\rho) =\sup_\lambda \big\{ \rho \lambda - p(\lambda) \big\}.
\end{equation*}

The \emph{relative entropy} $S(\nu|\mu)$ of the probability $\nu$ with
respect to $\mu$ is defined by
\begin{equation}
\label{relen}
S(\nu|\mu) = \int \! dP \: \frac{d\nu}{d\mu} \log
\frac{d\nu}{d\mu}.
\end{equation}
Observe that  if we choose $\mu$ as the uniform probability then
$S(\nu|\mu)$ is the Gibbs entropy.

Fix two chemical potentials $\lambda_0$ and $\lambda_1$. We claim that
\begin{equation}
\label{ler}
\lim_{\Lambda \uparrow {\boldmath Z}^d } \, \frac 1{|\Lambda|}
S\big( P_{\Lambda}^{\lambda_0} \big|P_{\Lambda}^{\lambda_1}\big)
= \big[
f(\bar\rho_0) - f(\bar\rho_1) - \lambda_1 (\bar\rho_0-\bar\rho_1)
\big],
\end{equation}
where $\bar\rho_0$ and $\bar\rho_1$ are the densities associated to
$\lambda_0$ and $\lambda_1$.  In view of \eqref{Veq} this
implies that in the thermodynamic limit the relative
entropy per unit volume is proportional to the function $
V_{\lambda_1,0}(\bar\rho_0)$ per unit  volume.  To prove \eqref{ler},
observe that in view of \eqref{relen} and the Gibbsian form
\eqref{gibbs},
\begin{equation*}
  \begin{split}
    &\frac 1 {|\Lambda|} \, S \big( P_{\Lambda}^{\lambda_0} \big|
    P_{\Lambda}^{\lambda_1} \big)
    \\
    &\;
    = \frac 1{|\Lambda|} \log \frac{
      Z_{\Lambda}(\lambda_1)} {Z_\Lambda(\lambda_0)} +
    (\lambda_0 -\lambda_1) \sum_{\eta} P_{\Lambda}^{\lambda_0}
    (\eta) \: \frac 1{|\Lambda|} \sum_{i\in \Lambda} \eta(i).
  \end{split}
\end{equation*}
By definition of the pressure, the first term converges to
$ [ p(\lambda_1) - p(\lambda_0)]$, while the second one converges to
$ (\lambda_0-\lambda_1)\bar\rho_0$.  The identity \eqref{ler} then
follows by Legendre duality.

The relationship \eqref{ler} between the relative entropy and the
quasi-potential extends, exactly with the same formulation, to
non-equilibrium states. Recall that $\Lambda\subset \mb R^d$ is the
macroscopic volume, and denote by $\Lambda_\epsilon$ the corresponding
subset of the lattice with spacing $\epsilon$, so that the number of
sites in $\Lambda_\epsilon$ is approximately $\epsilon^{-d}
|\Lambda|$.  Given the chemical potential $\lambda$ of the boundary
reservoirs and the external field $E$, let
$P^{\lambda,E}_{\Lambda_\epsilon}$ be the stationary distribution of a
driven stochastic lattice gas.

Given $(\lambda_0,E_0)$ and $(\lambda_1,E_1)$, we claim that
\begin{equation}
\label{wp=rel}
\lim_{\epsilon\to 0} \epsilon^{d}
\, S\big( P^{\lambda_0,E_0}_{\Lambda_\epsilon} \big|
P^{\lambda_1,E_1}_{\Lambda_\epsilon} 
\big) =  V_{\lambda_1,E_1}(\bar\rho_0),
\end{equation}
where  $\bar\rho_0$ is the stationary profile
corresponding to $(\lambda_0,E_0)$.

In the case of the zero-range processes, as discussed before, the
stationary ensemble has an explicit form. It is thus possible to prove
\eqref{wp=rel} by direct computation as in the equilibrium case.  For
other models, \eqref{wp=rel} has been derived in \cite{MR2999559}
under the assumptions that the stationary ensembles satisfy a strong
form of local equilibrium that holds for the boundary driven
symmetric simple exclusion process \cite{MR2740401, MR3162546}.  For
this model, in the special situation in which
$P^{\lambda_1,E_1}_{\Lambda_\epsilon}$ is an equilibrium ensemble, the
finite size corrections to the identity \eqref{wp=rel} have been
analyzed in \cite{MR2311899}.

\medskip
The connections between equilibrium statistical mechanics and classical
thermodynamics can be expressed in many ways. 
The argument of this section for equilibrium states shows that the identity
\eqref{ler} between the relative entropy per unit  volume and the
availability is another possibility. 
In view of \eqref{wp=rel}, if we take such a relationship as a general 
statement, it applies also to non-equilibrium states provided 
we replace the availability with the quasi-potential.

\section{Conclusions and outlook}

The MFT provides a unified treatment of
the thermodynamics of driven diffusive systems and their
fluctuations.
Its formulation has required an adroit balancing of thermodynamic and
statistical mechanics arguments.
The outcome is a purely macroscopic theory which can be used
as a phenomenological description requiring as input only the
transport coefficients which are measurable.
New variational principles are naturally formulated within the
MFT.
These principles
allow to solve concrete problems
as shown by the various applications of the theory discussed in this
article.

While the MFT has been developed for driven diffusive systems, the
case of hyperbolic systems can be recovered by considering the formal
limit of strong driving field.

The early derivation and development of the Macroscopic Fluctuation
Theory benefited from the explicit microscopic computations in
\cite{MR762032,DLS,DLS1}.
In particular, the result in \cite{DLS,DLS1} for the
boundary driven symmetric simple exclusion process has been obtained,
in a rather straightforward way, from the Hamilton-Jacobi equation for
quasi-potential in \cite{BDGJL02}.  It is remarkable that such a
perfect agreement has been always found between the results obtained
by the MFT and by exact microscopic
computations.

As the Boltzmann-Einstein formula, the MFT
provides an interface between thermodynamics and the underlying
microscopic world. It can thus be used in different ways.  At the
level of continuum mechanics it introduces, for non-equilibrium
states, the orthogonal splitting of the current that is realized
through the introduction of the quasi-potential.  With respect to
Onsager theory, this is a further step in the formulation of a
non-equilibrium thermodynamics for stationary states.  
At the level of microscopic ensembles, the fundamental formula gives
the asymptotic probability for fluctuations of the density and
current.  It has been used to predict the asymptotics of the current
cumulants \cite{BD04,MR2054160,ABDS13}.  and - quite surprisingly -
also their finite size corrections \cite{ADLW}.  The analysis of the
macroscopic variational principles introduced in the MFT has revealed
the occurrence of phase transitions peculiar of non-equilibrium
states \cite{lpt,noiprlcurrent,bdcurrpt}. Among the most recent
developments we mention \cite{kra0,kra1,merkra}

The fundamental formula is not restricted to the stationary
ensembles, it has indeed been applied also to non-stationary infinite
systems \cite{DG1,DG2,1742-5468-2013-12-P12011,PhysRevE.89.010101}.

The MFT so far has been supported by the analysis of stochastic
lattice gases and by numerical simulations. It is clear that the next
stage should be an experimental test of its predictions. This appears
a rather challenging task as fluctuations of thermodynamic system are
very improbable and the experimenter has to circumvent this
difficulty. One possibility is to rely on an active interpretation of
the fluctuation formulas: among the external fields that produce the
given fluctuation, choose the one which minimize the energy
dissipated.  Another possibility is to measure higher order
correlations of the thermodynamic variables in the stationary regime.
While the two-point correlations, as already mentioned, have been
measured \cite{DKS} and correspond to not too large (Gaussian)
fluctuations, for higher order correlation the MFT gives new
predictions.

\medskip

A fundamental problem in non-equilibrium physics is the turbulent
behavior of viscous fluids. Natural approaches to the problem of fully
developed turbulence lie within the broad topics of statistical
physics. Most of the attempts in this direction borrow basic concepts
from dynamical systems and equilibrium statistical mechanics.
Recently, Ruelle proposed a view of turbulence which appear to fit
well within the class of systems analyzed by the MFT \cite{ruelle}.
The implementation of Ruelle's ideas is an important problem for the
future.

Most challenging potential applications of the Macroscopic Fluctuation
Theory may lie in biology. Indeed, many of the processes in living
beings can be described as diffusive systems in stationary or
quasi stationary states depending on the time scales considered. On
the other hand, the understanding of the full biological significance
of these physical processes would require a formulation in
mathematical language of the main properties of living systems. 
In the words of a well-known mathematician \cite{grom},
\begin{quote}
\textit{``You feel there must be a new world of mathematical structures
  shadowing what we see in Life, a new language we do not know yet,
  something in the spirit of the language of calculus we use when
  describing physical systems.''}
\end{quote}

\begin{acknowledgments}
  The MFT includes several facets to which different groups, besides
  the present authors, have contributed.  We wish to mention in
  particular the work of T.\ Bodineau, B.\ Derrida, J.L.\ Lebowitz,
  E. R. Speer and their collaborators which has been a constant source
  of inspiration for us.  In a broader perspective, our understanding
  of non-equilibrium stationary states benefited from fruitful
  discussions with C.\ Bahadoran, G.\ Basile, C.\ Bernardin, A.\ De
  Masi, A.\ Faggionato, G.\ Gallavotti, T.\ Komatsu, C.\ Maes,
  K. Mallick, M.\ Mariani, N.\ Nakagawa, S.\ Olla, E.\ Presutti, S.\
  Sasa, G.\ Sch\"utz, H.\ Spohn, H.\ Tasaki, C.\ Toninelli, S.R.S.\
  Varadhan, M.E.\ Vares, H.T.\ Yau.  To all these colleagues and
  friends we express our deep gratitude.  We finally thank the
  referees for comments and suggestions that helped to improve the
  review.
\end{acknowledgments}

\end{document}